\documentclass[]{aastex631}
\usepackage{amsmath}
\usepackage{graphicx}
\usepackage{url}
\usepackage{multirow}
\usepackage{ragged2e}
\usepackage{booktabs}
\usepackage{subfigure}
\usepackage{natbib}

\newcommand{\HeII}{He\,{\sc ii}}

\begin{document}
\defcitealias{2024MNRAS.529..499L}{L24}
\defcitealias{2022ApJ...925...70Z}{Z22}
\title{A Morphological Study on AGN-host Dwarf Galaxies}
\author[0009-0001-3432-480X]{Jie Tian}
\affiliation{Yunnan Observatories, Chinese Academy of Sciences, Kunming 650216, People's Republic of China}
\affiliation{University of Chinese Academy of Sciences, Beijing 100049, People's Republic of China}
\author[0000-0002-9128-818X]{Yinghe Zhao}
\affiliation{Yunnan Observatories, Chinese Academy of Sciences, Kunming 650216, People's Republic of China}
\affiliation{Key Laboratory of Radio Astronomy and Technology (Chinese Academy of Sciences), A20 Datun Road, Chaoyang District, Beijing, 100101, People's Republic of China}
\author[0009-0009-4030-5208]{Xiejin Li}
\affiliation{Yunnan Observatories, Chinese Academy of Sciences, Kunming 650216, People's Republic of China}
\affiliation{University of Chinese Academy of Sciences, Beijing 100049, People's Republic of China}
\author{Jinming Bai}
\affiliation{Yunnan Observatories, Chinese Academy of Sciences, Kunming 650216, People's Republic of China}
\correspondingauthor{Yinghe Zhao}
\email{zhaoyinghe@ynao.ac.cn}

\begin{abstract}
The study of morphology in galaxies offers a convenient and quantitative method to measure the shapes and characteristics of galaxy light distribution that reflect the evolutionary history. For AGN-host dwarf galaxies, however, there is a lack of detailed studies on their morphologies. In this work, we compile a relatively large sample ($\sim$400 members) of local AGN-host dwarf ($M_{\star}\leq10\sp{9.5}~$M$_{\odot}$ and $z<0.055$) galaxies selected via various methods. We use the \textit{grz} bands images from DESI DR10 and the Python package \texttt{statmorph} to measure non-parametric coefficients. We also carry out visual inspection with the assistance of deep learning to classify these galaxies into early-type (ETGs), late-type (LTGs) galaxies, and mergers, and find that about 37\%, 44\%, and 13\% of the total sample sources are ETGs, LTGs, and mergers, respectively. In comparison to normal dwarf galaxies, AGN-host dwarfs have a higher probability to be LTGs, and a lower merger rate, indicating that mergers/interactions are not the primary driver of AGN activities. Among the subsamples selected with different methods, the BPT sample has the highest fraction of ETGs, the variability sample consists of the largest fraction of LTGs, and the mid-IR sample contains the most mergers. 
\end{abstract}

\section{INTRODUCTION} \label{sec:intro}
Active galactic nucleus (AGN) was previously believed to primarily arise from galaxies harboring supermassive black holes, formed through dynamic interactions or merging between galaxies \citep{2005Natur.433..604D,2005MNRAS.361..776S}. It was anticipated that these AGN-host galaxies would exhibit more irregular and asymmetric structures, such as tidal tails or other direct evidence of merger or interaction, compared to normal galaxies. Evidence suggests that the most luminous AGNs have experienced significant recent mergers \citep{2008ApJ...674...80U,2012ApJ...758L..39T}. However, subsequent research failed to provide evidence supporting this relationship in less luminous AGNs \citep{2005ApJ...627L..97G,2011ApJ...726...57C,2014MNRAS.439.3342V,2017MNRAS.466..812V}. The confusing mechanism underlying AGN may depend on its level of activity \citep{2009ApJ...694..599H,2012ApJ...751...72D}. The most active quasars only come from merging galaxies or other external events, whereas less luminous AGNs may be sustained by internal secular procession \citep{2008ARA&A..46..475H,2009ApJ...699..638H}. 

The morphology of a galaxy can give critical insights into its evolution and structural characteristics, which may also reveal the interaction of galaxies with their environments and internal perturbation, and thus has also been the subject of extensive study on AGN-host galaxies, especially for massive objects  (e.g., \citealt{2014ApJ...784L...9F,2017MNRAS.466..812V,2017MNRAS.470..755H,2022MNRAS.514..607G,2023ApJ...946L..14K}). For example, \citet[][hereafter Z22]{2022ApJ...925...70Z}, found that nuclear activity is not mainly driven by mergers and interactions, and the level of AGN activity does not correlate with asymmetry even among the major merger candidates after investigating the optical morphological asymmetry for a large sample of nearby AGN-host massive ($M_\star>10\sp{10}~$M$_\odot$) galaxies. For sources at higher redshifts, \cite{2014MNRAS.439.3342V} investigated the relation between morphological properties of AGN-hosts at $z\sim0.7$ and luminosities of AGN and host galaxy. They suggested that major mergers either exert a minor influence on the triggering of AGN within the examined luminosity range or that the time delays are too long for merger features to remain detectable. Furthermore, the study on a sample of $\sim$80 AGN-hosts ($\sim$90\% having $M_\star>10\sp{9.5}~$M$_\odot$; \citealt{2022ApJ...925..157Z}) at $z \lesssim 4.26$ demonstrates that these galaxies do not show clear morphological preference, but the merger rate increases with AGN luminosity. 

On the one hand, however, little attention has been paid to study the morphology of AGN-host dwarf galaxies (i.e., stellar mass comparable to the Large Magellanic Clouds, $M_{\star} \leq 3\times10\sp{9}~$M$_{\odot}$; \citealt{2013ApJ...775..116R}) until recent years (e.g., \citealt{2019MNRAS.489L..12K, 2021ApJ...911..134K, 2022MNRAS.511.4109D, 2024MNRAS.532..613B}), which might be due to the limited number of sources that have photometric observations deep enough for morphological analysis. Another reason is that there were fewer observations available (in a statistical manner) of dwarfs using a spatially resolved spectroscopy approach \citep{2024MNRAS.528.5252M}, where the AGN signature could be identified more easily. On the other hand, low-mass dwarf galaxies, which constitute the largest population of galaxies in the present-day universe  (e.g., \citealt{1993MNRAS.264..201K}) and serve as the building blocks from which more massive systems are formed through merging \citep{1993MNRAS.264..201K}, could be the best place to search for low-mass black holes (BHs) according to the correlation between galaxy and BH masses (e.g., \citealt{2013ARA&A..51..511K}), and they may also host BHs similar to the first seed BHs due to their relatively quiescent merger histories (\citealp{2011ApJ...742...13B}). Combining this with the fact that dwarfs are generally metal-poor \citep{1998ARA&A..36..435M}, these galaxies are allowing us to study AGN triggering and feedback (e.g., \citealt{2020ApJ...903...58C,2021RAA....21..204C,2024RAA....24f5006L}) in a near-primordial environment and at the low-luminosity end of AGNs. 

For normal dwarf galaxies, there exist numerous studies on their morphologies in the very local Universe (e.g., see \citealp[][hereafter L24]{2024MNRAS.529..499L}, and references therein). \cite{2003ApJS..147....1C} showed that, unlike their massive counterparts, typical morphological classes (e.g., early-type (ETGs) vs late-type (LTGs)) of dwarf galaxies are indistinguishable from each other using light concentration. \citetalias{2024MNRAS.529..499L} further suggested that it is considerably less effective by employing non-parametric coefficients to separate the types of dwarf galaxies than their massive counterparts, which may be caused by the different evolutionary behavior of dwarf ETGs compared to massive ones (\citealt{2023MNRAS.520.2109L}). By investigating a complete sample up to $z\sim0.08$ via visual inspection of extremely deep optical images in low-density environments, \citetalias{2024MNRAS.529..499L} revealed that about 43\%, 45\%, and 10\% of dwarfs are ETGs, LTGs, and featureless class, respectively.

For AGN-host dwarfs, a series of work via visual inspection of deep optical images (\citealt{2019MNRAS.489L..12K, 2022MNRAS.511.4109D, 2024MNRAS.532..613B}) found that the fraction of AGN-hosts ($0.1<z<0.5$) with interacting/merging sign ($f_\mathrm{merger}=6.4^{+1.3}_{-1.3}\%, 6.4^{+3.9}_{-1.8}\%$ and $14^{+3}_{-3}\%$ respectively) is indistinguishable from the controls, though the samples were selected with different methods (infrared color, radio emission and SED fitting, respectively) and differ in size by a factor of up to $\sim$2.5. These authors suggested that interactions/mergers do not play a significant role in triggering AGN activity in their samples. Using a smaller sample of 41 nearby sources ($z<0.055$) selected with optical emission line ratios (\citealt{2013ApJ...775..116R}) and performing 2D surface brightness fitting to the Hubble Space Telescope near-infrared images, \cite{2021ApJ...911..134K} found a wide range of morphologies of these AGN-host dwarfs, including LTGs with regular (29/41) and irregular (4/41) appearance, ETGs (6/41) and interactions/mergers (2/41), indicating a similar $f_\mathrm{merger}$ to those in \cite{2019MNRAS.489L..12K} and \cite{2022MNRAS.511.4109D}.

However, no single AGN selection technique is complete due to the diversity of the AGN population. In order to reduce/overcome the influence of selection effect, and to further investigate the morphological properties of AGN-host dwarf galaxies, the current work compiles a relatively large ($>$400 members; see \S\ref{subsec:sample}) and heterogeneous sample of AGNs, which are selected with various techniques using data from large spectroscopic and photometric surveys, in a consistent way. For a subset ($>$50\%) of the sample having optical spectroscopic observations we have studied their stellar populations (\citealt{2020ApJ...903...58C, 2024RAA....24f5006L}), thus allowing us to explore the connection between galactic structure and star formation history and AGN activity, which will be presented in a future paper. 
This work endeavors to carry out a detailed study on the morphology of AGN-host dwarfs, which may shed some light on the triggering mechanism of AGNs in the low-mass regime, by taking advantage of non-parametric measurements of galaxy structure and visual classification of morphological types. The non-parametric method, such as the \textit{CAS} system (\citealt{2003ApJS..147....1C}) and the $Gini-M_{20}$ diagram (\citealt{2004AJ....128..163L}), measures the galaxy image automatically, and thus can be applied to big data from large surveys. The results directly and objectively reveal the light distribution of the target source and contain many details about its structure. Visual inspection can produce accurate classifications \citep{2011MNRAS.410..166L,2014MNRAS.440.2944K}, especially for identifying strongly perturbed systems (see, e.g., \citealt{2010MNRAS.401.1552D}), but is time-consuming.

The structure of this article is organized as follows: In Section 2, we outline the sample and image data used in the current study, and introduce the procedures undertaken before the non-parametric coefficients derived from \texttt{statmorph}\footnote{\url{https://statmorph.readthedocs.io/en/latest/overview.html}} \citep{2019MNRAS.483.4140R}. Meanwhile, we also perform visual inspection, combining with the results from deep learning presented in \cite{2023MNRAS.526.4768W}, to further check the efficiency and capability of morphological parameters in separating typical galaxy types of dwarf galaxies. Section 3 presents an analysis of the data across different bands and subsamples, and compares our results to normal dwarf and AGN-host massive galaxies. In Section 4, we provide a summary of this study and present our conclusions. Where required we adopt a Hubble constant of $H_0 = 73$~km~s$^{-1}$~Mpc$^{-1}$ as used in our previous work.

\section{SAMPLE, DATA AND METHOD} \label{sec:data}
\subsection{Sample} \label{subsec:sample}

Here we compile a sample of AGN-host dwarf galaxies from the literature that employs various methods, as described briefly in the following, to identify AGNs using spectroscopic and photometric data in the local Universe.
 \begin{itemize}
     \item \textit{The X-ray Sample (hereafter X-ray)}: \cite{2020MNRAS.492.2268B} conducted a matching of a sample of dwarf galaxies ($M_{\star}\leq10\sp{9.5}~$M$_{\odot}$ and $z<0.25$) from the MPA-JHU catalog (based on Sloan Digital Sky Survey (SDSS) Data Release 8 (DR8)) with the central X-ray samples from 3XMM DR7. They employed the following criteria to search the AGNs: (1) the position-error-normalized separation, the ratio between the separation (between the optical and X-ray objects) and the error in X-ray position, must be less than 3.5, and the extent of the X-ray source must be less than 10\arcsec\ to ensure point-like emission consistent with an AGN \citep{2016A&A...590A...1R}; and (2) there must be an X-ray excess relative to the contribution estimated from X-ray binaries estimated by the star formation rate (SFR), stellar mass and redshift ($z$) \citep{2016ApJ...825....7L} and hot gas in the interstellar medium estimated by SFR \citep{2012MNRAS.426.1870M}. Notably, off-nuclear emissions may also be classified as AGN, resulting in the identification of 61 AGNs.
     
     \item \textit{The Optical Sample (hereafter BPT and \HeII)}: \cite{2013ApJ...775..116R} employed BPT diagram, utilizing data from SDSS DR8 \citep{2011ApJS..195...26A}, to identify AGNs within emission-line galaxies, constraining their mass to $M_{\star}\leq10\sp{9.5}~$M$_{\odot}$, as derived from NASA-Sloan Atlas (NSA v0\_1\_2) catalog which limits the redshift $z<0.055$. They identified 136 objects, comprising 35 AGN-dominated and 101 composite systems. \cite{2015MNRAS.454.3722S} utilized the BPT diagram and the \HeII~$\lambda$4868 emission-line diagnostics \citep{2012MNRAS.421.1043S} to select AGNs from the same dwarf sample as that adopted in the mid-IR method (see below), resulting in 48 and 121 BPT- and \HeII-based sources, respectively. The optical sample presented here consists of the samples from these two studies.
     
     \item \textit{The Mid-Infrared (IR) Sample (hereafter mid-IR)}: \cite{2015MNRAS.454.3722S} cross-matched the OSSY (Oh-Sarzi-Schawinski-Yi; \citealt{2011ApJS..195...13O}) and the MPA-JHU catalogs, and obtained a parent sample of dwarf galaxies with $M_{\star}\leq10\sp{9.5}~$M$_{\odot}$ at $z<0.1$. Subsequently, they identified 189 objects after applying the following color criteria based on the Wide-field Infrared Survey Explorer (WISE; \citealt{2010AJ....140.1868W}) photometric system: (1) $[W1-W2] \geq 0.8$ (\citealt{2012ApJ...753...30S}); and/or (2) $2.2 <[W2-W3]<4.2 $ \&  $  (0.1 \times [W2-W3] + 0.38)<[W1-W2]<1.7$ (\citealt{2011ApJ...735..112J}). 
     
     \item \textit{The Radio Sample (hereafter radio)}: Based on the NSA v0\_1\_2 catalog, \cite{2020ApJ...888...36R} performed sensitive, high-resolution observations for 111 (out of 151) nearby dwarfs ($10^7 \lesssim M_\star/M_\odot \leq 10\sp{9.5}$ and $z<0.055$) detected by the Faint Images of the Radio Sky at Twenty centimeters Survey \citep{1995ApJ...451L...1S}, using the Karl G. Jansky Very Large Array. The authors identified 13 out of 39 compact radio sources as AGNs, as alternative explanations cannot explain the radio emission. Therefore, these 13 objects constitute the ``radio" sample.
     
     \item \textit{The Variability-selected Sample (hereafter variability)}: \cite{2022ApJ...936..104W} constructed a sample of variability-selected AGNs with a stellar mass cutoff of $M_{\star}<10\sp{9.75}~$M$_{\odot}$ and redshifts of $0.02<z<0.35$ from the NSA v1\_0\_1 catalog\footnote{\url{https://www.sdss4.org/dr17/manga/manga-target-selection/nsa}}, using optical photometry from the Zwicky Transient Facility \citep{2019PASP..131a8002B,2019PASP..131g8001G,2020PASP..132c8001D} and forward-modeled mid-IR photometry of time-resolved WISE co-added images. After trialing a number of variability statistics and performing manual inspection, they identified 44 AGN candidates exhibiting optical variability and 148 AGN candidates having mid-IR variability. In our paper, the ``variability" sample refers to these optically/mid-IR variable AGNs.
 \end{itemize}
 
To keep consistency of the basic galactic parameters (e.g., $z$, $M_\star$) across various samples, we first cross-match the above samples with the NSA v1\_0\_1 catalog, requiring a match radius $\leq$5\arcsec. Then, we limit the redshifts to $z<0.055$, mainly due to the fact that it is the limitation of NSA v0\_1\_2 catalog. For our purpose, however, such a low redshift cut is also needed to ensure the image qualities (e.g., high enough signal-to-noise ratios, and/or good physical resolutions) and to reduce selection bias. We further apply a stellar mass (``ELPETRO\_MASS" in the NSA v1\_0\_1 catalog) cutoff of $M_{\star}\leq10\sp{9.5}~$M$_{\odot}$. After removing overlapping sources and taking the range of images into consideration (please see \S\ref{subsec:method} for details), the final total sample comprises 416 unique objects, and Table \ref{tab:number} lists the number of objects of each subsample. 

\begin{figure}[t]
    \centering
    \includegraphics[width=1\linewidth]{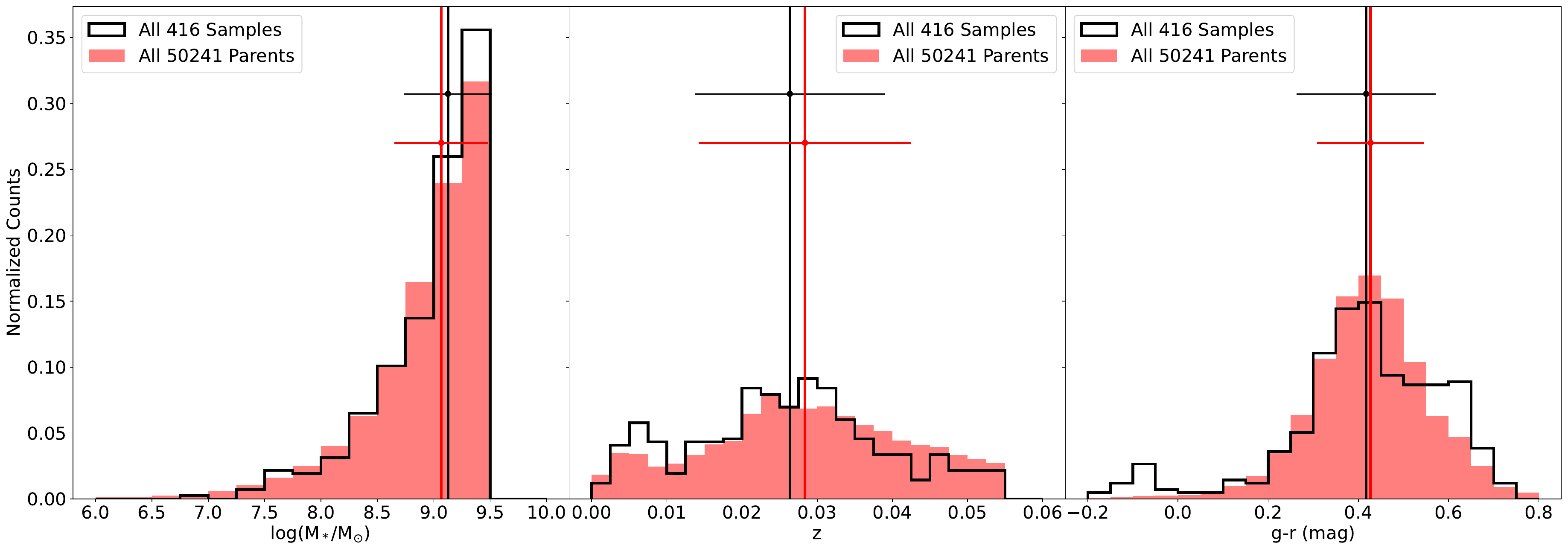}
    \caption{The distribution for AGN-host dwarf (open histograms) and the whole dwarf (filled histograms) samples. The whole dwarf sample consists of 50241 galaxies with stellar mass $M_{\star}\leq10\sp{9.5}~$M$_{\odot}$ and $z<0.055$ from the NSA v1\_0\_1 catalog. The vertical lines indicate the corresponding median values with their associated uncertainties.}
    \label{fig:partent}
\end{figure}

The current sample is not complete in any sense (although the minimum stellar mass is about $10^{6.8}~$M$_\odot$), which is common in the studies of AGN-host dwarfs in the literature (but see \citealt{2024MNRAS.532..613B} for a mass complete sample down to $M_\star\sim10^8~$M$_\odot$), since it is difficult to achieve a complete sample for faint galaxies (however, see recent efforts to construct a complete sample of dwarf galaxies from \citealt{2005ApJ...631..208B}, \citealt{2008ApJS..178..247K}, \citealt{2009ApJ...692.1305L}, \citetalias{2024MNRAS.529..499L}, and \citealt{2025MNRAS.538..153K}). Fortunately, our sample is a heterogeneous sample compiled from the six most frequently used methods of identifying AGNs, and spans a large range of galactic parameters including stellar mass and $g-r$ color, as plotted in Figure \ref{fig:partent}. From the figure we can see that it covers the full redshift range of the total dwarf sample (i.e., $M_{\star}\leq10\sp{9.5}~$M$_{\odot}$ and $z<0.055$; filled histogram) presented in the NSA v1\_0\_1 catalog, and generally shows a similar distribution (open histogram) in $M_\star$, indicating that our sample could be representative of the total dwarf sample since the physical properties of nearby galaxies are strongly correlated with their stellar masses (e.g., \citealt{2004MNRAS.353..713K,2011ApJ...742...96W,2013ApJ...765..140A}). Nevertheless, it needs a more complete sample to better probe the dwarf population of AGN-hosts in future work.

\begin{table}[t!]
\movetableright=-15mm
\caption{The number of different types of sample in \textit{grz} bands.}
\label{tab:number}
\begin{center}
\begin{tabular}{cccccc}\hline\hline 
Sample                  & Band & \textbf{Sample A} & \textbf{Sample B} & Fail Sample & Total                \\\hline
\multirow{3}{*}{X-ray}  & g    & 30       & 6        & 4 & \multirow{3}{*}{40} \\
                        & r    & 30       & 5        & 5 &                     \\
                        & z    & 30       & 5        & 5 &                     \\\hline
\multirow{3}{*}{BPT}    & g    & 96       & 5        & 9 & \multirow{3}{*}{110}\\
                        & r    & 94       & 7        & 9 &                     \\
                        & z    & 93       & 6        & 11&                     \\\hline
\multirow{3}{*}{\HeII}  & g    & 78       & 4        & 6 & \multirow{3}{*}{88} \\
                        & r    & 78       & 4        & 6 &                     \\
                        & z    & 77       & 3        & 8 &                     \\\hline
\multirow{3}{*}{Mid-IR} & g    & 84       & 8        & 8 & \multirow{3}{*}{100}\\
                        & r    & 82       & 10       & 8 &                     \\
                        & z    & 69       & 2        & 29&                     \\\hline
\multirow{3}{*}{Radio}  & g    & 12       & 1        & 0 & \multirow{3}{*}{13} \\
                        & r    & 11       & 1        & 1 &                     \\
                        & z    & 11       & 1        & 1 &                     \\\hline
\multirow{3}{*}{Variability}     & g    & 81       & 6        & 4 & \multirow{3}{*}{91} \\
                        & r    & 83       & 4        & 4 &                     \\
                        & z    & 78       & 5        & 8 &                     \\\hline
\multirow{3}{*}{Total}  & g    & 363      & 26       & 27& \multirow{3}{*}{416}\\
                        & r    & 360      & 28       & 28&                     \\
                        & z    & 337      & 22       & 57&                     \\\hline
\end{tabular}
\end{center}
\tablecomments{ The \textit{z}-band experiences a greater loss of samples compared to the other two bands, with further details available in \S\ref{subsec:measure}. NSAID 337898 falls outside the range of DESI DR10, resulting in a final sample comprising 416 unique objects. \textbf{Sample A} includes those objects that meet all established criteria. \textbf{Sample B} consists of certain objects that did not satisfy criterion 2 in \S\ref{subsec:measure}. However, upon review, a majority of the data from these samples can still be utilized, except for $A_\mathrm{S}$. Failed sample refers to those objects that do not meet the selection criteria in \S\ref{subsec:measure}.}
\end{table}

\subsection{Multi-band Images and Preprocessing} \label{subsec:method}
We acquire images from the Dark Energy Spectroscopic Instrument (DESI) Data Release 10 (DR10)\footnote{\url{https://www.legacysurvey.org/dr10/description/}}, which encompasses the Dark Energy Camera Legacy Survey (DECaLS), Beijing-Arizona Sky Survey (BASS), and Mayall \textit{z}-band Legacy Survey (MzLS), all of which are now complete. While I-band images are initially provided, their coverage is smaller compared to other bands, resulting in the loss of some images from AGN-host dwarfs. Consequently, we choose the \textit{grz} bands data to measure the morphological parameters. The images of DR10 originate from the DECaLS \textit{grz} bands observations, with depth in optical bands measured at \textit{g} = 24.0, \textit{r} = 23.4, and \textit{z} = 22.5 mag \citep{2019AJ....157..168D}, defined as the optimal-extraction (forced-photometry) depth near the limits of DESI, which is approximately 2 mags deeper than SDSS imaging. Notably, NSAID 337898 lies outside the range of DESI DR10, resulting in a final sample of 416 unique objects. We designated the right ascension and declination from the NSA v1\_0\_1 catalog as the center of the images. The cutouts measure 512 pixels, with a pixel scale of 0.262\arcsec\ across all three bands. It is important to note that 9 objects exceed the dimensions of the images, prompting us to extend the cutout size to 1024 pixels.

DESI DR10 also supplies the point spread function (PSF) for individual exposures, computed independently for each CCD using \texttt{PSFEx} \citep{2011ASPC..442..435B}. Beginning with DR9, the PSF model was enhanced to account for and subtract the extended wings of bright stars from DECam images. This model is a linear combination of a flexible inner PSF and a fixed outer PSF. Further details regarding the PSF can be found on the DESI website \footnote{\url{https://www.legacysurvey.org/dr10/psf/}}. The full width at half maximum (FWHM) of the PSF is approximately 1\arcsec\ in the optical data \citep{2019AJ....157..168D}. 

We employ \texttt{SEXtractor} \citep{1996A&AS..117..393B} to isolate the target from the background, subsequently masking the contamination far from the center. To effectively eliminate contamination, we establish a detection threshold for objects $1.5-3\sigma$ above the background. Typically, targets are positioned at the center of the image, therefore, we designate a clean zone spanning 240 to 270 pixels to ensure that targets remain unaffected. Unfortunately, some contamination is situated near or even within the target's image, potentially compromising the measurements. After conducting a visual inspection, we mask these sources with an appropriate radius to guarantee the complete absence of contamination. The mask radius ranges from 3 to 6 times the object's semi-major and semi-minor axes. 

Following the masking process, we utilize \texttt{Photutils} \citep{2020zndo...4049061B} to generate the segmentation map, setting the threshold $2-3\sigma$ above the background, with $npixels = 10$. Some targets may still be affected by close contamination, requiring adjustments to the thresholds for both the mask and segmentation map.

\subsection{Non-parametric Measurements} \label{subsec:measure}
The non-parametric coefficients of morphology contain many details about the light distribution of galaxies. These parameters also facilitate the exploration of the evolutionary and structural characteristics of galaxies. Here we adopt the Python package \texttt{statmorph}, which provides the following measurements:

Concentration parameter (\textit{C}): described as \citep{2000AJ....119.2645B,2003ApJS..147....1C}
\begin{equation}
    C = 5 \times \log_{10} \left( \frac{r_{80}}{r_{20}} \right),
    \label{sec: function 1}
\end{equation}
where $r_{20}$ and $r_{80}$ stand for the radii of the circular apertures including 20 and 80 percent of the total flux of the galaxy, respectively. \texttt{statmorph} uses 1.5~\textit{$r_\mathrm{petro}$} \citep{1976ApJ...209L...1P} which is set to $\eta$ = 0.2 to describe the total flux \citep{2003ApJS..147....1C,2004AJ....128..163L} .

Asymmetry parameter (\textit{A}): the symmetry of the whole galaxy \citep{1995ApJ...451L...1S,1996MNRAS.279L..47A,2000ApJ...529..886C}
\begin{equation}
     A = \frac{\sum_{i,j} \lvert I_{ij} - I_{ij}^{180} \rvert}{\sum_{i,j} \lvert I_{ij} \rvert} - A_\mathrm{bgr}.
     \label{sec: function 2}
\end{equation}
Use the original image to subtract the image that is rotated by 180 degrees, where \textit{$I_{ij}$} and \textit{$I\sp{180}_{ij}$} stand for pixel flux values of the original and rotated images, respectively, and $A_\mathrm{bgr}$ is the average asymmetry of the background. We also use 1.5~\textit{$r_\mathrm{petro}$} to get the rotation center, which is determined by minimizing \textit{A}.

Smoothness parameter (\textit{S}; also referred to as ``Clumpiness"), obtained by subtracting the galaxy image smoothed with a boxcar filter of width $\sigma$ from the original image \citep{2003ApJS..147....1C} 
\begin{equation}
    S = \frac{\sum_{i,j}  I_{ij} - I_{ij}^{S} }{\sum_{i,j}  I_{ij}} - S_\mathrm{bgr}.
    \label{sec: function 3}
\end{equation}
\textit{$I_{ij}$} is same as \textit{A}, and \textit{$I\sp{S}_{ij}$} is the pixel flux value of the smoothed images. Meanwhile, $S_\mathrm{bgr}$ is also the average smoothness of the background. Note that the range of the measurement is between the $\sigma$ = 0.25~\textit{$r_\mathrm{petro}$} and 1.5~\textit{$r_\mathrm{petro}$} and any pixel whose $I_{ij}-I\sp{S}_{ij}<$ 0 is excluded from the measurement.

Shape asymmetry parameter ($A_\mathrm{S}$) is mentioned by \citep{2019MNRAS.483.4140R} and has the same function as \textit{A} but in a different binary detection mask. This parameter is considered the edge of the galaxy, so it is very easy to be influenced by the background noise, and 
 \citetalias{2022ApJ...925...70Z} suggested $A_\mathrm{S}$ could not be used in any type of galaxy.

The Gini coefficient (\textit{G}): computed as \citep{2004AJ....128..163L}
\begin{equation}
    G=\frac{1}{\overline{\lvert X \lvert}n(n-1)} \sum_{i=1} ^n (2i-n-1)\lvert X_i \lvert,
    \label{sec: function 4}
\end{equation}
and \textit{$\overline{\lvert X \lvert}$} is the mean of  \textit{$\lvert X_i\lvert$} which represent absolute values of the pixels. The values of the \textit{G} are between 0 and 1, and if the \textit{G} is bigger, the flux is more concentrated. The \textit{G} is sensitive to the pixels in the calculation, so \textit{G} needs the Gini-segmentation map only depending on \textit{$r_\mathrm{petro}$}. We then select pixels from the original image above the convolved image's mean surface brightness at \textit{$r_\mathrm{petro}$} to be assigned to the galaxy.

The $M_{20}$ statistic \citep{2004AJ....128..163L} represents the second moment of a galaxy's brighter regions, including 20 percent of the whole flux, relative to the whole second-order central moment $\mu_\mathrm{tot}$, which is calculated as:
\begin{equation}
    \mu_\mathrm{tot} = \sum_{i=1}^n \mu_i
    \equiv \sum_{i=1}^n I_i[(x_i-x_c)^2-(y_i-y_c)^2],
    \label{sec: function 5}
\end{equation}
and $M_{20}$ is 
\begin{equation}
    M_{20} \equiv \log_{10} \left( \frac{\sum_i \mu_i}{\mu_\mathrm{tot}} \right),  \text{while}  \sum_i I_i < 0.2I_\mathrm{tot},
    \label{sec: function 6}
\end{equation}
where $I_\mathrm{tot}$ is the total flux of the regions identified by the segmentation map. The value of $M_{20}$ is negative, and the lower it is, the higher the concentration within the galaxy.

The bulge statistic $F(G, M_{20})$: defined as \citep{2015MNRAS.454.1886S}
\begin{equation}
    F(G, M_{20}) = -0.693M_{20} + 4.95G-3.96,
    \label{sec: function 7}
\end{equation}
which means the line intersects at $(G_0, M_{20,0})=(0.533, -1.75)$ and is also perpendicular to the line separating merging and non-merging galaxies. \texttt{statmorph} here selects $G>0.14M_{20}+0.80$ as the function of $F(G, M_{20})$ and $G>-0.14M_{20}+0.33$ \citep{2008ApJ...672..177L} as the function of $S(G, M_{20})$, and if $F(G, M_{20})>0$ means the galaxy is ETG, otherwise, the galaxy is LTG and also is closely related to Concentration index (\textit{C}) and S\'{e}rsic index.

The merger statistic $S(G, M_{20})$: defined as \citep{2015MNRAS.451.4290S}:
\begin{equation}
    S(G, M_{20}) = 0.139M_{20} + 0.990G-0.327,
    \label{sec: function 8}
\end{equation}
which means the line separating merging and non-merging galaxies. If the $S(G, M_{20})>0$, the galaxy is merging; otherwise, the galaxy is non-merging.

To ensure the reliability of all data, we establish the following criteria for selection:
\begin{enumerate}
    \item \texttt{flag} == 0 and \texttt{flag\_sersic} == 0. (\texttt{statmorph} provides these two criteria to mark good data.)
    \item The Mask part does not take up too much area of the whole galaxy. (It ensures all the contamination can be removed cleanly, or does not affect the measurement of the target.)
    \item $r_{20}$ is bigger than half of the FWHM of the PSF. 
    \item sn\_per\_pixel is bigger than 2.5. 
\end{enumerate}

After applying the criteria, we ultimately acquire the data, with the various types of samples detailed in Table \ref{tab:number}. Based on the number of reliable parameters, we categorize samples into \textbf{Sample A} and \textbf{Sample B}. \textbf{Sample A} comprises objects that meet all established criteria, while \textbf{Sample B} consists of certain sources that do not pass criterion 2. However, upon further examination, most of the data remains useful. Although these objects exhibit a higher degree of contamination compared to \textbf{Sample A}, the majority of measurements remain unaffected, except for $A_\mathrm{S}$, which relies on the galaxy's edge, indicating $A_\mathrm{S}$ is unusable in \textbf{Sample B}. Failed sample refers to those objects that do not meet the selection criteria, and after visual inspection, these objects contain contamination which affects the measurement.

In Table \ref{tab:number}, we present the sizes of the final sample across the \textit{grz} bands: \textbf{Sample A} comprises 363, 360, and 337, while \textbf{Sample B} consists of 26, 28, and 22, respectively. The mean and median values of the parameters are detailed in Table \ref{tab:mean} in Appendix \ref{subsec:A1}. Notably, the \textit{z}-band data have a loss of 13.7\% of the total sample, a figure nearly doubles that of the other two bands, which can be mainly attributed to its relatively lower S/N (criterion 4) that fails in 25 objects. Furthermore, the median $r_{20}$ of \textit{z}-band images is about 6$-$10\% smaller than that of $g$- and $r$-band, and 15, 10 and 8 objects are removed from the samples of $z$-, $r$- and $g$-band, respectively, due to criterion 3.

\subsection{Morphological Classification via Visual Inspection} \label{subsec:visual}

\begin{figure}[t]
    \centering
    \subfigure[]{
        \includegraphics[width=0.4\textwidth]{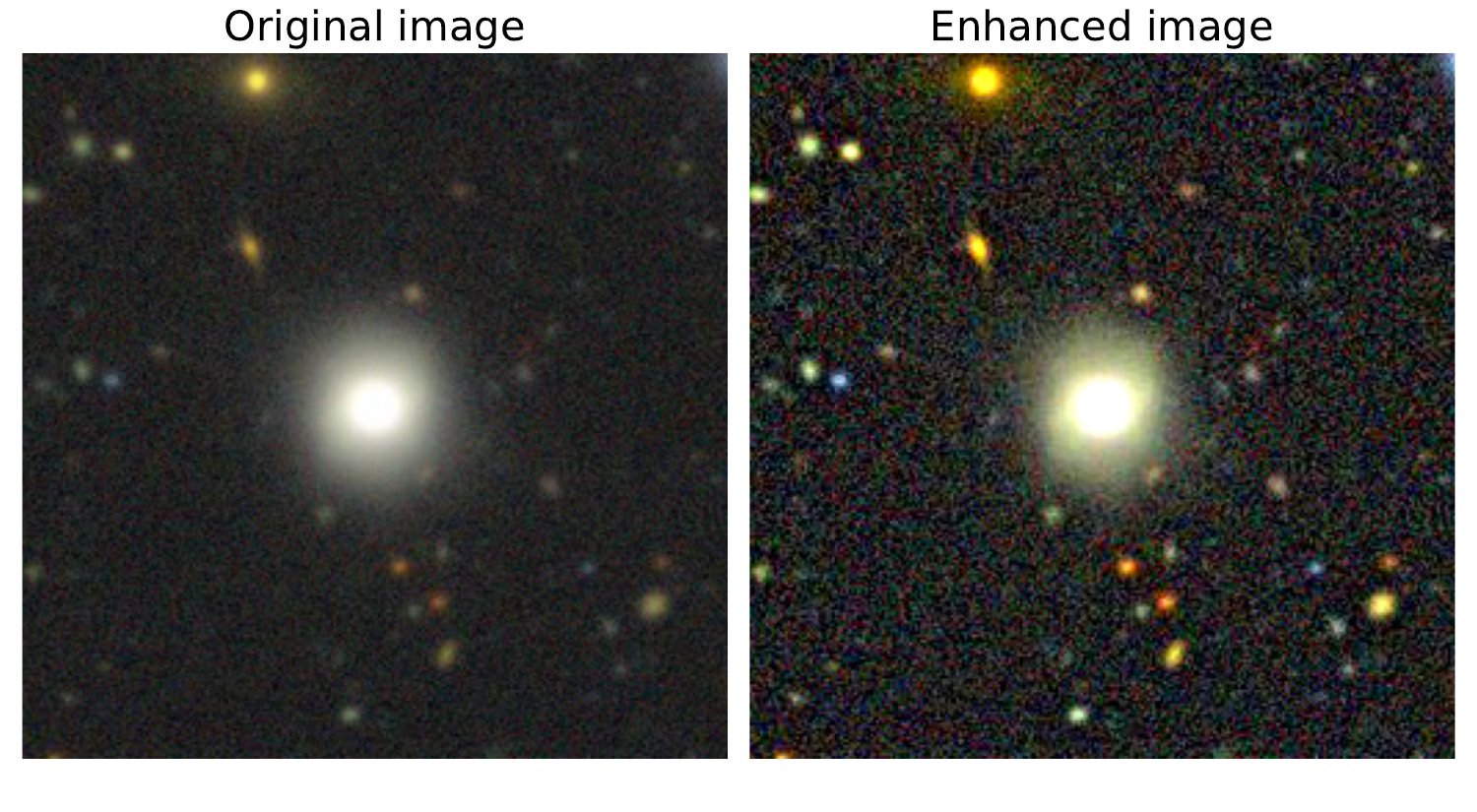}
    }
    \subfigure[]{
        \includegraphics[width=0.4\textwidth]{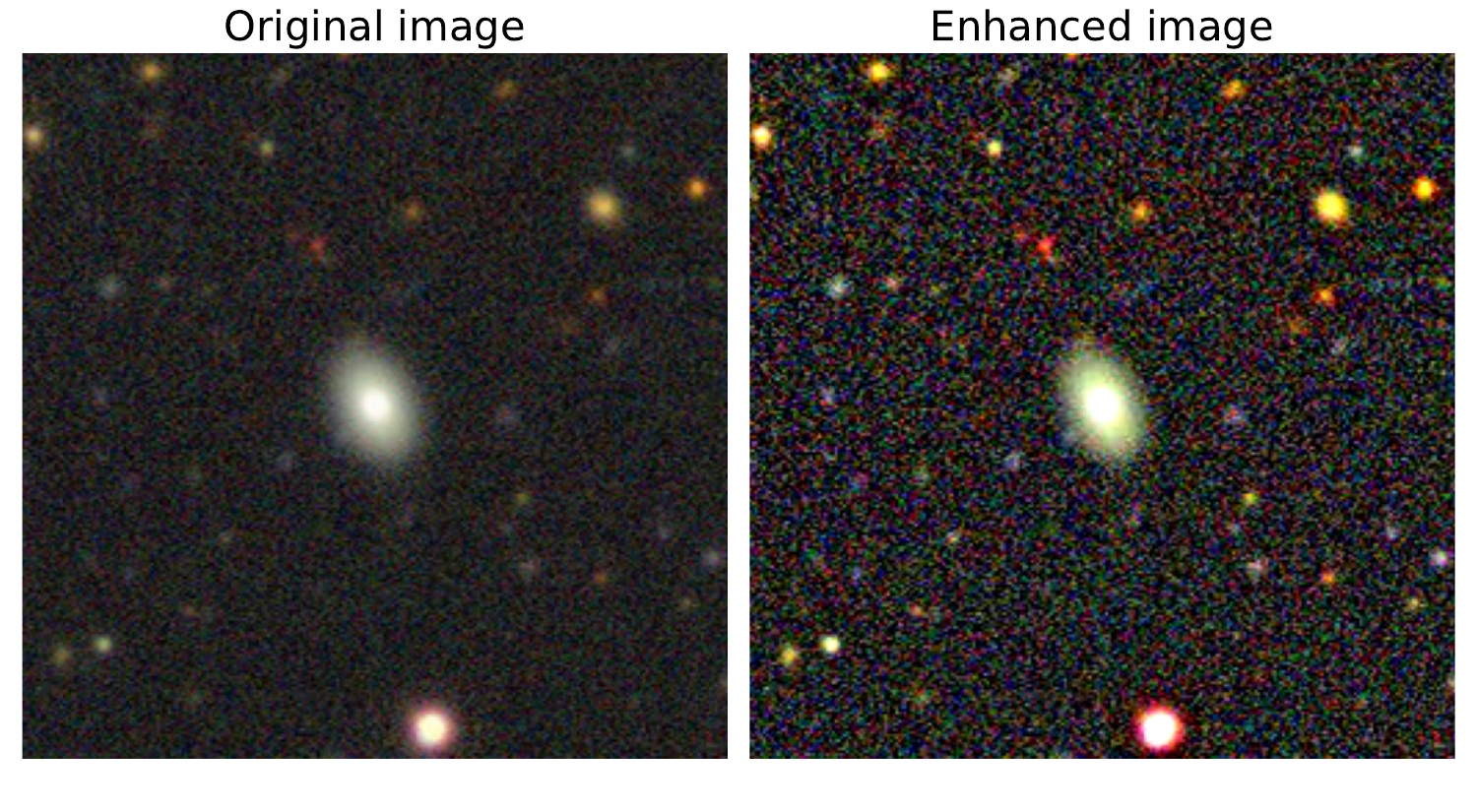}
    }
    \subfigure[]{
        \includegraphics[width=0.4\textwidth]{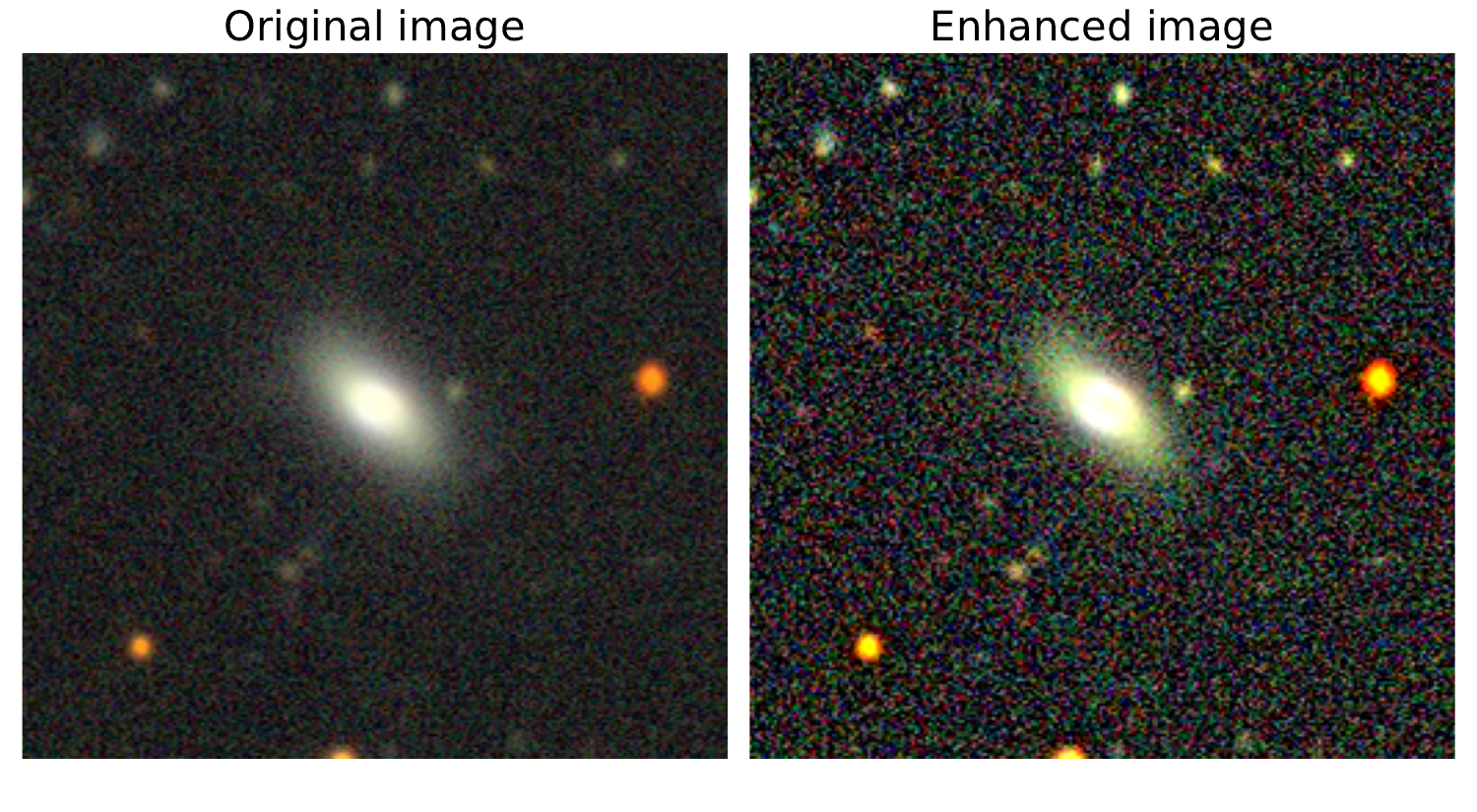}
    }
    \subfigure[]{
        \includegraphics[width=0.4\textwidth]{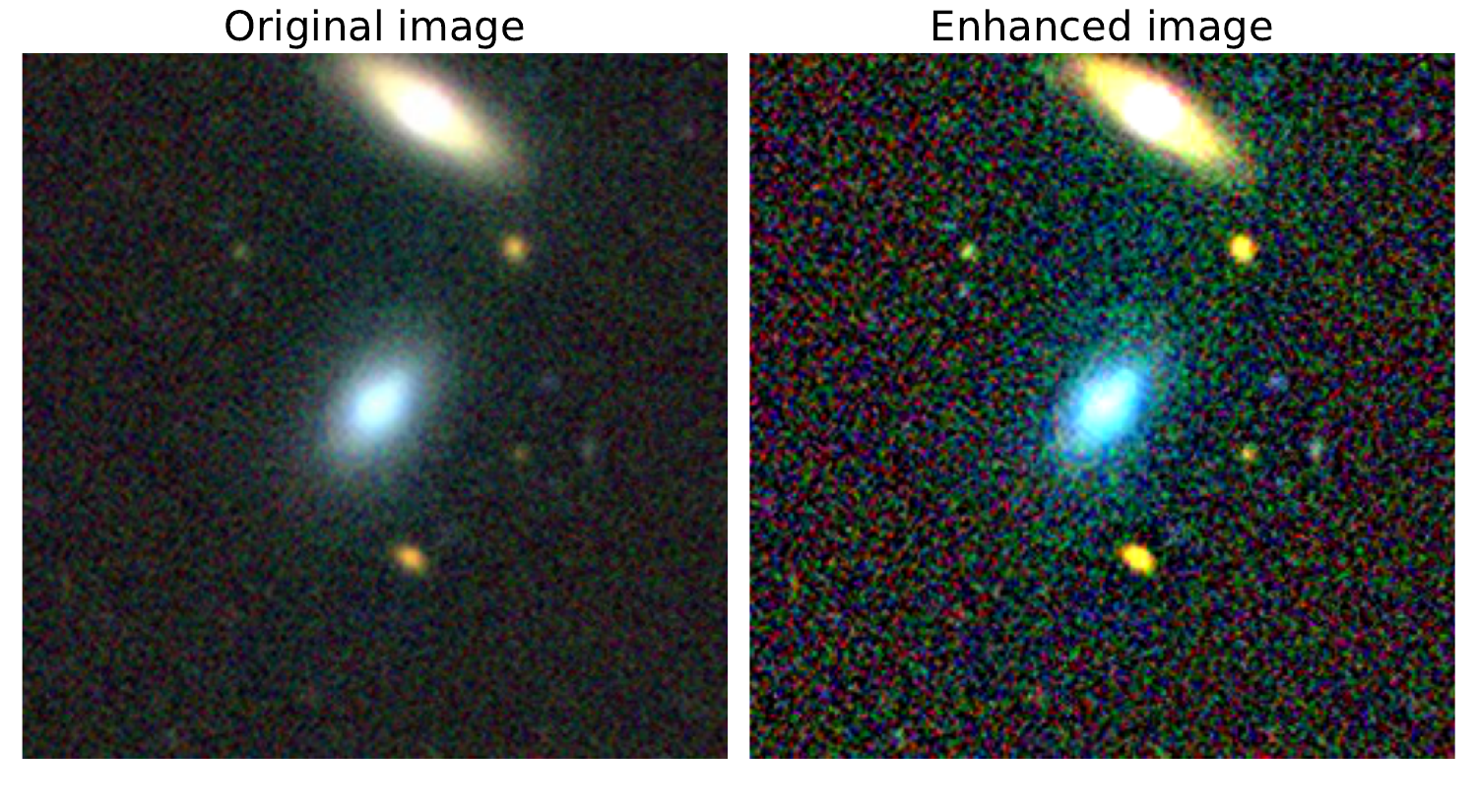}
    }
    
    \subfigure[]{
        \includegraphics[width=0.4\textwidth]{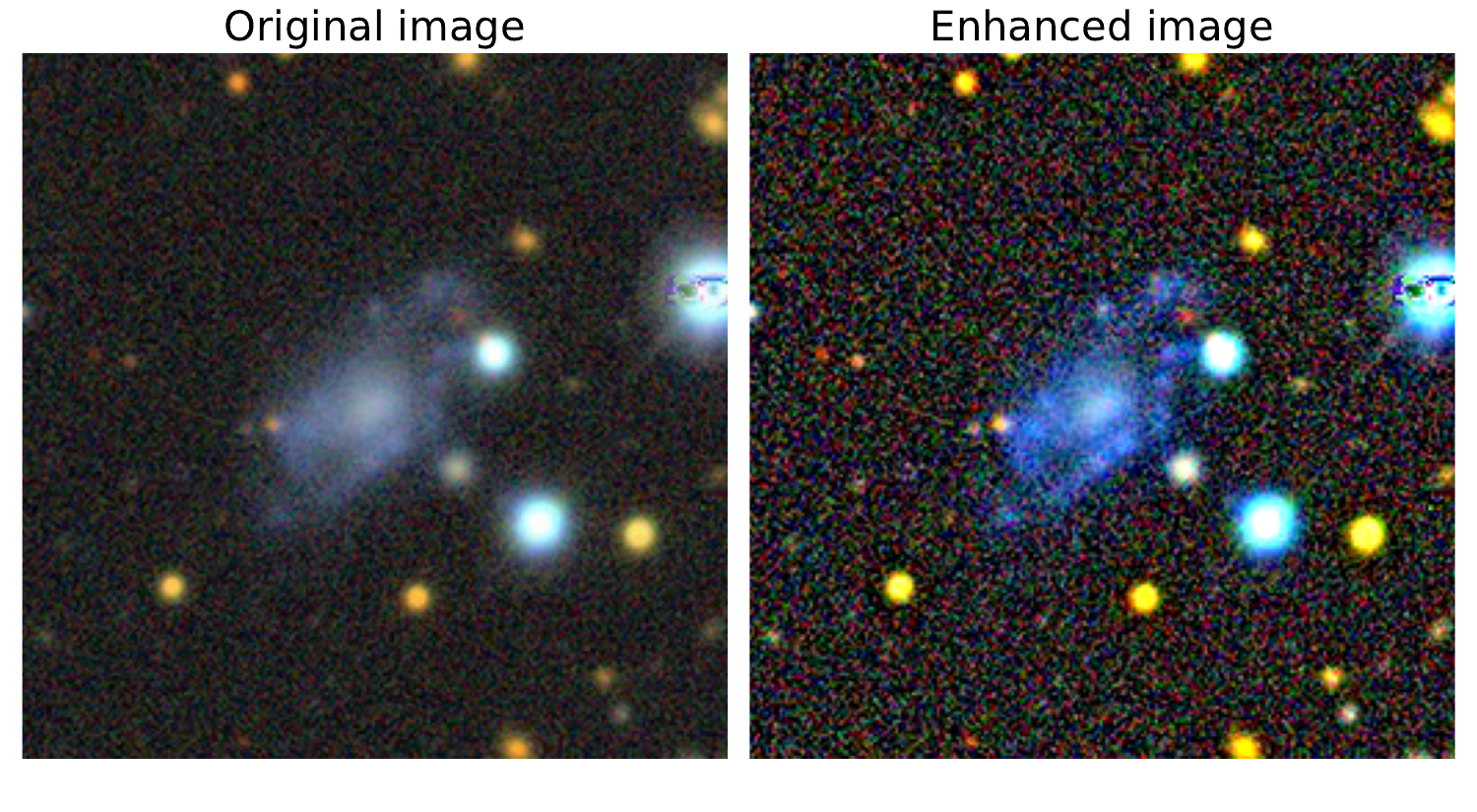}
    }
    \subfigure[]{
        \includegraphics[width=0.4\textwidth]{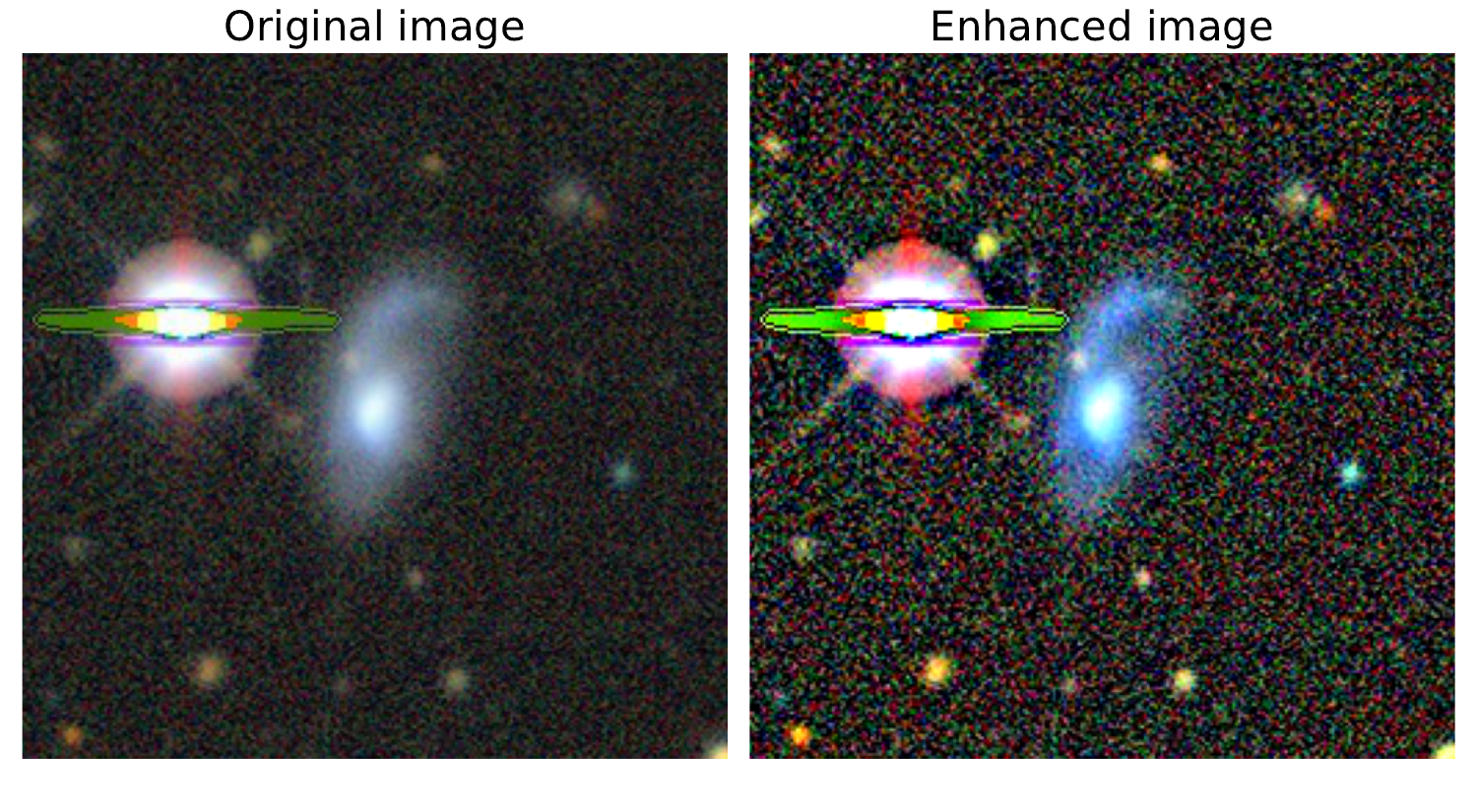}
    }
    \subfigure[]{
        \includegraphics[width=0.4\textwidth]{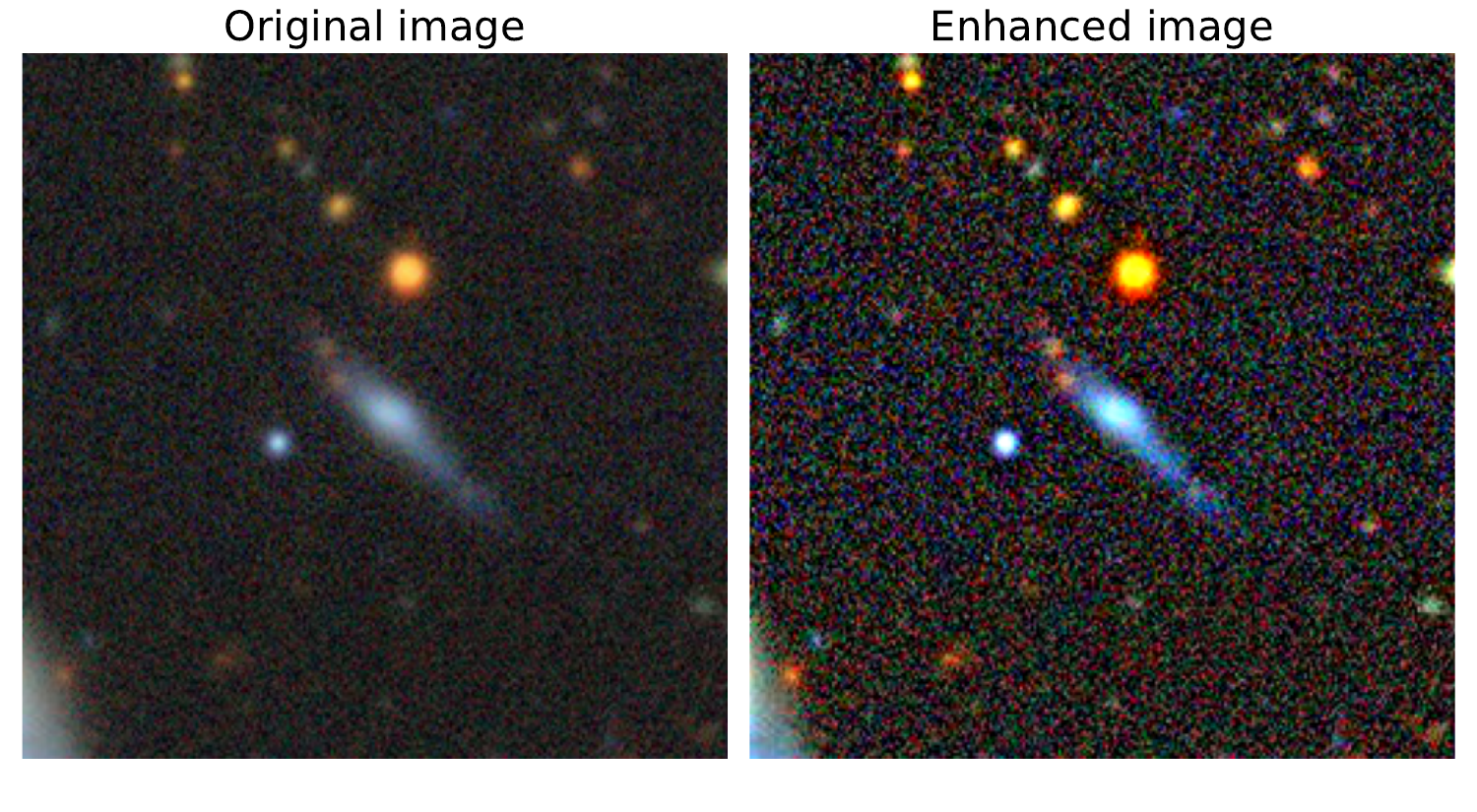}
    }
    \subfigure[]{
        \includegraphics[width=0.4\textwidth]{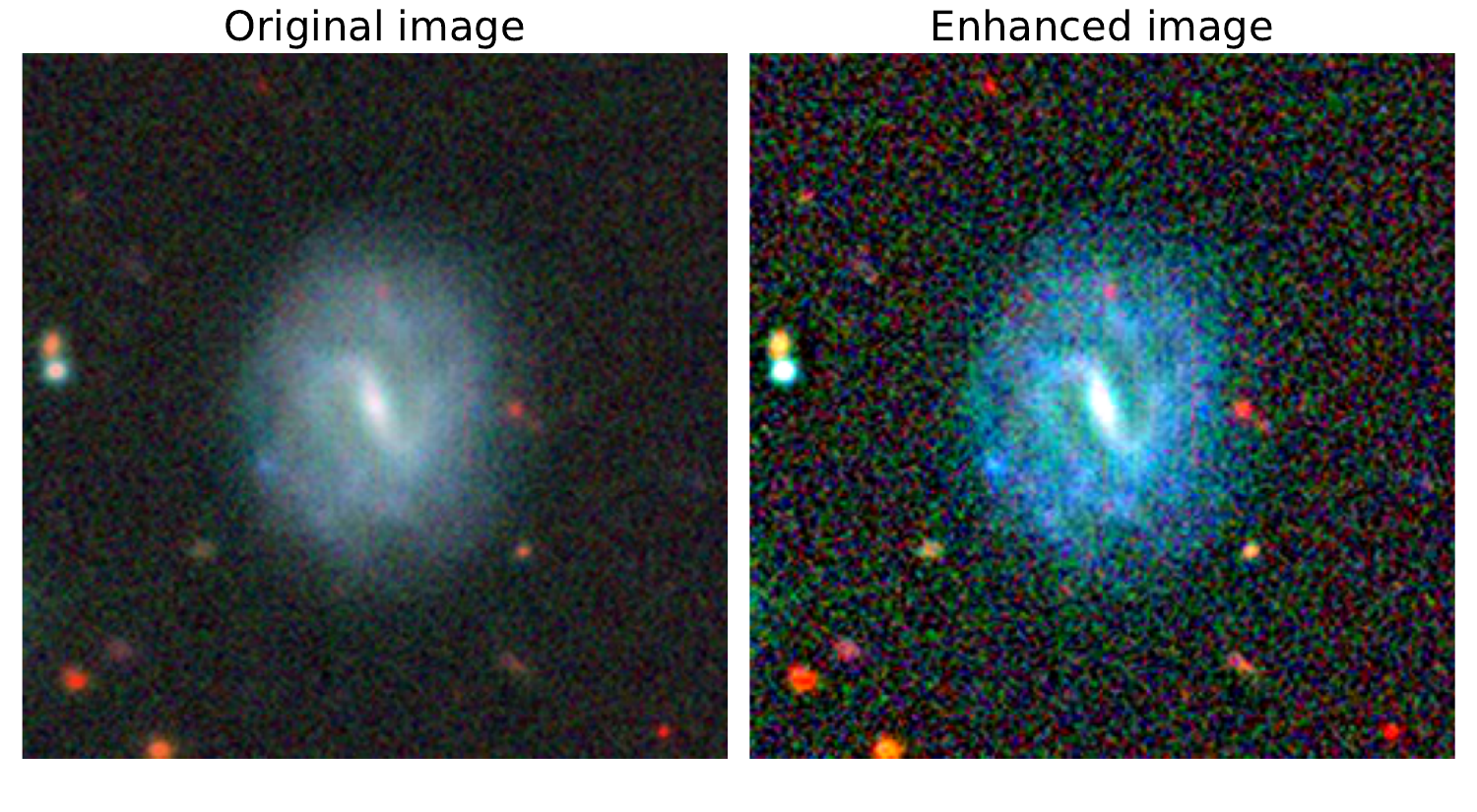}
    }
    \caption{Examples of different types of AGN-host dwarf galaxies via visual Classification.
    ETGs in Row(1)-(2): The left-hand panel is the original image from DESI DR10, and the right-hand panel is the enhanced image calculated by Unsharp-masking. (a)-(c) exhibit a bright core in their center, and (d) has a different shape.
    LTGs in Row(3)-(4): (e) contains two distinct spiral arms; (f) only has one spiral arm; (g) is an edge-on galaxy with a possible spiral arm; (h) has a very clear bar in its center.
    Mergers in Row(5)-(6): (i) has a shell outside; (j) shows two cores and an irregular shape; (k) exhibits a clear tidal feature; (l) has a visible large-scale asymmetry within the main body of the galaxy.
    Unsure in Row(7)-(8): (m) and (n) are edge-on galaxies, and (o) and (p) exhibit small dots, which are very difficult to classify.
    }
\label{fig:type}
\end{figure}

\begin{figure}[t]
\figurenum{2}
    \centering
    \setcounter{subfigure}{8}
     \subfigure[]{
        \includegraphics[width=0.4\textwidth]{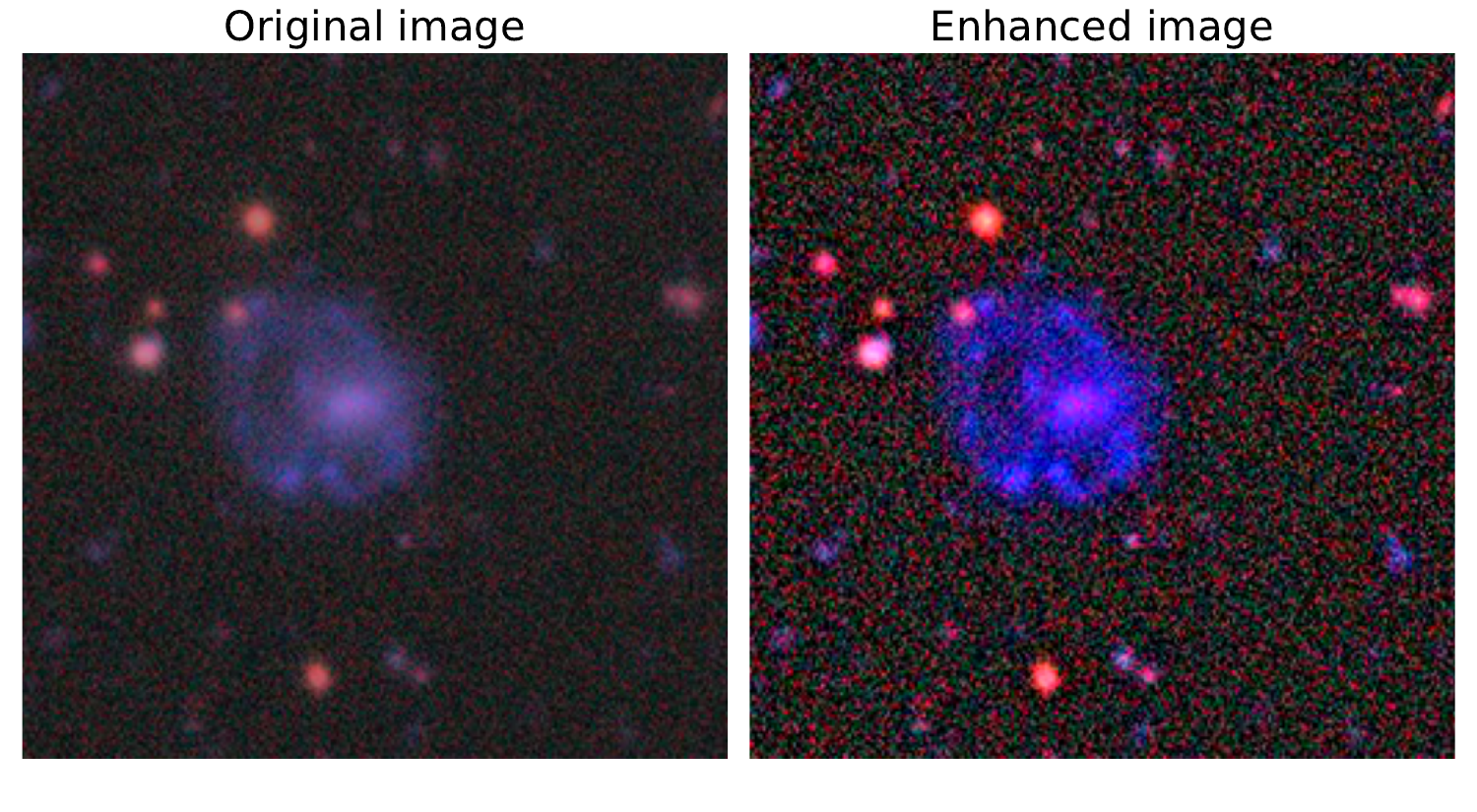}
    }
    \setcounter{subfigure}{9}
    \subfigure[]{
        \includegraphics[width=0.4\textwidth]{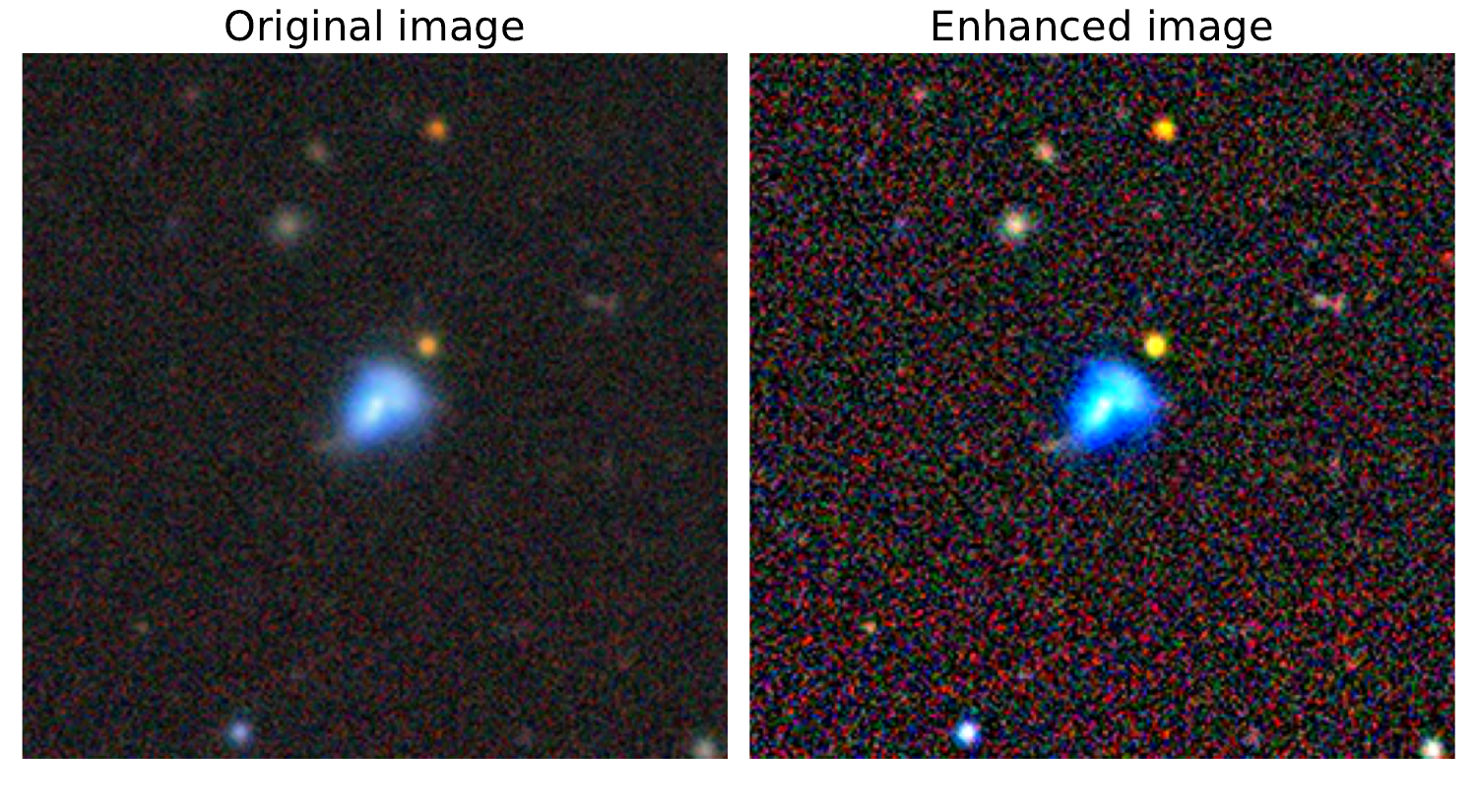}
    }
    \setcounter{subfigure}{10}
    \subfigure[]{
        \includegraphics[width=0.4\textwidth]{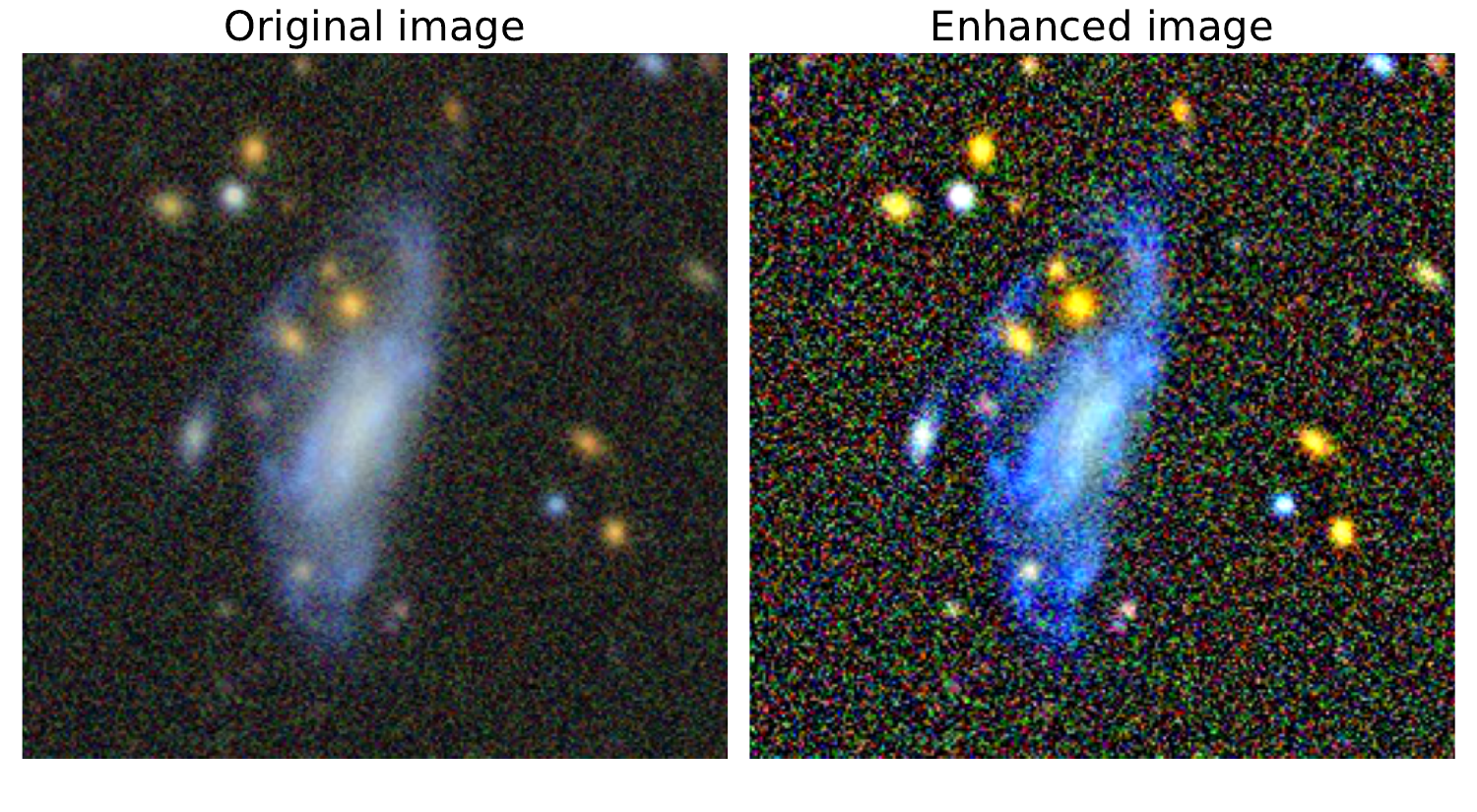}
    }
    \setcounter{subfigure}{11}
    \subfigure[]{
        \includegraphics[width=0.4\textwidth]{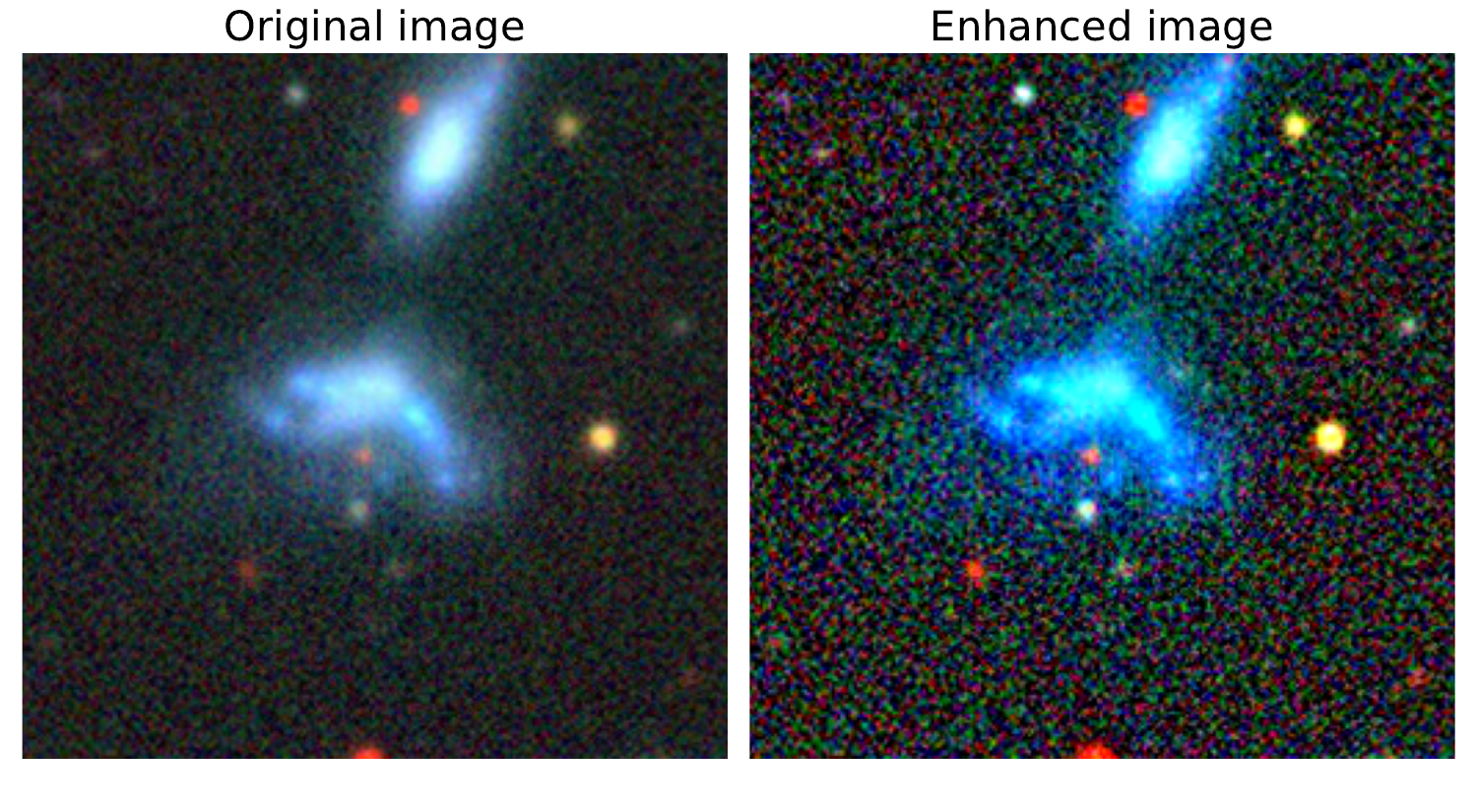}
    }
    \setcounter{subfigure}{12}
    \subfigure[]{
        \includegraphics[width=0.4\textwidth]{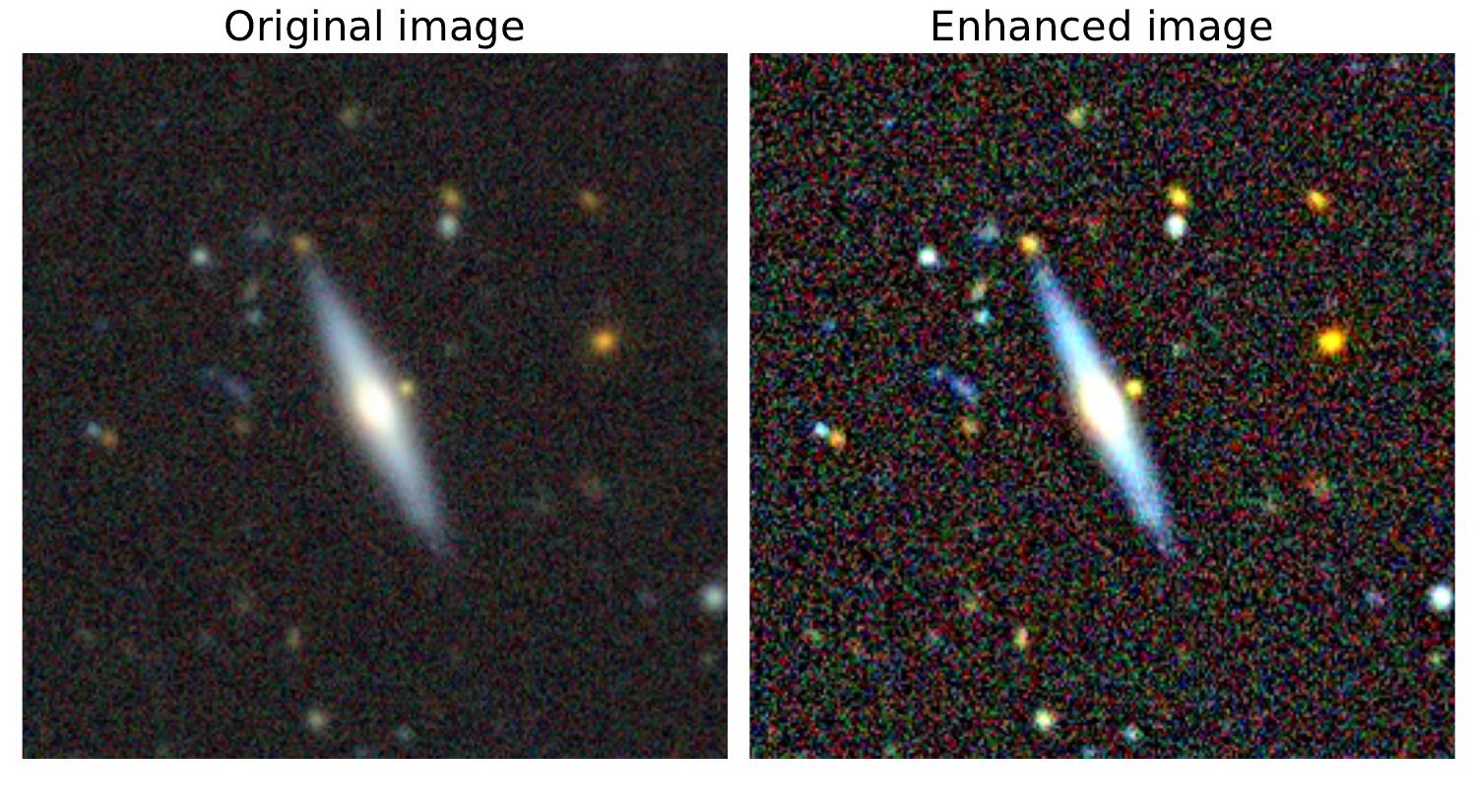}
    }
    \setcounter{subfigure}{13}
    \subfigure[]{
        \includegraphics[width=0.4\textwidth]{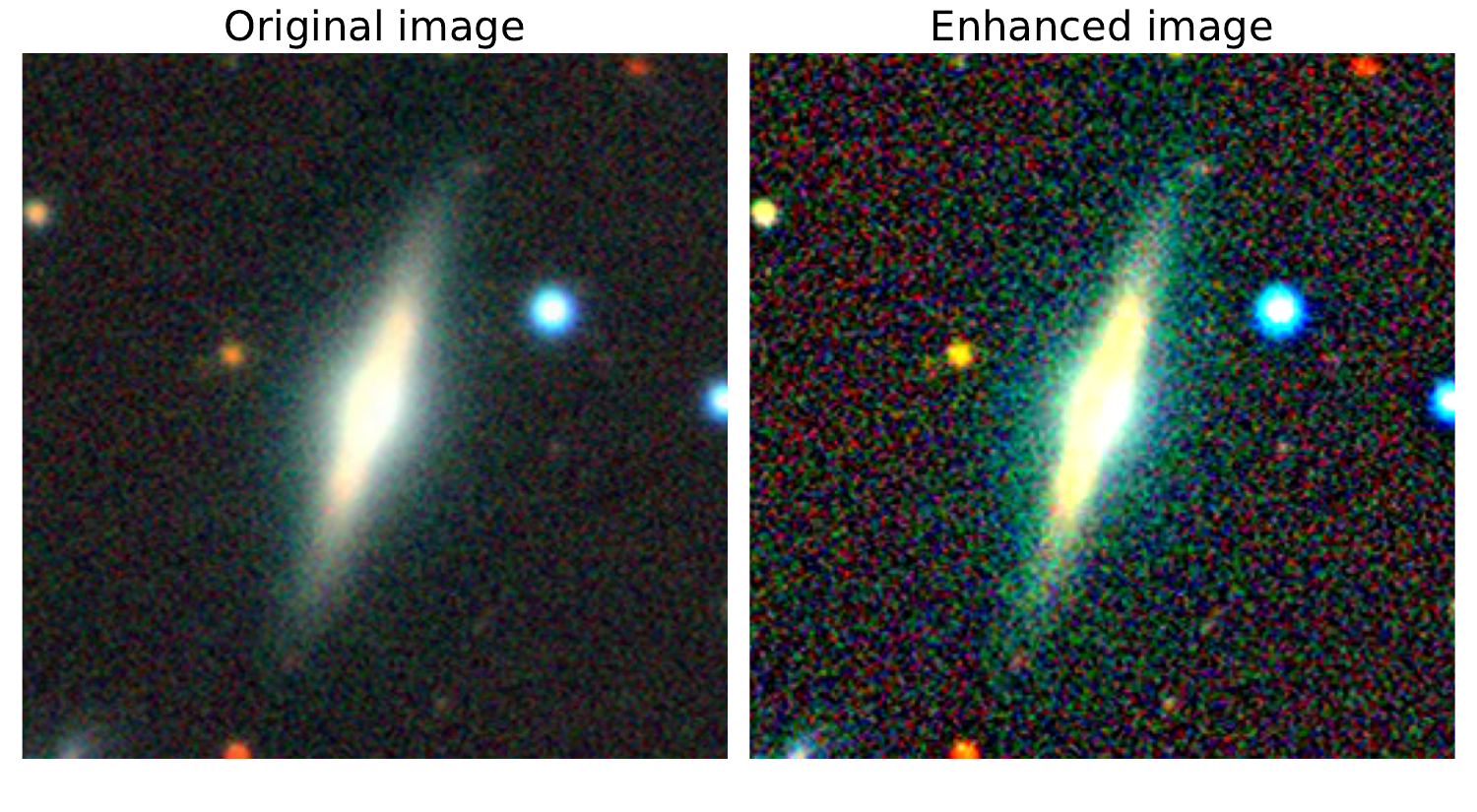}
    }
    \setcounter{subfigure}{14}
    \subfigure[]{
        \includegraphics[width=0.4\textwidth]{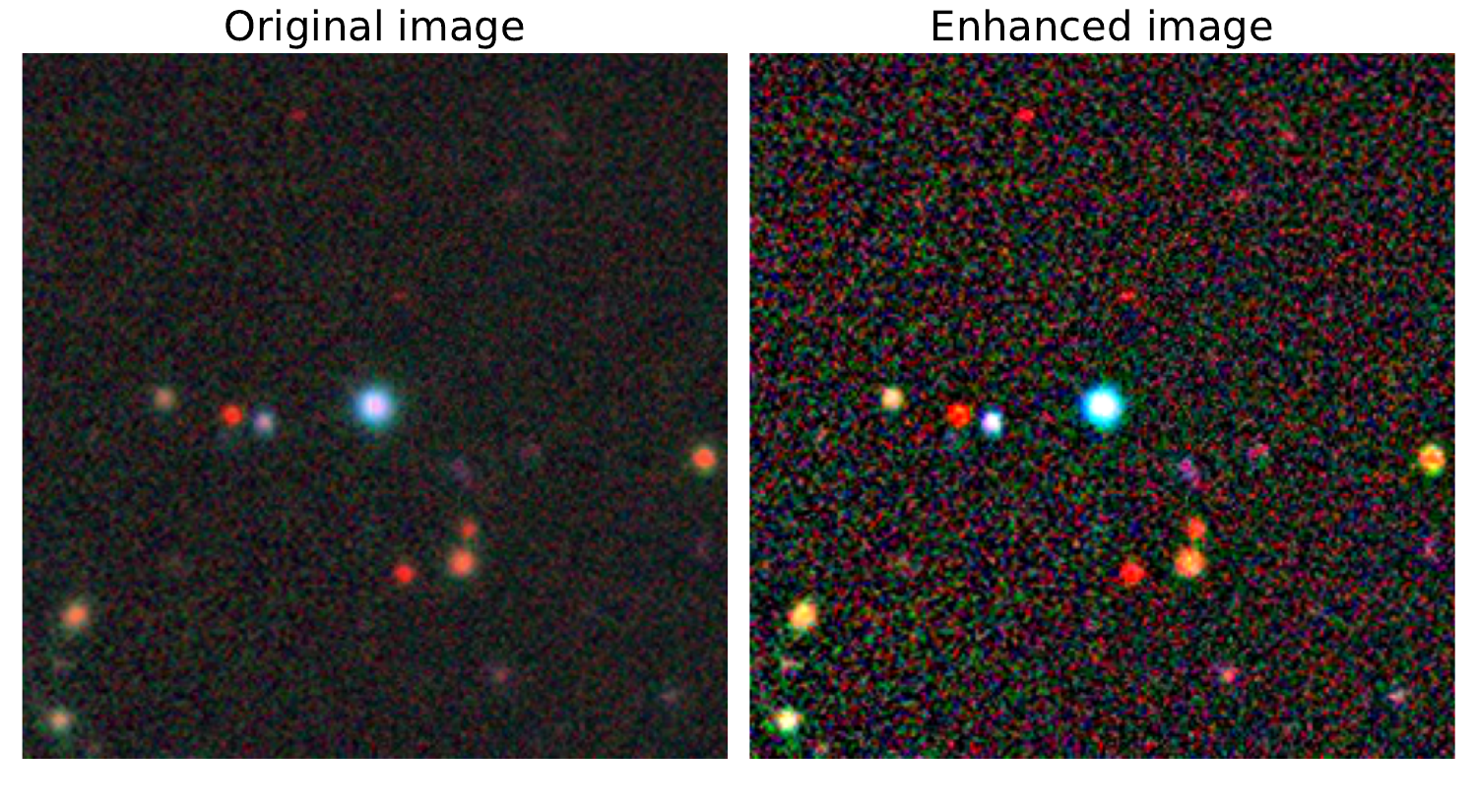}
    }
    \setcounter{subfigure}{15}
    \subfigure[]{
        \includegraphics[width=0.4\textwidth]{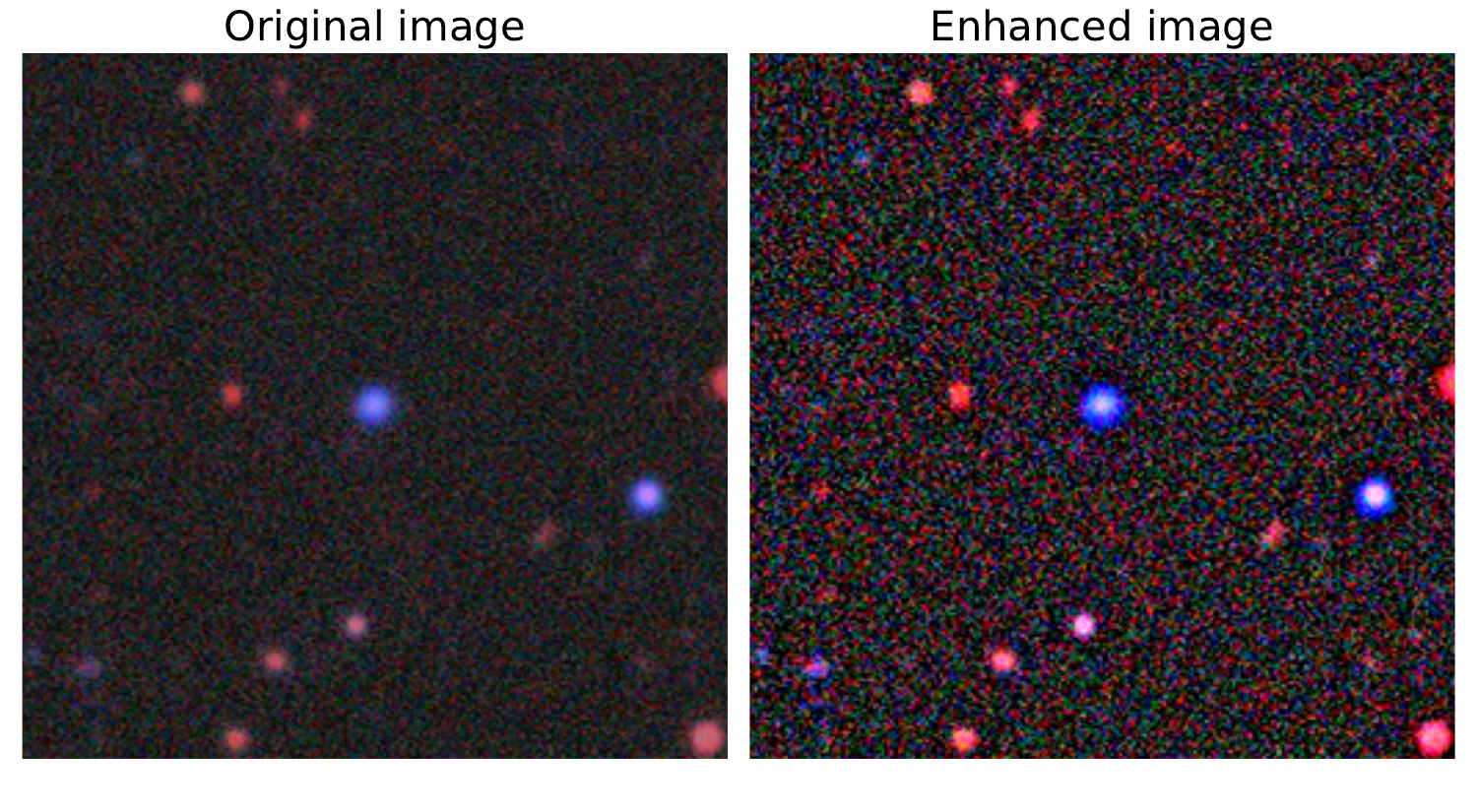}
    }
    \caption{(Continued.)}
\end{figure}

The non-parametric method, which has been used in massive galaxies to separate the galaxy types, might not be effective in classifying dwarf galaxies (e.g., \citetalias{2024MNRAS.529..499L}). Therefore, we also carry out visual inspection to further check the capability of non-parametric methods to distinguish classical types of dwarf galaxies. To get more details in the image for better classification, we also create an ``unsharp masked" version of all the images. Unsharp masking \citep{1977AASPB..16...10M} is a linear image processing technique that adds sharp details, identified as a difference between the original and its blurred version, back to the original image. The amount and radius represent the multiple of the sharp details and the sigma parameter of the Gaussian filter, respectively. We have tried many different combinations of these two parameters, and find that most of the resulting images are similar. Therefore, we use a fixed amount and fixed radius to generate the ``unsharp masked" images. 

Here we employ the same characteristics found in massive galaxies to separate the galaxy type: the criteria for identifying mergers are similar to those established in \citetalias{2024MNRAS.529..499L}: tidal features, the visible large-scale asymmetry within the main body of the galaxy, or shells; ETGs are characterized by a luminous core in their center and also with a regular shape, while LTGs exhibit spiral arms or bars, or some features that are very different from ETGs and mergers; part of the edge-on galaxies which are very difficult to figure out the classes definitively and a small number of point galaxies are classified as unsure galaxies. We show examples of ETGs, LTGs, mergers, and unsure objects in Figure \ref{fig:type}.

To assist our visual classification in some confusing cases, we also exploit the probability catalog provided by \cite{2023MNRAS.526.4768W}, who used deep learning models trained on Galaxy Zoo volunteer votes to measure morphologies for 8.67 million galaxies in the DESI Legacy Imaging Surveys. We find that most of our AGN-host dwarfs have the probability of features in their catalog. For the purpose of classifying dwarf galaxies into typical types (e.g., ETGs, LTGs, and mergers), here we use the following four probabilities: smooth fraction (SFSM); featured or disk fraction (SFFDF); fraction of no spiral arm (SANo); and fraction of non-merger (MNo). 

At first, we analyze the above four probabilities of the galaxies we have already visually determined their types, and find that most of the galaxies can be distinguished with the following criteria: ETGs ($SFSM \geq 0.8$), LTGs ($SFFDF \geq 0.8$ or $SANo \leq 0.35$), and mergers ($MNo \leq 0.35$). Then, we adopt the above criteria to classify those sources that are confused to separate by visual inspection. Finally, there are 24 objects (including 20 edge-on and 4 point galaxies), which will be removed in the following discussion, can not be determined their types neither by visual inspection nor with the assistance of the deep learning probabilities. The numbers and fractions of each morphological type for the total and various subsamples of our AGN-host dwarfs are listed in Table \ref{tab:fraction}, in which the $1\sigma$ uncertainty of the fraction is computed with the method presented in \cite{Cameron_2011}.

\begin{table}[tb]
\movetableright=-15mm
\centering
\caption{Numbers and fractions in different morphological types for various subsamples.}
\label{tab:fraction}
\begin{tabular}{ccccccc}\\\hline \hline
\multirow{1}{*}{Samples}&\multicolumn{1}{c}{ETGs}&\multicolumn{1}{c}{LTGs}&\multicolumn{1}{c}{Mergers}&\multicolumn{1}{c}{Unsure}&\multicolumn{1}{c}{Total}\\\cline{1-6}             
\multirow{1}{*}{X-ray}    &13/$37.1\pm{8.0}$\%  &18/$51.4\pm{8.2}$\% &3/$8.6\pm{5.0}$\%   &1/$2.9\pm{3.5}$\%   &35/100\%\\  
\multirow{1}{*}{BPT}      &63/$62.4\pm{4.8}$\%  &23/$22.8\pm{4.2}$\% &3/$3.0\pm{1.8}$\%   &12/$11.9\pm{3.2}$\% &101/100\%\\                 
\multirow{1}{*}{\HeII}    &35/$42.7\pm{5.4}$\%  &34/$41.5\pm{5.4}$\% &8/$9.8\pm{3.3}$\%   &5/$6.1\pm{2.7}$\%   &82/100\%\\
\multirow{1}{*}{Mid-IR}   &21/$22.8\pm{4.4}$\%  &37/$40.2\pm{5.1}$\% &31/$33.7\pm{4.9}$\% &3/$3.3\pm{2.0}$\%   &92/100\%\\  
\multirow{1}{*}{Radio}    &3/$25.0\pm{11.9}$\%   &6/$50.0\pm{13.4}$\%  &2/$16.7\pm{10.7}$\%  &1/$8.3\pm{8.9}$\%   &12/100\%\\
\multirow{1}{*}{Variability} &21/$24.1\pm{4.6}$\%  &56/$64.4\pm{5.1}$\% &6/$6.9\pm{2.8}$\%  &4/$4.6\pm{2.4}$\%   &87/100\%\\\hline    \multirow{1}{*}{Total}    &142/$36.6\pm{2.4}$\% &170/$43.8\pm{2.5}$\%&52/$13.4\pm{1.7}$\% &24/$6.2\pm{1.2}$\% &388/100\%\\\hline
\end{tabular}
\tablecomments{We calculate the $1\sigma$ uncertainty of each fraction using the method provided in \cite{Cameron_2011}.}
\end{table}

\section{RESULTS AND DISCUSSION} \label{subsec:result}
\subsection{Structural Parameters}\label{subsec:struc}

\begin{figure}[t]
    \centering
    \includegraphics[width=1\linewidth]{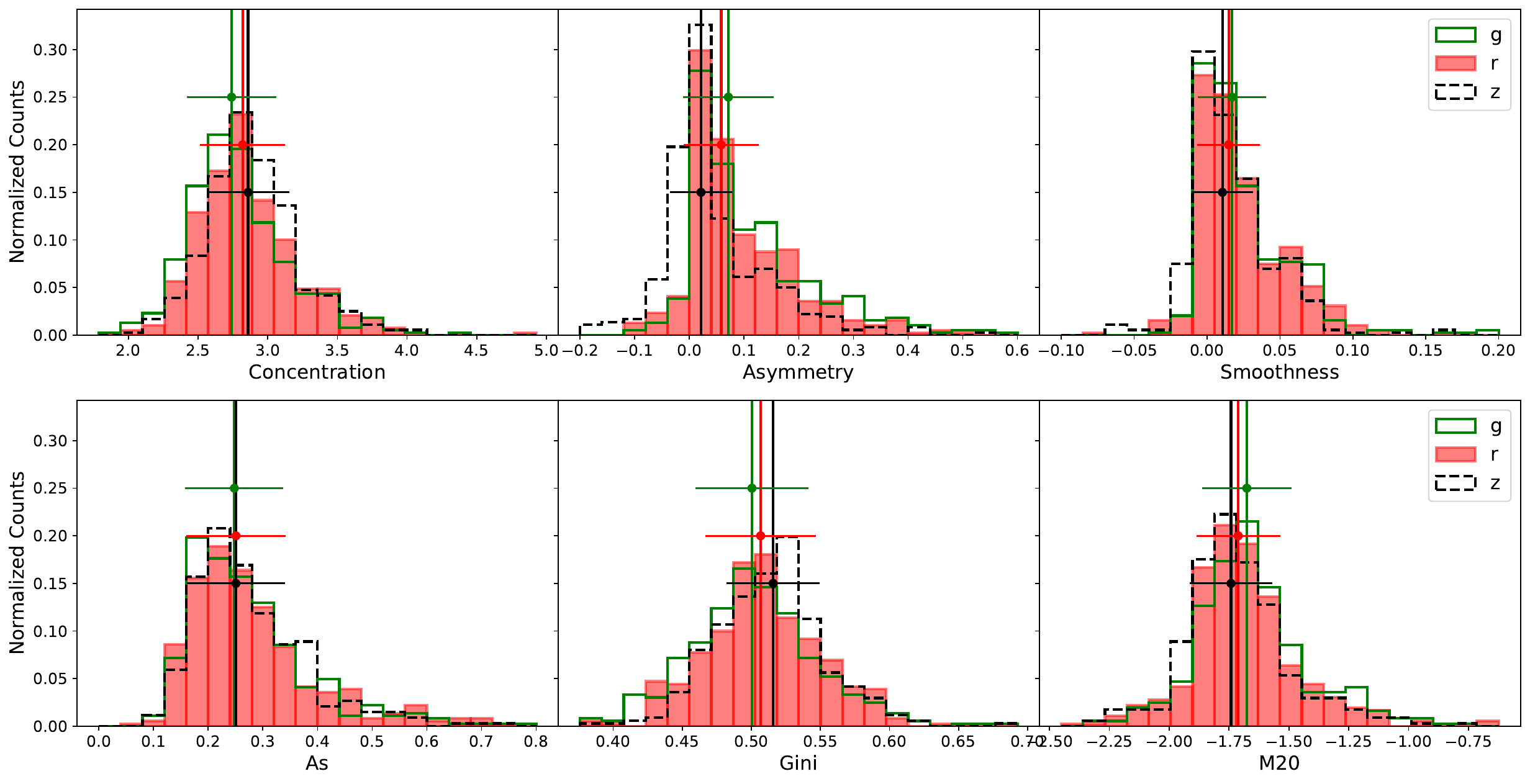}
    \caption{Distributions of non-parametric coefficients obtained from the $g$- (green histogram), $r$- (red histogram), and $z$-band (black histogram) images for the whole sample of AGN-host dwarf galaxies. The vertical lines indicate the corresponding median values with their associated uncertainties.}
    \label{fig:band}
\end{figure}

From Table \ref{tab:mean} in Appendix \ref{subsec:A1}, we can see that the mean and median values of each non-parametric coefficient obtained from different bands only show a slight difference for the whole sample of AGN-host dwarf galaxies. Furthermore, these parameters from different bands also show similar distributions, as illustrated in Figure \ref{fig:band}. These results suggest that the structural parameters are almost independent of the band used for the measurement within this wavelength range (i.e, $4700-9300$ \AA). Therefore, we will mainly discuss the results based on the \textit{r}-band images in the following sections since the \textit{r}-band data contain nearly the same number of sources as \textit{g}-band but suffer from less extinction.

\begin{table}[tb]
\movetableright=-15mm
\centering
\caption{Result of Kolmogorov-Smirnov test.}
\label{tab:KS}
\begin{tabular}{ccccccc}\\\hline \hline
\multirow{1}{*}{Bands}&\multicolumn{1}{c}{$C$}&\multicolumn{1}{c}{$A$}&\multicolumn{1}{c}{$S$}&\multicolumn{1}{c}{$A_\mathrm{S}$}&\multicolumn{1}{c}{$G$}&\multicolumn{1}{c}{$M_{20}$}\\\cline{1-7}             
\multirow{1}{*}{\textit{g-r}}    &0.004   &0.068  &0.372  &0.901 &0.067  &0.021\\                   
\multirow{1}{*}{\textit{r-z}}    &0.211   &$<10^3$&0.004  &0.814 &0.011  &0.202\\                       
\multirow{1}{*}{\textit{g-z}}    &$<10^3$ &$<10^3$&$<10^3$&0.693 &$<10^3$&$<10^3$\\\hline
\end{tabular}
\tablecomments{The \textit{p}-value of the Kolmogorov-Smirnov test on different parameters from \textit{grz} bands. Row (1): Comparison between \textit{g}- and \textit{r}-band. Row (2): Comparison between \textit{g}- and \textit{z}-band. Row (3): Comparison between \textit{r}- and \textit{z}-band.}
\end{table}

However, we also note that there exists a slight trend of increasing $C$ and $G$ but decreasing $A$, $S$, and $M_{20}$ towards longer wavelengths, suggesting a shallower but more asymmetric light distribution at bluer bands. To validate the trend, we perform the Kolmogorov–Smirnov (K-S) test on \textit{g}-, \textit{r}- and \textit{z}-band distributions, suggesting that more than half of the \textit{p}-values are lower than 0.05 (Table \ref{tab:KS}), and the \textit{p}-value between \textit{g}- and \textit{z}-band are lower than 0.05, expect $A_\mathrm{S}$. These low values indicate that the samples from \textit{grz} bands have different distributions in most parameters.

It is known that the morphology and quantitative structures at $g$-band traces the distribution of younger stellar populations, whereas those at $z$-band trace older stellar populations that dominate the total stellar mass. Therefore, the above results suggest that (1) these AGN-hosts only contain a relatively low-level star formation; and (2) older stars are more likely to reside in the center, whereas younger populations appear to be distributed more prevalent in outer regions, supporting that AGNs in dwarf galaxies might suppress the current star formation of the host galaxies (e.g., \citealt{2020ApJ...903...58C,2021RAA....21..204C,2024RAA....24f5006L}) or these dwarfs exhibit a similar inside-out formation scenario to massive galaxies \citep{2015A&A...581A.103G,2015MNRAS.452.1128M,2017MNRAS.465.4572Z}, but in contrast to the outside-in mode found in nearby dwarf galaxies \citep{2012AJ....143...47Z}. However, the median values from different bands are also consistent with each other within the associated uncertainties, and thus it needs more data to confirm these trends.

\subsubsection{The \textit{CAS} System}\label{subsec:CAS}
As mentioned in \S\ref{subsec:measure}, the \textit{CAS} parameters describe quantitatively the light distribution within galaxies. For massive galaxies, it is found that LTGs generally show a lower light concentration (\textit{C}) than ETGs (see, e.g., \citealt{2003ApJS..147....1C,2007ApJS..172..406S,2013MNRAS.434..282F,2018MNRAS.480.2266M,2020MNRAS.491.1408M,2022MNRAS.511..607J,2023ApJ...954..113Y}), which may be linked to galaxy formation history, such as merging (lower $C$) or secular evolution (higher $C$). For the total sample of our AGN-host dwarfs, the mean concentration is $\langle C \rangle=2.9\pm0.4$, smaller than those of massive galaxies presented in \cite{2003ApJS..147....1C}, who obtained $\langle C \rangle =4.4\pm0.3$, $3.9\pm0.5$ and $3.1\pm0.4$ for ellipticals, early-type spirals (Sa and Sb) and late-type spirals (Sc and Sd), respectively, in the representative sample of bright, well-resolved, Hubble-type galaxies (\citealp{1996AJ....111..174F}). We note that $\langle C \rangle$ for our AGN-host dwarfs is similar to that of dwarf irregulars (dIs; $\langle C \rangle =2.9\pm0.3$), though it is a bit larger than the value ($2.5\pm0.3$) of dwarf ellipticals (dEs).   

For the asymmetry ($A$), the mean value for our total sample is $\langle A \rangle=0.09\pm0.12$, close to that of early-type spirals ($\langle A \rangle=0.07\pm0.04$; \citealt{2003ApJS..147....1C}), but is $>$4 times higher than that of (dwarf) ellipticals. This asymmetry is smaller than those of late-type spirals ($\langle A \rangle=0.15\pm0.06$) and dIs ($\langle A \rangle=0.17\pm0.10$). However, our AGN-host dwarfs have a much larger ($2-6\times$) scatter in $\langle A \rangle$ than the above types of galaxies, except for dwarf irregulars, which have a comparable scatter (\citealt{2003ApJS..147....1C}). 

The clumpiness ($S$) parameter, which can be used to trace a galaxy's star formation properties as argued in \cite{2003ApJS..147....1C}, shows a significant difference ($4-20\times$) between our AGN-host dwarfs ($\langle S \rangle = 0.02\pm0.03$) and massive spirals ($\langle S \rangle=0.08\pm0.08$ for Sa and Sb; $\langle S \rangle=0.29\pm0.13$ for Sc and Sd), and their irregular counterparts ($\langle S \rangle=0.40\pm0.20$). We argue that this difference should not be mainly caused by the difference of physical resolutions between our sample and those in \cite{2003ApJS..147....1C}, since (1) we find no dependence of $S$ on redshift for our sample; and (2) the change in $S$ due to lower resolution, if exists, would be much smaller than that observed here according to the simulating results presented in \cite{2003ApJS..147....1C} and \cite{2004AJ....128..163L}. 

On the other hand, the mean value of clumpiness for our dwarfs is only slightly larger than those of dwarf ($\langle S \rangle=0.00\pm0.06$) and massive ($\langle S \rangle=0.00\pm0.04$) ellipticals. It is known that spiral galaxies show modest star-forming activities, and these dIs are dominated by star formation (\citealt{2000AJ....119.2757V}). Combining this with the fact that both massive and dwarf ellipticals have little star formation, the above results suggest that our AGN-host dwarfs may generally contain little star formation, consistent with the stellar population results that these sources have a median light-weighted stellar age of $>10^8$~yr (\citealt{2024RAA....24f5006L}).     
\subsubsection{\texorpdfstring{The \textit{G}$-M_{20}$ Method}{The G-M20 Method}}\label{subsec:GM20}
As discussed in \cite{2004AJ....128..163L}, the $G$ and $M_{20}$ parameters are correlated with concentration $C$ but $G$ does not depend on the large-scale distribution of the galaxy's light, whereas $M_{20}$ is more sensitive than $C$ to merging signatures such as multiple nuclei and bright tidal tails. Therefore, the combination of $G$ and $M_{20}$ can be used to distinguish different types of morphology among massive galaxies (e.g., \citealp{2004AJ....128..163L,2015MNRAS.451.4290S,2015MNRAS.454.1886S}). 

For the whole sample of our AGN-host dwarfs, we obtain the mean values of $G$ and $M_{20}$ of $\langle G \rangle=0.51\pm0.04$ and $\langle M_{20} \rangle=-1.69\pm0.22$, very close to those ($\langle G \rangle=0.51\pm0.04$ and $\langle M_{20} \rangle=-1.76\pm0.21$) of the Sc-Sd galaxies in the \cite{1996AJ....111..174F} catalog measured by \cite{2004AJ....128..163L}. Whereas massive ETGs and Sa-Sb spirals have higher $\langle G \rangle$ ($0.59\pm0.04$ and $0.54\pm0.05$, respectively) but lower $\langle M_{20} \rangle$ ($-2.45\pm0.15$ and $-2.11\pm0.38$, respectively) values. Compared to dIs, which have $\langle G \rangle=0.51\pm0.03$ and $\langle M_{20} \rangle=-1.31\pm0.24$ measured with $B$-band images (\citealp{2004AJ....128..163L}), AGN-host dwarfs have a similar $\langle G \rangle$ but a smaller  $\langle M_{20}\rangle$. 

Therefore, AGN-host dwarfs usually have shallow light profiles comparable to late-type spirals and dIs, but are more symmetric (similar to early-type spirals). Compared to dIs and massive spirals, however, they show much less clumpiness. The mean $S$ of the AGN-host dwarfs is only slightly larger than those of massive ETGs and dEs. Do these results suggest that AGN-hosts lack star-forming activities or AGNs prefer to reside in early-type objects? To further explore these questions, we will compare our results to those of massive counterparts and normal dwarfs with the assistance of our visual classification in the following section.

\subsection{Visual Classification}\label{subsec:class}

\begin{figure}[t]
\includegraphics[width=0.45\textwidth]{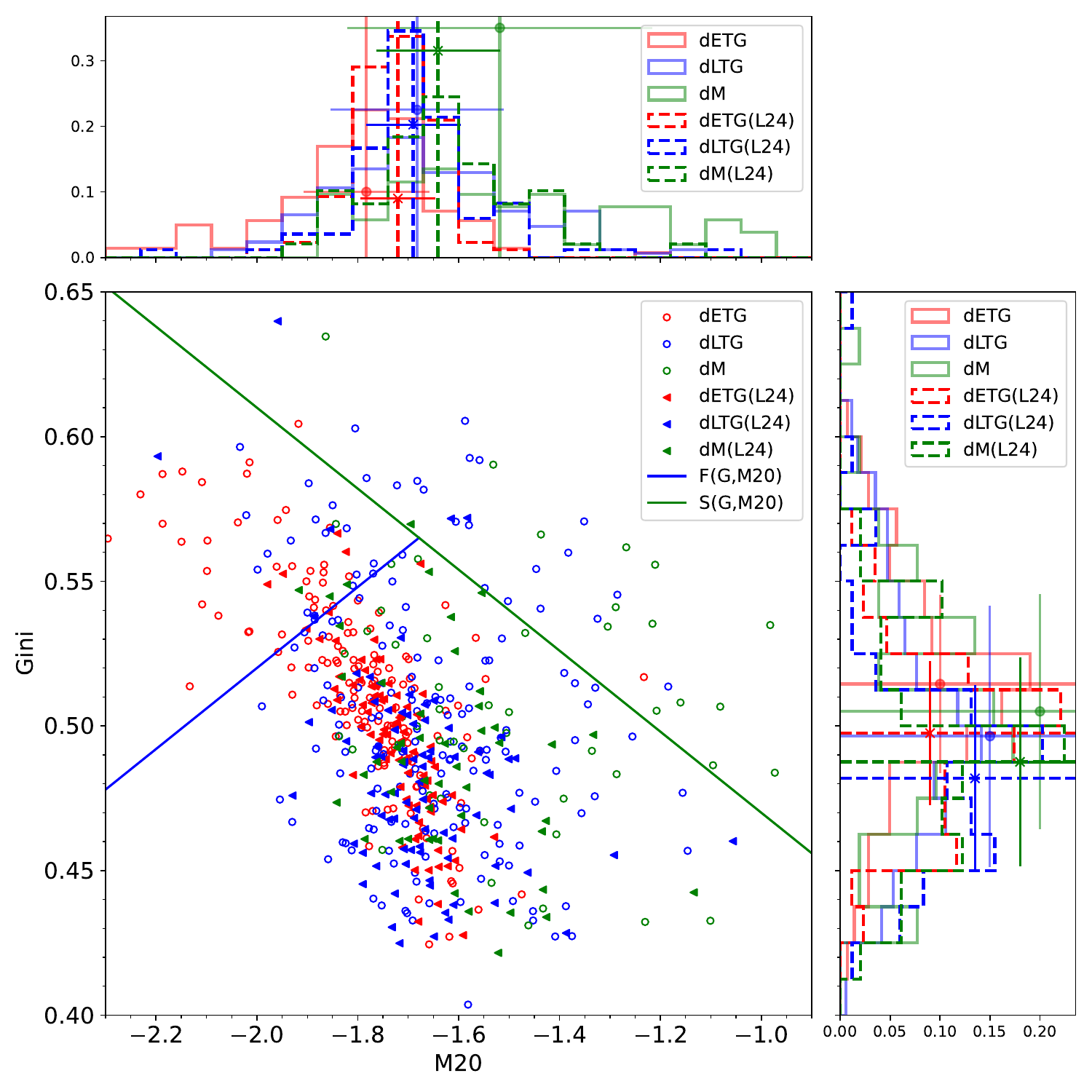}
\centering
\caption{\textit{G} versus $M_{20}$ for our AGN-host (open circles) and the \citetalias{2024MNRAS.529..499L} normal (solid triangles) dwarfs. The blue and green lines correspond to $F(G, M_{20})$ and $S(G, M_{20})$, respectively. Distributions of parameters and median values (together with their associated uncertainties) are shown on the sides of the main panel. Here, ``d'' represents the dwarf regime. ETG = early-type, LTG = late-type, M = merger.}
\label{fig:GM20}
\end{figure}

After visual inspection, we find that about 37\%, 44\%, and 13\% of the total sample comprise ETGs, LTGs, and mergers, respectively, as listed in Table \ref{tab:fraction}. This merging fraction ($f_{\mathrm{merger}}$) is similar to that in \cite{2024MNRAS.532..613B}, who have estimated $f_{\mathrm{merger}}$ ($\equiv 0.14\pm0.03$) for a sample of 162 AGN-hosts ($10\sp{8}<M_\star/M_\odot<10\sp{10}$ and $0.1<z<0.3$) selected via SED-fitting. However, our $f_{\mathrm{merger}}$ is two times larger than those in IR-selected (\citealt{2019MNRAS.489L..12K}) and radio-selected (\citealt{2022MNRAS.511.4109D}) AGN-host dwarfs at $0.1<z<0.5$.  Nevertheless, these merging fractions indicate that mergers/interactions are not the primary driver of AGN activities in dwarf galaxies.  

However, the fractions obtained via the \textit{G}$-M_{20}$ parametric method \citep{2004AJ....128..163L}, as shown in Figure \ref{fig:GM20}, are much different from our visual results. For the visually classified sample, the majority of AGN-host dwarfs ($\sim$74\%) are located in the LTG region, 16\% are classified as ETGs, and about 10\% are treated as mergers. This result further confirms that the \textit{G}$-M_{20}$ method, which has been used to separate galaxy types of massive objects, may not be effective in separating morphological classes (especially ETGs vs LTGs) for dwarf galaxies as noted in \citetalias{2024MNRAS.529..499L}. It is worth noting that for our sample galaxies, the merging fractions identified by visual inspection and \textit{G}$-M_{20}$ method are apparently consistent, although only 40\% of \textit{G}$-M_{20}$-based mergers are visually confirmed. As a compromise solution, therefore, the \textit{G}$-M_{20}$ method seems still usable for estimating the merging fraction of AGN-host dwarfs when the sample size is too large to perform visual classification.

\subsection{Comparison to literature results} \label{subsec:compare}
\subsubsection{Comparison with Normal Dwarf Galaxies} \label{subsec:normal}
Using deep Hyper Suprime-Cam Subaru Strategic Program imaging data, \citetalias{2024MNRAS.529..499L} studied the morphological properties for a complete, unbiased sample of 257 dwarf galaxies ($8<\log (M_\star/M_\odot)<9.5$) at $z < 0.08$ in the COSMOS field. They visually classified their sample sources into four types, i.e., early-type, late-type, featureless class, and irregular, among which some objects could contain interacting/merging signs. Furthermore, the authors also employed \texttt{statmorph} to measure the same non-parametric coefficients as adopted in our study. Therefore, it is an ideal sample that can be used to compare the morphological properties of our AGN-hosts with the general dwarf population in the nearby Universe. 

Before the comparison, however, we need to re-divide the objects in \citetalias{2024MNRAS.529..499L} into the same three types (ETGs, LTGs, and mergers) as our visual classification. To do so, we re-assign irregulars to LTGs and sources with interacting/merging signs to mergers, and keep the remaining ETGs and LTGs unchanged. The featureless galaxies are different from any galaxy morphology (e.g., \citetalias{2024MNRAS.529..499L}), so we remove these galaxies from the following comparison. The resulting fractions of ETGs, LTGs and mergers in \citetalias{2024MNRAS.529..499L} dwarf sample are $f_{\mathrm{ETG}}=(39.3\pm3.3)\%$, $f_{\mathrm{LTG}}=(38.4\pm3.3)\%$ and $f_{\mathrm{merger}}=(22.4\pm2.8)\%$, respectively.  

Compared to the above fractions, we find that the differences between AGN-host and normal dwarfs are $\Delta f_{\mathrm{ETG}} = (-2.7\pm4.1)\%$ and $\Delta f_{\mathrm{LTG}} = (5.4\pm4.1)\%$, indicating that AGNs tend to have higher probabilities to reside in LTGs than ETGs. Regarding the merging fraction, it is found to be $f_{\mathrm{merger}}=(13.4\pm1.7)$\% for our AGN-host dwarfs. This merging fraction is smaller by a factor of 1.7 than that of normal dwarfs in \citetalias{2024MNRAS.529..499L}, and thus further confirms that AGN activities in dwarf regime are not mainly triggered by mergers/interactions, but dominated by secular processes, as evidenced by the recent simulation \citep{2024MNRAS.52710855S}.

\begin{figure}[t]
\centering
\subfigure[Concentration vs Asymmetry]{
       \includegraphics[width=0.45\textwidth]{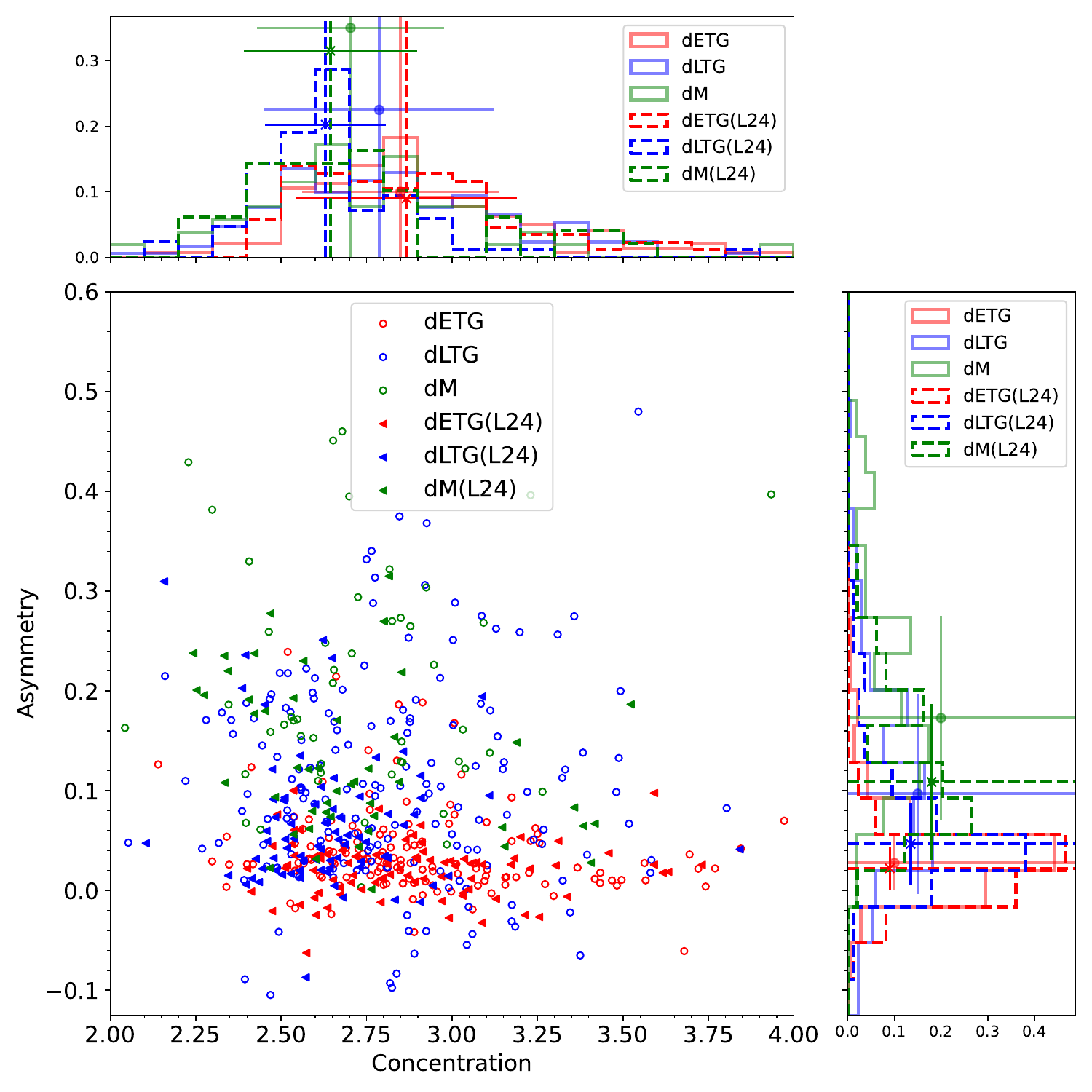}
       }
\subfigure[Concentration vs Clumpiness]{
       \includegraphics[width=0.45\textwidth]{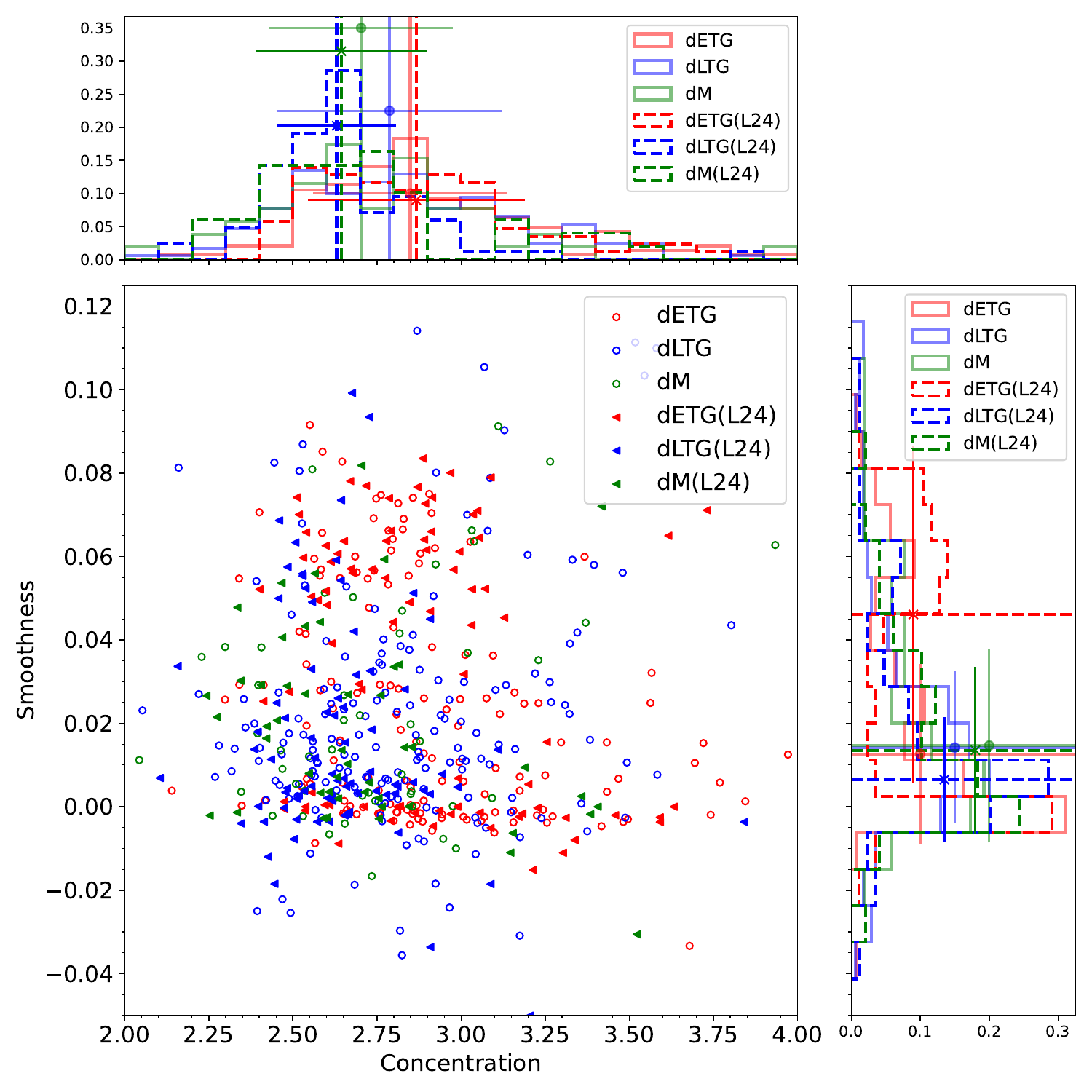}
       }
\subfigure[Asymmetry vs Clumpiness]{
       \includegraphics[width=0.45\textwidth]{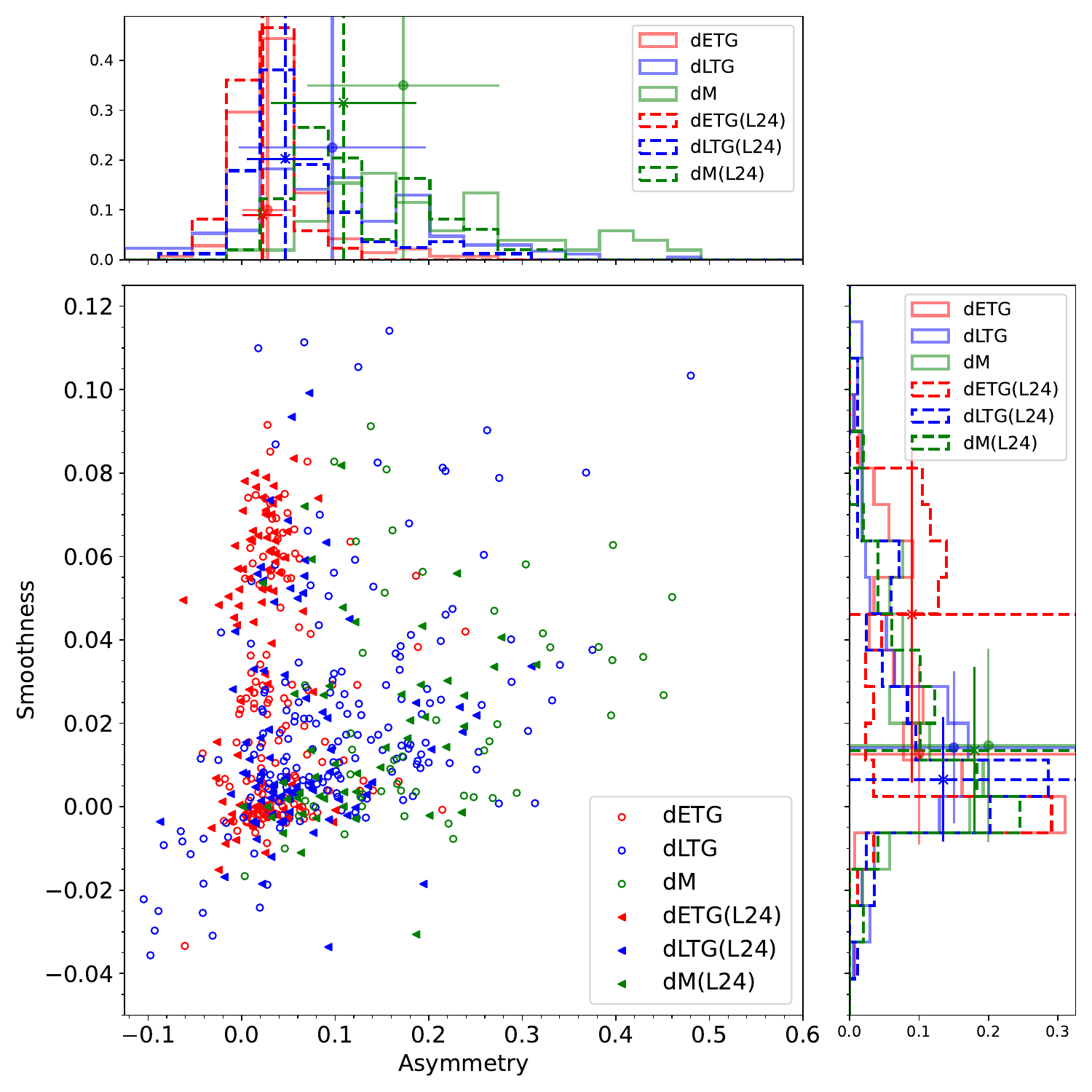}
       }
\caption{Concentration vs Asymmetry (a),  Concentration vs Clumpiness (b), and Asymmetry vs Clumpiness (c) for AGN-host (open circles) and normal (solid triangles; \citetalias{2024MNRAS.529..499L}) dwarfs. Distributions (shown by the histograms) of parameters and median values (shown by the lines, together with their associated uncertainties) are also plotted on the sides of the main panel. 
}
\label{fig:CAS}
\end{figure}

To have a quantitative view of the difference in the morphological information between our AGN-host and normal dwarfs, the following will further compare the measured parametric coefficients between our and L24's samples for different types of objects. As shown in Figure \ref{fig:GM20}, AGN-host dwarfs (open circles) have a broader spread than their normal counterparts (solid triangles) in the \textit{G}$-M_{20}$  plane. For each type, AGN-hosts show a larger median $G$ (please see the histograms alongside the main panel in Figure \ref{fig:GM20}) than normal dwarfs, but the difference is tiny, i.e., less than 5 per cent. Regarding $M_{20}$, the differences of the median values are also slight, i.e., from $-3\%$ (ETGs) to 8\% (mergers). However, the scatter of $M_{20}$ for AGN-hosts is $1.7-2.5$ times larger than normal dwarfs, indicating more complicated structures.

Compared with the normal dwarf sample, it is also interesting to note that there is a larger fraction of AGN-hosts located in the ``ETG" region (i.e., the left side of the blue $F(G,M_{20})$ line; $\sim$16\%), as well as that in the ``merger" region (i.e., above the green $S(G,M_{20})$ line; $\sim$10\%), as demonstrated in Figure \ref{fig:GM20}. For \citetalias{2024MNRAS.529..499L} dwarf sample, only $<$5\% objects are classified as ETGs, and $\sim$2\% sources are mergers. These differences may suggest that AGNs could have affected the structures of their hosts. 

Figure \ref{fig:CAS} plots different combinations of the $CAS$ parameters for AGN-hosts in comparison to normal dwarfs from \citetalias{2024MNRAS.529..499L}. In each subfigure, the distributions of parameters and median values are also shown on the sides of the main panel. (a) of Figure \ref{fig:CAS} illustrates the Concentration versus Asymmetry plane, from which we can see that both AGN-hosts and normal dwarfs are well mixed. Similar to what is found in $G$, all types of our sample galaxies have a larger median $C$ than their normal counterparts, but the differences are less than 6 per cent. However, the scatter for our LTGs is about 2 times larger. 

Meanwhile, there exists a much larger difference of asymmetry between AGN-host and normal dwarfs: the median $A$ for AGN-host ETGs, LTGs, and mergers are about 1.3, 2.1, and 1.6 times, respectively, higher than their normal counterparts, indicating that AGN-host dwarfs may have a more irregular morphology. It is noteworthy that numerous AGN-host LTGs exhibit high \textit{A} comparable to mergers, and many of them possess well-defined spiral arms or pronounced star-forming regions, as illustrated in (h) of Figure \ref{fig:type}. In contrast, ETGs form an obvious locus (i.e., $A\sim0.02$) with an increasing scatter when $C \lesssim 3$, and most ETGs are located between the $A = -0.02$ and $A=0.06$ lines.

(b) and (c) of Figure \ref{fig:CAS} show the Clumpiness versus Concentration and Asymmetry planes. It seems that there is a gap around $S=0.04$, and ETGs have two peaks at $S\sim0.0$ and $S\sim0.06$, for both AGN-host and normal dwarf samples. However, we find no significant difference in median $S$ among ETGs, LTGs, and mergers, except for normal ETGs, which have a $\gtrsim$3$\times$ higher median $S$. Nevertheless, the clumpiness of dwarf galaxies is significantly smaller in comparison to massive late-type spirals and dIs, as already discussed in \S\ref{subsec:CAS} and \citetalias{2024MNRAS.529..499L}, which might be indicative of lacking luminous star-forming regions in these dwarfs. 
\cite{2022MNRAS.514..607G} investigated the effect of AGN on the morphological parameters of their host galaxies, and found that all parameters are affected to some extent, depending on the fractional AGN contribution ($x_{\mathrm{AGN}}$). The above differences (e.g., $G$, $M_{20}$, and $C$) between AGN-host and normal dwarfs are consistent with those obtained by \cite{2022MNRAS.514..607G} with $x_{\mathrm{AGN}}\lesssim5\%$. Indeed, this overall AGN contribution to the total luminosity for these dwarfs agrees well with our previous results obtained through stellar populations (\citealt{2020ApJ...903...58C,2024RAA....24f5006L}). However, the difference in asymmetry between AGN-host and normal dwarfs is much larger than the simulated result presented in \cite{2022MNRAS.514..607G} for $x_{\mathrm{AGN}}\lesssim5\%$, suggesting that AGNs may prefer residing in more irregular hosts.

\subsubsection{Comparison with AGN-host Massive Galaxies} \label{subsec:massive}

\begin{figure*}[t]
    \centering
    \includegraphics[width=0.7\linewidth]{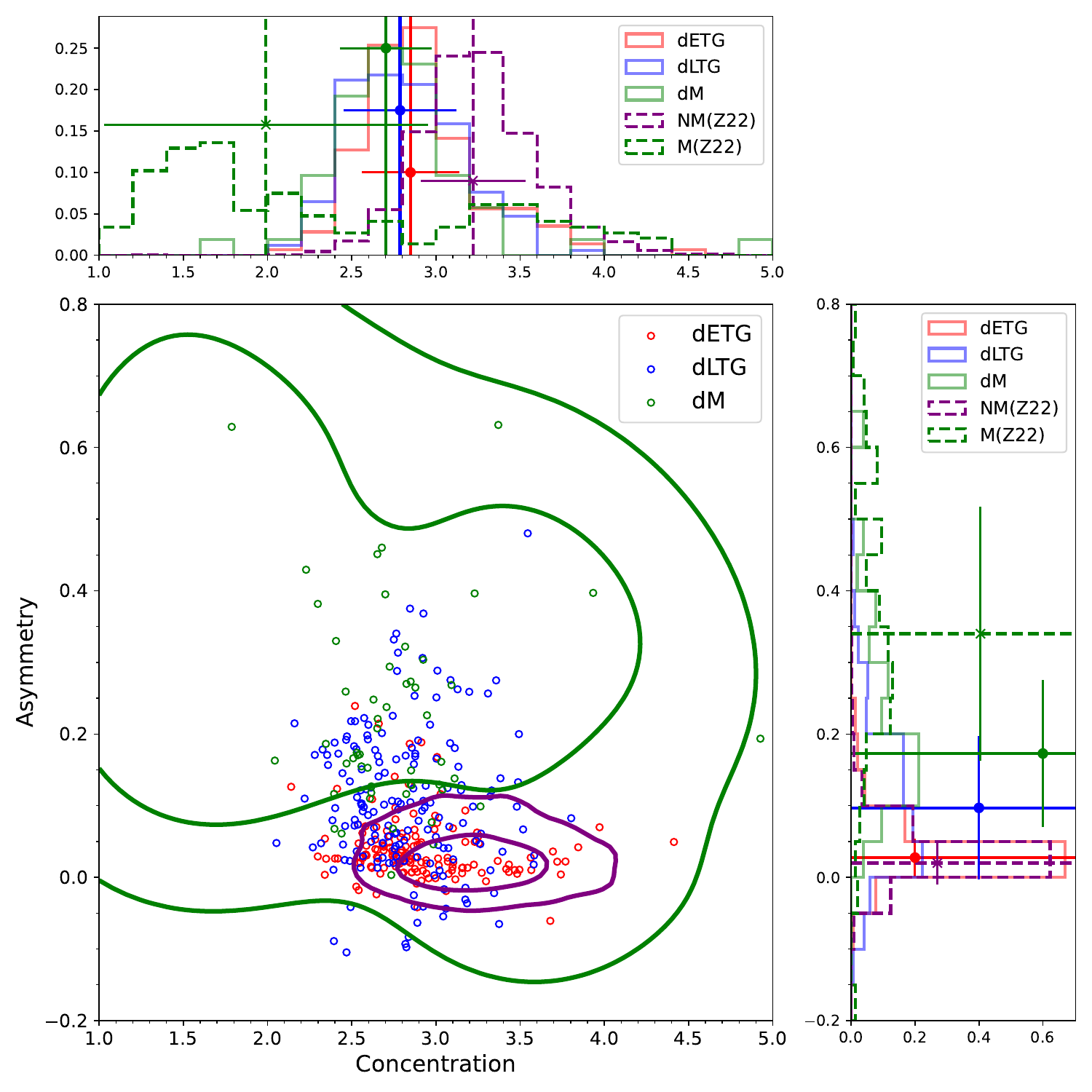}
    \caption{Concentration versus Asymmetry for AGN-host dwarfs (open circles) and massive (contours) galaxies from \citetalias{2022ApJ...925...70Z}. Distributions of parameters and median (dashed lines for massive galaxies, together with their associated uncertainties) values are shown on the sides of the main panel. The green and purple lines illustrate the 68\% and 95\% contours for mergers and non-merging galaxies, respectively. Here, ``d" represents the dwarf regime. ETG = early-type, LTG = late-type, M = merger. ``NM" and ``M" represent non-merging galaxies and mergers in \citetalias{2022ApJ...925...70Z}, respectively. }
    \label{fig:massive}
\end{figure*}

\citetalias{2022ApJ...925...70Z} employed \textit{i}-band image from Pan-STARRS to derive the non-parametric coefficients (including \textit{C}, \textit{A}, and $A_\mathrm{S}$) in low redshift ($0.04<z<0.15$) AGN-host massive galaxies, both for type 1 (AGN1s) and type 2 AGNs (AGN2s). To explore whether there is any difference in morphology between AGN-host massive and dwarf galaxies, here we compare our results with those from \citetalias{2022ApJ...925...70Z}.

\begin{figure}[t]
\centering
\subfigure[X-ray]{
\includegraphics[width=0.7\textwidth]{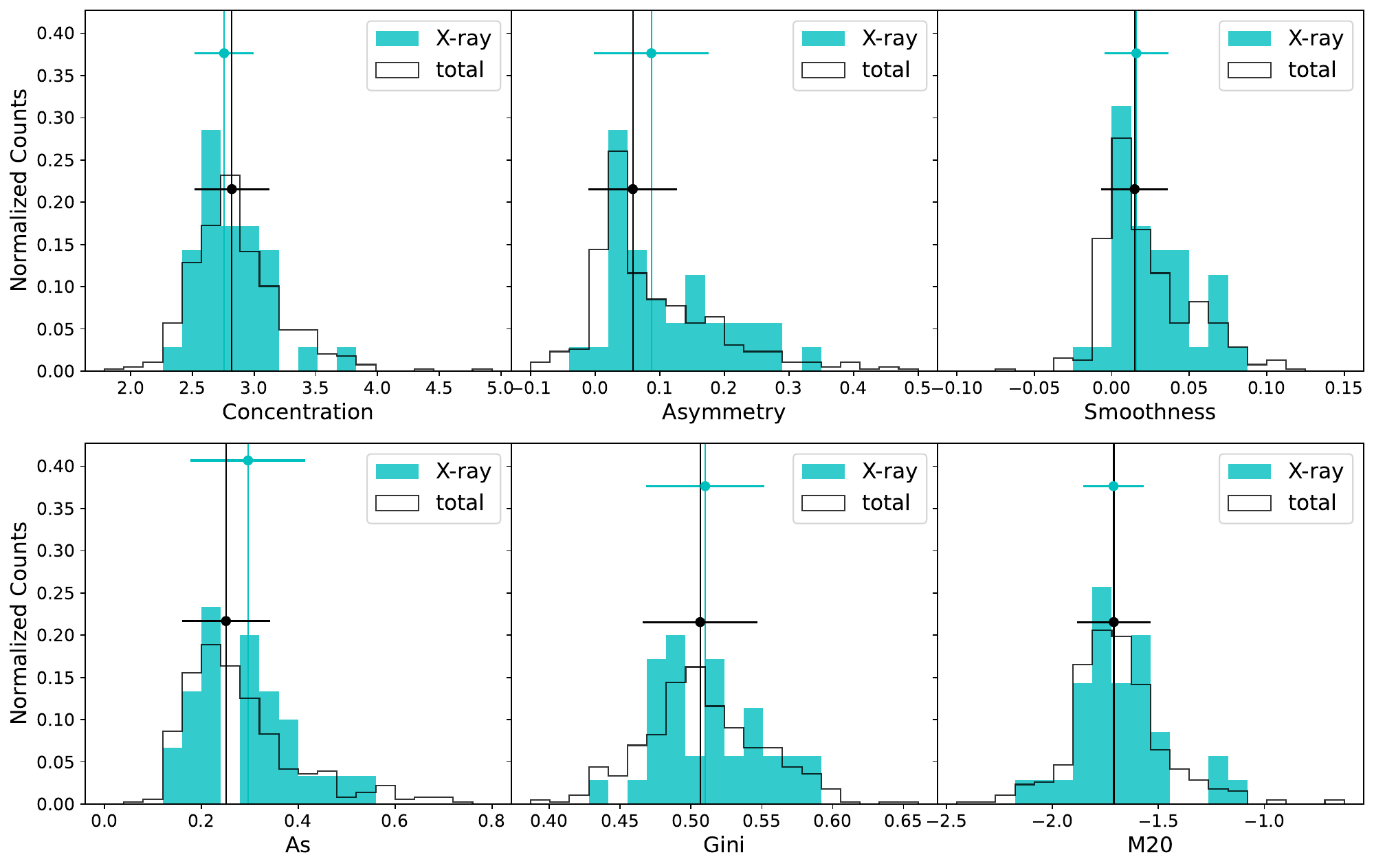}
}

\subfigure[\HeII]{
\includegraphics[width=0.7\textwidth]{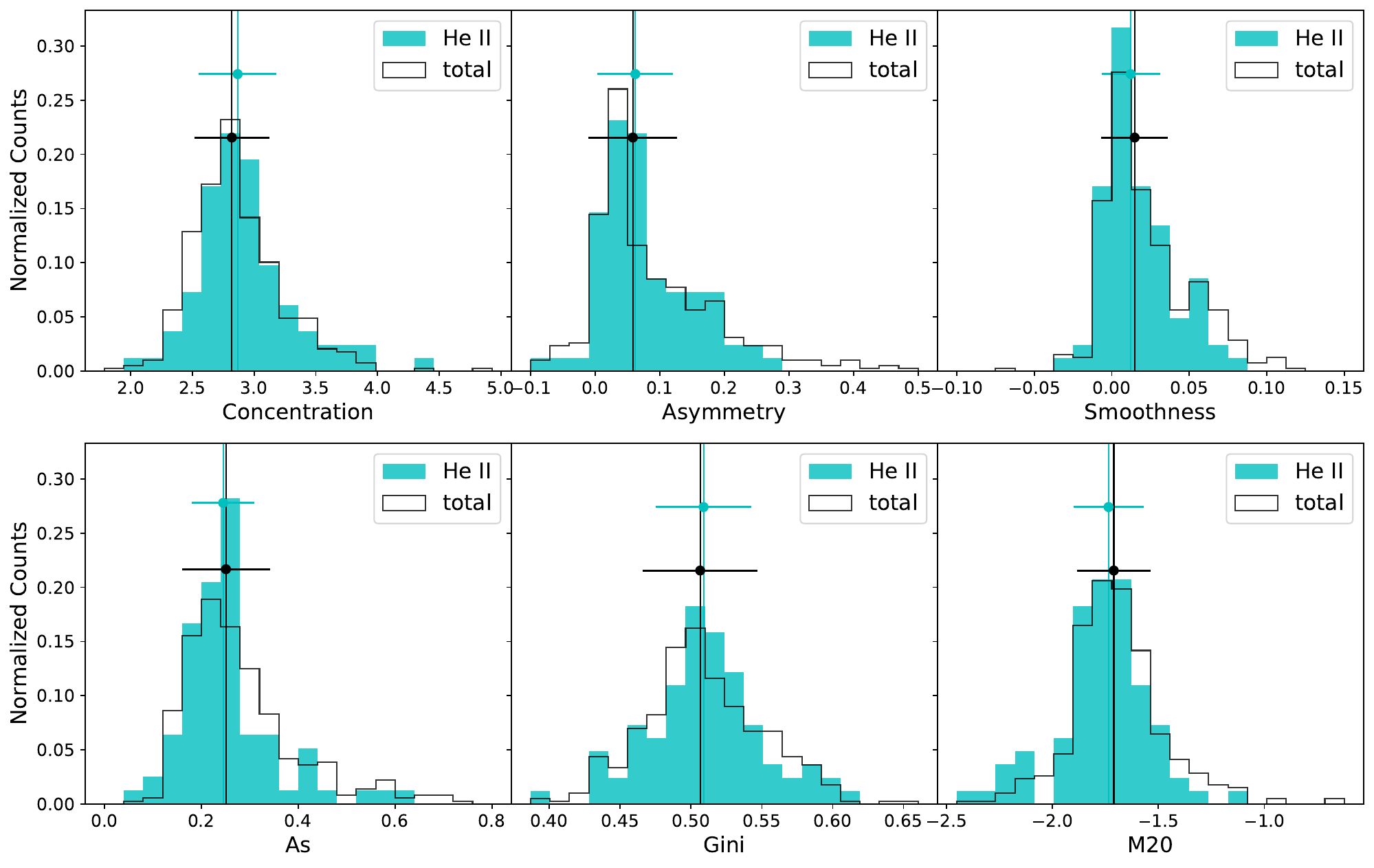}
}

\caption{The distributions of parameters for the subsamples (solid histogram), overlaid with the total sample (open histogram). The vertical lines represent the median values with their associated uncertainties. Note that $A_\mathrm{S}$ are only for \textbf{Sample A}.}
\label{fig:subsample}
\end{figure}

To select merging galaxies, \citetalias{2022ApJ...925...70Z} assumes that any extended source projected within 2$r_\mathrm{petro}$ of the primary target containing at least 25\% of its flux is a possible companion galaxy, and such systems are considered as major merger candidates. They found that the merger fractions are 3.1\% for AGN1s and 3.7\% for AGN2s, much lower than what is found in AGN-host dwarf galaxies, which might be caused by the different methods used to select merger candidates. Nevertheless, both results would suggest that secular processes dominate AGN triggering from low to the highest AGN luminosities (e.g., \citealt{2017MNRAS.466..812V}).

In Figure \ref{fig:massive}, we further plot \textit{C} versus \textit{A} for our AGN-host dwarfs (from \textit{r}-band; open circles) and \citetalias{2022ApJ...925...70Z}'s massive (from \textit{i}-band; contours) galaxies. The green and purple lines illustrate the 68\% and 95\% contours for mergers and non-merging galaxies, respectively. We can see that non-merging galaxies from \citetalias{2022ApJ...925...70Z} have a median \textit{C} close to late-type spirals but higher than all AGN-host dwarfs. The massive mergers, by contrast, have a much lower median value of \textit{C}, and the distribution is very broad, indicating shallower but more diverse light profiles in massive mergers. 

Regarding asymmetry, it is interesting to note that the distribution of \textit{A} for non-merging galaxies from \citetalias{2022ApJ...925...70Z} closely resembles our AGN-host dwarf ETGs, with most values concentrated around $A=0-0.05$. However, \textit{A} for massive mergers is considerably higher (about 2 times larger) than that of their dwarf counterparts, suggesting that major mergers in massive galaxies may transform a galaxy's morphology more violently and/or that ongoing mergers have a larger asymmetry than merger remnants.

\subsection{Comparison between Different Samples}\label{subsec:subsample}

\begin{figure}[t]
\figurenum{7}
\centering
\setcounter{subfigure}{2}
\subfigure[radio]{
\includegraphics[width=0.7\textwidth]{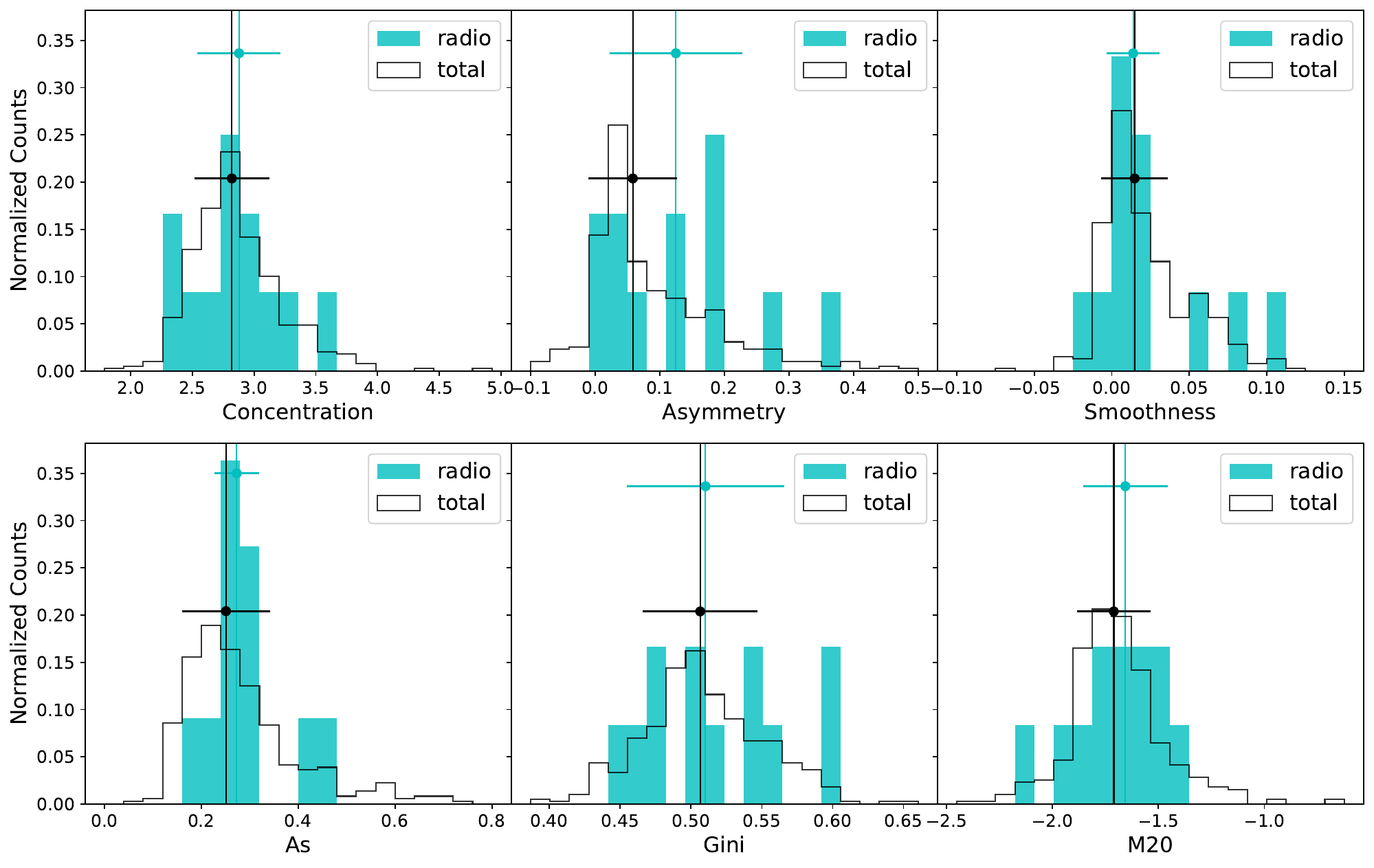}
}
\setcounter{subfigure}{3}
\subfigure[BPT]{
\includegraphics[width=0.7\textwidth]{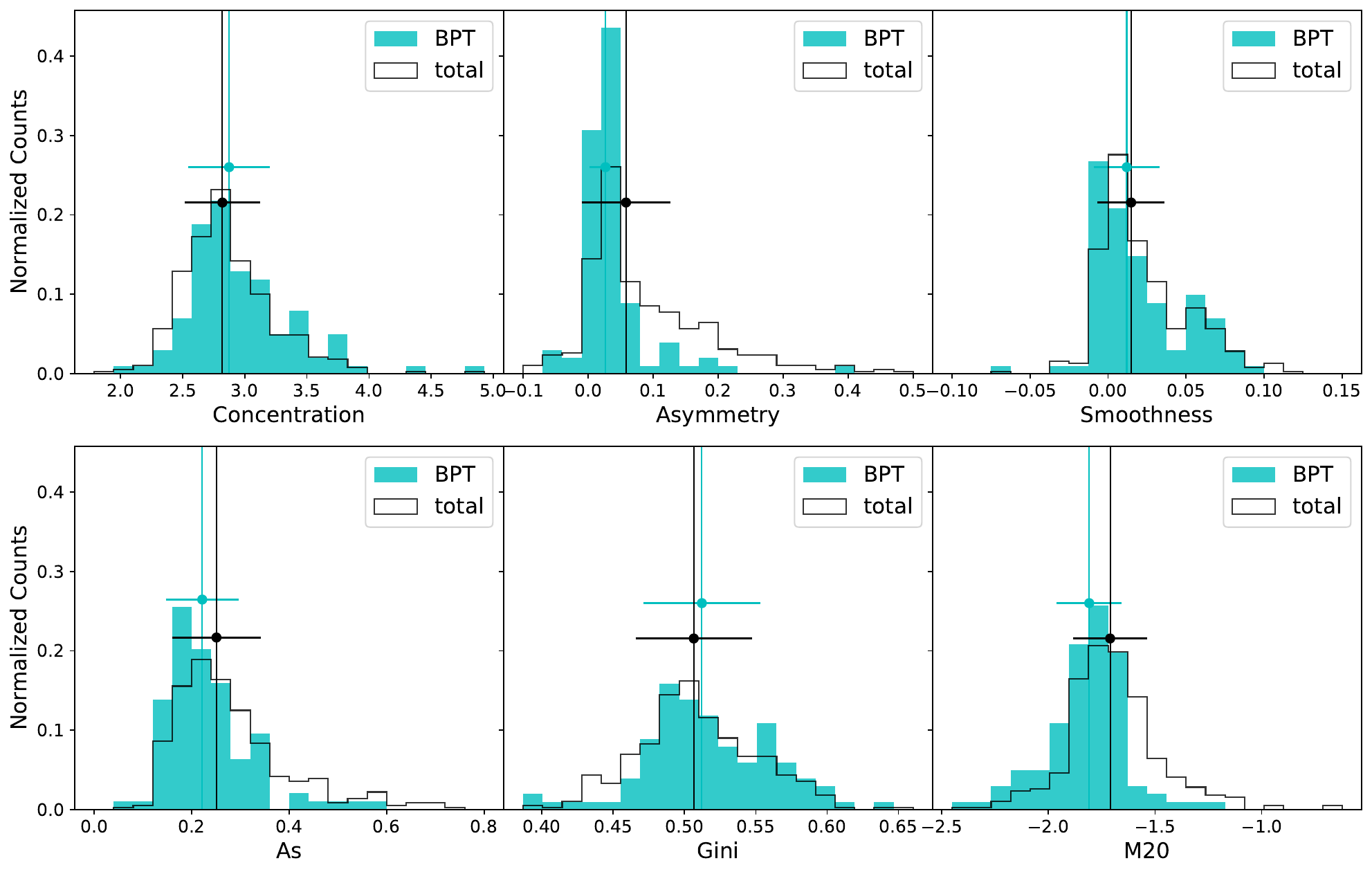}
}

\caption{(Continued.)}
\end{figure}

\begin{figure}[t]
\figurenum{7}
\centering
\setcounter{subfigure}{4}
\subfigure[variability]{
\includegraphics[width=0.7\textwidth]{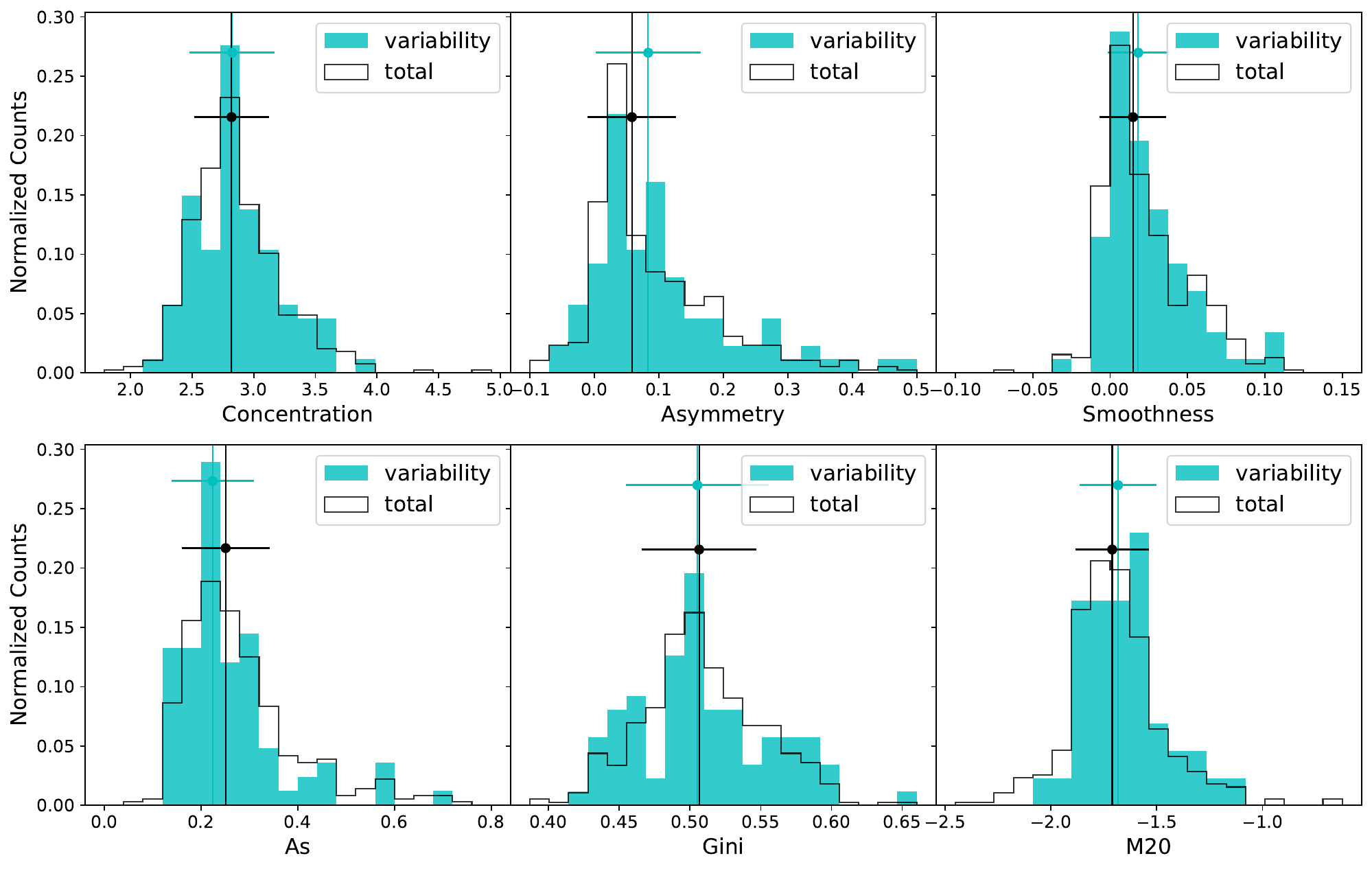}
}
\setcounter{subfigure}{5}
\subfigure[mid-IR]{
\includegraphics[width=0.7\textwidth]{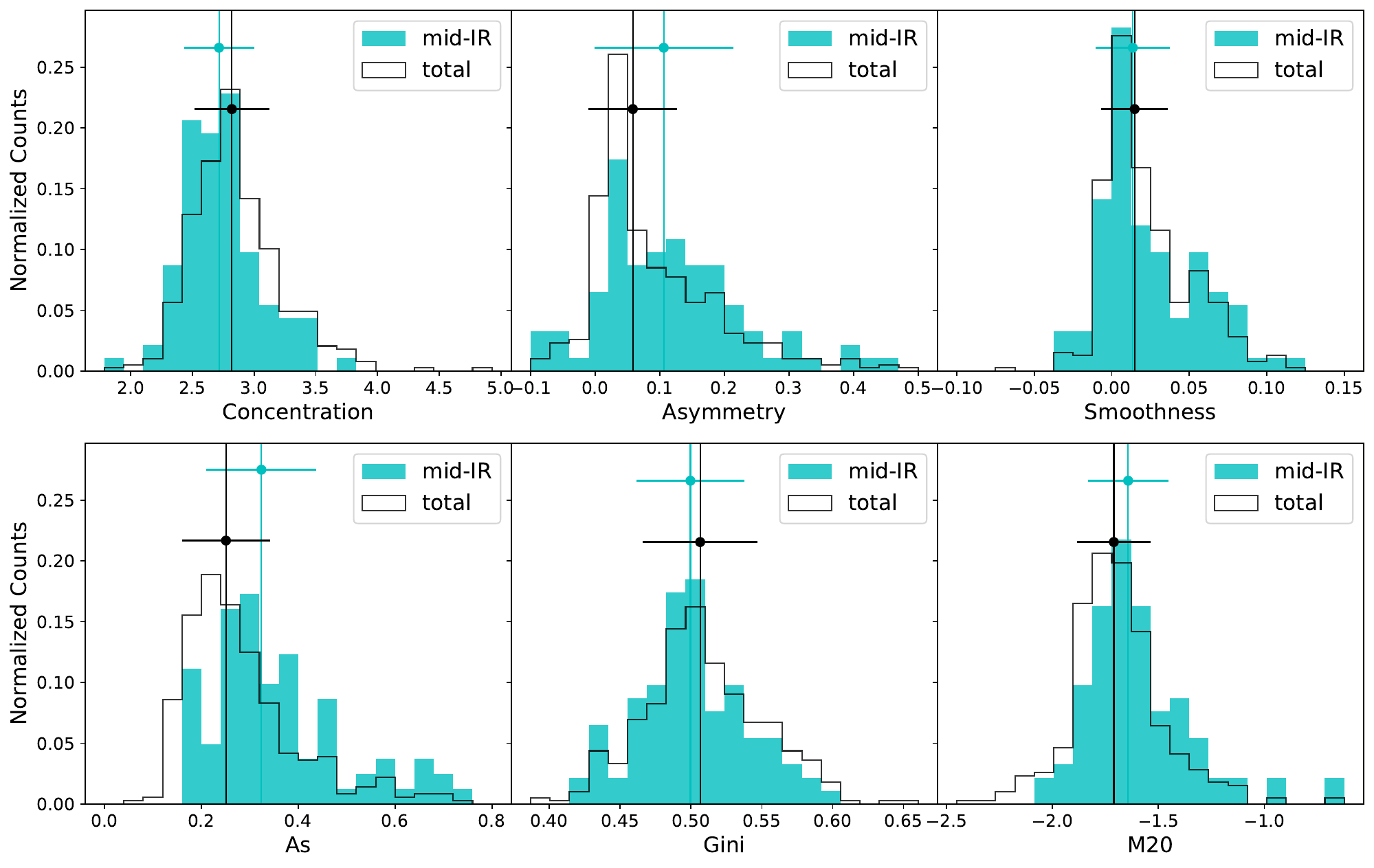}
}
\caption{(Continued.)}
\end{figure}

The results for different subsamples selected with various methods are also listed in Tables \ref{tab:fraction} and \ref{tab:mean} in Appendix \ref{subsec:A1}. We also show the distributions of these parameters from \textit{r}-band in Figure \ref{fig:subsample}. In general, all samples show similar distributions of these parameters to the total sample except for the radio sample, which consists of only 13 members, and the BPT sample, which may suffer from a selection effect. The BPT sample exhibits larger values for \textit{C} and \textit{G} in (d) of Figure \ref{fig:subsample}, and lacks sources towards higher \textit{A} and $M_{20}$, indicating that the BPT-selected galaxies possess a more concentrated central region similar to dwarf ETGs. Indeed, about 62\% of the BPT sample consists of ETGs, the highest proportion in Table \ref{tab:fraction}. Furthermore, only 3\% (the lowest among our samples) of the BPT sample galaxies are visually classified into mergers. 

From Table \ref{tab:mean} in Appendix \ref{subsec:A1}, we can see that, in contrast to the BPT sample, the mid-IR sample almost has the largest $A$ and smallest $C$. Visual classification further demonstrates that $\sim$40\% and 34\% of the mid-IR sample comprises LTGs and mergers, respectively. This merger fraction is the highest among our samples and is about 3 times larger than that of the total sample. Notably, mergers from the mid-IR sample account for about 60\% of the total mergers. Consequently, only about 23\% of this sample are ETGs, the lowest fraction among our samples. 

We note that the variability sample consists of the highest fraction ($\sim$64\%) of LTGs. This sample includes 17 optical and 70 infrared variability-selected objects. Specifically, the fractions of LTGs (ETGs) are 59\% (29\%) and 66\% (21\%), respectively for the optical and infrared variability-selected sources. These fractions suggest that variability-selected AGNs prefer residing in LTGs, different from their massive counterparts at higher redshifts presented by \cite{2022ApJ...925..157Z}, who found that their optical variability-selected AGNs have no clear morphological preference. However, their merger fraction (9.8\%), which was determined by visual inspection, is well consistent with that (6.9\%) for our variability sample.

\section{SUMMARY}\label{sec:summary}
In this study, we establish a various and comprehensive set of AGN-host dwarf galaxies in the local Universe, utilizing the image data from DESI DR10 including \textit{grz} bands. We employ the Python package \texttt{statmorph} to measure non-parametric morphological parameters, including \textit{CAS}, $A_\mathrm{S}$, \textit{G}, $M_{20}$, $F(G, M_{20})$, $S(G, M_{20})$, to investigate the role of the AGN in the dwarf galaxies. Meanwhile, we also employ visual inspection with the assistance of deep learning to perform morphology classification. We also compare the results among various subsamples, as well as to normal dwarf and AGN-host massive galaxies. Our main results are: 
\begin{enumerate}
    \item The morphology of AGN-host dwarf galaxies shows about 37\% early-type and 44\% late-type galaxies, which is different from that of normal dwarfs, suggesting that AGNs tend to reside in late-type dwarf galaxies.    
    \item The merger fraction is about 13\% in AGN-host dwarfs, much lower than that found in normal dwarfs, indicating that mergers/interactions are not the primary driver of AGN activities. 
    \item There are no significant differences in most of the non-parametric coefficients between AGN-host and normal dwarfs. However, the larger asymmetry and higher scatter of $M_{20}$ of the AGN-host dwarfs suggest that they tend to have more irregular and diverse morphologies.
    \item  The BPT sample has the highest fraction of ETGs but the lowest fraction of mergers, the mid-IR sample consists of the largest fraction of mergers, and the variability sample contains the most LTGs.
\end{enumerate}

\begin{acknowledgments}
The authors thank the anonymous referee for valuable comments/suggestions. We acknowledge support from the China Manned Space Project (Nos. CMS-CSST-2021-A06) and the NSFC grant (12173079). All the authors acknowledge the work of the Dark Energy Spectroscopic Instrument (DESI) team. The DESI Legacy Imaging Surveys consist of three individual and complementary projects: the Dark Energy Camera Legacy Survey (DECaLS), the Beijing-Arizona Sky Survey (BASS), and the Mayall \textit{z}-band Legacy Survey (MzLS). DECaLS, BASS, and MzLS together include data obtained, respectively, at the Blanco telescope, Cerro Tololo Inter-American Observatory, NSF’s NOIRLab; the Bok telescope, Steward Observatory, University of Arizona; and the Mayall telescope, Kitt Peak National Observatory, NOIRLab. NOIRLab is operated by the Association of Universities for Research in Astronomy (AURA) under a cooperative agreement with the National Science Foundation. Pipeline processing and analyses of the data were supported by NOIRLab and the Lawrence Berkeley National Laboratory (LBNL). Legacy Surveys also uses data products from the Near-Earth Object Wide-field Infrared Survey Explorer (NEOWISE), a project of the Jet Propulsion Laboratory/California Institute of Technology, funded by the National Aeronautics and Space Administration. Legacy Surveys was supported by: the Director, Office of Science, Office of High Energy Physics of the U.S. Department of Energy; the National Energy Research Scientific Computing Center, a DOE Office of Science User Facility; the U.S. National Science Foundation, Division of Astronomical Sciences; the National Astronomical Observatories of China, the Chinese Academy of Sciences and the Chinese National Natural Science Foundation. LBNL is managed by the Regents of the University of California under contract to the U.S. Department of Energy. 
Software: \texttt{statmorph} \citep{2019MNRAS.483.4140R}; \texttt{Photutils} \citep{2020zndo...4049061B}; \texttt{SEXtractor} \citep{1996MNRAS.279L..47A}; \texttt{Unsharp-masking} \citep{2014PeerJ...2..453V}.
\end{acknowledgments}

\appendix
\section{Morphological parameters in \textit{r}-band.}\label{subsec:A1}

\begin{table}[t]
\movetabledown=30mm
\movetableright=-75mm
\begin{rotatetable}
\renewcommand{\arraystretch}{2} 
\caption{Mean and median values of parameters from different subsamples in \textit{r}-band.}
\label{tab:mean}
\centering
\resizebox{1.3\columnwidth}{!}{
\begin{tabular}{ccccccccccccccccc}\\\hline \hline
\multirow{2}{*}{Subsamples}&\multicolumn{2}{c}{Concentration}&\multicolumn{2}{c}{Asymmetry}&\multicolumn{2}{c}{Smoothness}&\multicolumn{2}{c}{$A_\mathrm{S}$}&\multicolumn{2}{c}{Gini}& \multicolumn{2}{c}{$M_{20}$}&\multicolumn{2}{c}{$F(G, M_{20})$}&\multicolumn{2}{c}{$S(G, M_{20})$} \\\cline{2-17}
                         &mean($\sigma$)&median($\sigma$)&mean$(\sigma)$&median($\sigma$)&mean($\sigma$)&median($\sigma$)& mean($\sigma$)&median($\sigma$)&mean($\sigma$)&median($\sigma$)&mean$(\sigma)$ &median($\sigma$)&mean($\sigma$)&median($\sigma$)&mean($\sigma$)&median($\sigma$)\\\hline
\multirow{1}{*}{X-ray}   &2.826(0.286) & 2.757(0.240)&0.116(0.091) & 0.087(0.089)&0.026(0.025) & 0.016(0.021)
                         &0.291(0.105) & 0.296(0.118)&0.512(0.036) & 0.510(0.042)&-1.685(0.210) & -1.712(0.141)
                         &-0.256(0.280) & -0.312(0.243)&-0.054(0.034) & -0.056(0.028)\\\hline
\multirow{1}{*}{BPT}     &2.974(0.443) & 2.874(0.328)&0.041(0.115) & 0.027(0.024)&0.020(0.043) & 0.012(0.021)
                         &0.238(0.091) & 0.222(0.075)&0.516(0.046) & 0.512(0.041)&-1.817(0.188) & -1.808(0.152)
                         &-0.146(0.320) & -0.193(0.313)&-0.069(0.038) & -0.073(0.024)\\\hline
\multirow{1}{*}{\HeII}   &2.936(0.398) & 2.867(0.314)&0.079(0.070) & 0.062(0.059)&0.018(0.022) & 0.012(0.019)
                         &0.256(0.099) & 0.245(0.064)&0.510(0.041) & 0.509(0.034)&-1.759(0.217) & -1.735(0.165)
                         &-0.217(0.328) & -0.237(0.253)&-0.067(0.029) & -0.068(0.025)\\\hline
\multirow{1}{*}{Mid-IR}  &2.738(0.316) & 2.715(0.284)&0.126(0.137) & 0.106(0.107)&0.024(0.032) & 0.014(0.024)
                         &0.360(0.139) & 0.324(0.113)&0.501(0.039) & 0.500(0.038)&-1.590(0.258) & -1.643(0.190)
                         &-0.377(0.295) & -0.373(0.257)&-0.052(0.045) & -0.061(0.035)\\\hline
\multirow{1}{*}{Radio}   &2.843(0.375) & 2.878(0.337)&0.131(0.110) & 0.125(0.102)&0.026(0.036) & 0.014(0.017)
                         &0.300(0.080) & 0.273(0.046)&0.519(0.051) & 0.510(0.055)&-1.692(0.224) & -1.656(0.200)
                         &-0.220(0.384) & -0.263(0.428)&-0.049(0.033) & -0.048(0.037)\\\hline
\multirow{1}{*}{Variability} &2.870(0.337) & 2.825(0.344)&0.109(0.113) & 0.084(0.081)&0.025(0.027) & 0.018(0.020)
                         &0.269(0.141) & 0.224(0.085)&0.510(0.048) & 0.505(0.051)&-1.654(0.193) & -1.683(0.181)
                         &-0.288(0.289) & -0.318(0.304)&-0.052(0.051) & -0.059(0.051)\\\hline
\multirow{1}{*}{Total}   &2.863(0.365) & 2.819(0.304)&0.093(0.117) & 0.058(0.068)&0.022(0.032) & 0.015(0.021)
                         &0.284(0.126) & 0.251(0.091)&0.509(0.042) & 0.507(0.041)&-1.694(0.224) & -1.710(0.173)
                         &-0.268(0.305) & -0.293(0.282)&-0.059(0.042) & -0.064(0.032)\\\hline
\end{tabular}
}

\tablecomments{Row (1)-(7): Different subsamples from the \textit{grz} bands. All the data come from \textbf{Sample A} and \textbf{Sample B}, except $A_\mathrm{S}$, which is only derived from \textbf{Sample A}. 1 $\sigma$ dispersion is estimated using 1.48 $\times$ MAD, where MAD is the median absolute deviation of the sample, and the dispersion of a mean value is estimated using the standard deviation of the sample throughout the paper.}
\end{rotatetable}
\end{table}

\bibliography{sample631}

\begin{thebibliography}{}
\expandafter\ifx\csname natexlab\endcsname\relax\def\natexlab#1{#1}\fi
\providecommand{\url}[1]{\href{#1}{#1}}
\providecommand{\dodoi}[1]{doi:~\href{http://doi.org/#1}{\nolinkurl{#1}}}
\providecommand{\doeprint}[1]{\href{http://ascl.net/#1}{\nolinkurl{http://ascl.net/#1}}}
\providecommand{\doarXiv}[1]{\href{https://arxiv.org/abs/#1}{\nolinkurl{https://arxiv.org/abs/#1}}}

\bibitem[{{Abraham} {et~al.}(1996){Abraham}, {Tanvir}, {Santiago}, {Ellis}, {Glazebrook}, \& {van den Bergh}}]{1996MNRAS.279L..47A}
{Abraham}, R.~G., {Tanvir}, N.~R., {Santiago}, B.~X., {et~al.} 1996, \mnras, 279, L47, \dodoi{10.1093/mnras/279.3.L47}

\bibitem[{{Aihara} {et~al.}(2011){Aihara}, {Allende Prieto}, {An}, {Anderson}, {Aubourg}, {Balbinot}, {Beers}, {Berlind}, {Bickerton}, {Bizyaev}, {Blanton}, {Bochanski}, {Bolton}, {Bovy}, {Brandt}, {Brinkmann}, {Brown}, {Brownstein}, {Busca}, {Campbell}, {Carr}, {Chen}, {Chiappini}, {Comparat}, {Connolly}, {Cortes}, {Croft}, {Cuesta}, {da Costa}, {Davenport}, {Dawson}, {Dhital}, {Ealet}, {Ebelke}, {Edmondson}, {Eisenstein}, {Escoffier}, {Esposito}, {Evans}, {Fan}, {Femen{\'\i}a Castell{\'a}}, {Font-Ribera}, {Frinchaboy}, {Ge}, {Gillespie}, {Gilmore}, {Gonz{\'a}lez Hern{\'a}ndez}, {Gott}, {Gould}, {Grebel}, {Gunn}, {Hamilton}, {Harding}, {Harris}, {Hawley}, {Hearty}, {Ho}, {Hogg}, {Holtzman}, {Honscheid}, {Inada}, {Ivans}, {Jiang}, {Johnson}, {Jordan}, {Jordan}, {Kazin}, {Kirkby}, {Klaene}, {Knapp}, {Kneib}, {Kochanek}, {Koesterke}, {Kollmeier}, {Kron}, {Lampeitl}, {Lang}, {Le Goff}, {Lee}, {Lin}, {Long}, {Loomis}, {Lucatello}, {Lundgren}, {Lupton}, {Ma}, {MacDonald}, {Mahadevan}, {Maia}, {Makler},
  {Malanushenko}, {Malanushenko}, {Mandelbaum}, {Maraston}, {Margala}, {Masters}, {McBride}, {McGehee}, {McGreer}, {M{\'e}nard}, {Miralda-Escud{\'e}}, {Morrison}, {Mullally}, {Muna}, {Munn}, {Murayama}, {Myers}, {Naugle}, {Neto}, {Nguyen}, {Nichol}, {O'Connell}, {Ogando}, {Olmstead}, {Oravetz}, {Padmanabhan}, {Palanque-Delabrouille}, {Pan}, {Pandey}, {P{\^a}ris}, {Percival}, {Petitjean}, {Pfaffenberger}, {Pforr}, {Phleps}, {Pichon}, {Pieri}, {Prada}, {Price-Whelan}, {Raddick}, {Ramos}, {Reyl{\'e}}, {Rich}, {Richards}, {Rix}, {Robin}, {Rocha-Pinto}, {Rockosi}, {Roe}, {Rollinde}, {Ross}, {Ross}, {Rossetto}, {S{\'a}nchez}, {Sayres}, {Schlegel}, {Schlesinger}, {Schmidt}, {Schneider}, {Sheldon}, {Shu}, {Simmerer}, {Simmons}, {Sivarani}, {Snedden}, {Sobeck}, {Steinmetz}, {Strauss}, {Szalay}, {Tanaka}, {Thakar}, {Thomas}, {Tinker}, {Tofflemire}, {Tojeiro}, {Tremonti}, {Vandenberg}, {Vargas Maga{\~n}a}, {Verde}, {Vogt}, {Wake}, {Wang}, {Weaver}, {Weinberg}, {White}, {White}, {Yanny}, {Yasuda}, {Yeche}, \&
  {Zehavi}}]{2011ApJS..195...26A}
{Aihara}, H., {Allende Prieto}, C., {An}, D., {et~al.} 2011, \apjs, 195, 26, \dodoi{10.1088/0067-0049/195/2/26}

\bibitem[{{Andrews} \& {Martini}(2013)}]{2013ApJ...765..140A}
{Andrews}, B.~H., \& {Martini}, P. 2013, \apj, 765, 140, \dodoi{10.1088/0004-637X/765/2/140}

\bibitem[{{Bellm} {et~al.}(2019){Bellm}, {Kulkarni}, {Graham}, {Dekany}, {Smith}, {Riddle}, {Masci}, {Helou}, {Prince}, {Adams}, {Barbarino}, {Barlow}, {Bauer}, {Beck}, {Belicki}, {Biswas}, {Blagorodnova}, {Bodewits}, {Bolin}, {Brinnel}, {Brooke}, {Bue}, {Bulla}, {Burruss}, {Cenko}, {Chang}, {Connolly}, {Coughlin}, {Cromer}, {Cunningham}, {De}, {Delacroix}, {Desai}, {Duev}, {Eadie}, {Farnham}, {Feeney}, {Feindt}, {Flynn}, {Franckowiak}, {Frederick}, {Fremling}, {Gal-Yam}, {Gezari}, {Giomi}, {Goldstein}, {Golkhou}, {Goobar}, {Groom}, {Hacopians}, {Hale}, {Henning}, {Ho}, {Hover}, {Howell}, {Hung}, {Huppenkothen}, {Imel}, {Ip}, {Ivezi{\'c}}, {Jackson}, {Jones}, {Juric}, {Kasliwal}, {Kaspi}, {Kaye}, {Kelley}, {Kowalski}, {Kramer}, {Kupfer}, {Landry}, {Laher}, {Lee}, {Lin}, {Lin}, {Lunnan}, {Giomi}, {Mahabal}, {Mao}, {Miller}, {Monkewitz}, {Murphy}, {Ngeow}, {Nordin}, {Nugent}, {Ofek}, {Patterson}, {Penprase}, {Porter}, {Rauch}, {Rebbapragada}, {Reiley}, {Rigault}, {Rodriguez}, {van Roestel}, {Rusholme}, {van
  Santen}, {Schulze}, {Shupe}, {Singer}, {Soumagnac}, {Stein}, {Surace}, {Sollerman}, {Szkody}, {Taddia}, {Terek}, {Van Sistine}, {van Velzen}, {Vestrand}, {Walters}, {Ward}, {Ye}, {Yu}, {Yan}, \& {Zolkower}}]{2019PASP..131a8002B}
{Bellm}, E.~C., {Kulkarni}, S.~R., {Graham}, M.~J., {et~al.} 2019, \pasp, 131, 018002, \dodoi{10.1088/1538-3873/aaecbe}

\bibitem[{{Bellovary} {et~al.}(2011){Bellovary}, {Volonteri}, {Governato}, {Shen}, {Quinn}, \& {Wadsley}}]{2011ApJ...742...13B}
{Bellovary}, J., {Volonteri}, M., {Governato}, F., {et~al.} 2011, \apj, 742, 13, \dodoi{10.1088/0004-637X/742/1/13}

\bibitem[{{Bershady} {et~al.}(2000){Bershady}, {Jangren}, \& {Conselice}}]{2000AJ....119.2645B}
{Bershady}, M.~A., {Jangren}, A., \& {Conselice}, C.~J. 2000, \aj, 119, 2645, \dodoi{10.1086/301386}

\bibitem[{{Bertin}(2011)}]{2011ASPC..442..435B}
{Bertin}, E. 2011, in Astronomical Society of the Pacific Conference Series, Vol. 442, Astronomical Data Analysis Software and Systems XX, ed. I.~N. {Evans}, A.~{Accomazzi}, D.~J. {Mink}, \& A.~H. {Rots}, 435

\bibitem[{{Bertin} \& {Arnouts}(1996)}]{1996A&AS..117..393B}
{Bertin}, E., \& {Arnouts}, S. 1996, \aaps, 117, 393, \dodoi{10.1051/aas:1996164}

\bibitem[{{Bichang'a} {et~al.}(2024){Bichang'a}, {Kaviraj}, {Lazar}, {Jackson}, {Das}, {Smith}, {Watkins}, \& {Martin}}]{2024MNRAS.532..613B}
{Bichang'a}, B., {Kaviraj}, S., {Lazar}, I., {et~al.} 2024, \mnras, 532, 613, \dodoi{10.1093/mnras/stae1441}

\bibitem[{{Birchall} {et~al.}(2020){Birchall}, {Watson}, \& {Aird}}]{2020MNRAS.492.2268B}
{Birchall}, K.~L., {Watson}, M.~G., \& {Aird}, J. 2020, \mnras, 492, 2268, \dodoi{10.1093/mnras/staa040}

\bibitem[{{Blanton} {et~al.}(2005){Blanton}, {Lupton}, {Schlegel}, {Strauss}, {Brinkmann}, {Fukugita}, \& {Loveday}}]{2005ApJ...631..208B}
{Blanton}, M.~R., {Lupton}, R.~H., {Schlegel}, D.~J., {et~al.} 2005, \apj, 631, 208, \dodoi{10.1086/431416}

\bibitem[{{Bradley} {et~al.}(2020){Bradley}, {Sip{\H{o}}cz}, {Robitaille}, {Tollerud}, {Vin{\'\i}cius}, {Deil}, {Barbary}, {Wilson}, {Busko}, {G{\"u}nther}, {Cara}, {Conseil}, {Bostroem}, {Droettboom}, {Bray}, {Andersen Bratholm}, {Lim}, {Barentsen}, {Craig}, {Pascual}, {Perren}, {Greco}, {Donath}, {De Val-Borro}, {Kerzendorf}, {Bach}, {Weaver}, {D'Eugenio}, {Souchereau}, \& {Ferreira}}]{2020zndo...4049061B}
{Bradley}, L., {Sip{\H{o}}cz}, B., {Robitaille}, T., {et~al.} 2020, {astropy/photutils: 1.0.1}, 1.0.1,  Zenodo, \dodoi{10.5281/zenodo.4049061}

\bibitem[{{Cai} {et~al.}(2020){Cai}, {Zhao}, {Zhang}, {Bai}, \& {Liu}}]{2020ApJ...903...58C}
{Cai}, W., {Zhao}, Y., {Zhang}, H.-X., {Bai}, J.-M., \& {Liu}, H.-T. 2020, \apj, 903, 58, \dodoi{10.3847/1538-4357/abb81c}

\bibitem[{{Cai} {et~al.}(2021){Cai}, {Zhao}, \& {Bai}}]{2021RAA....21..204C}
{Cai}, W., {Zhao}, Y.-H., \& {Bai}, J.-M. 2021, Research in Astronomy and Astrophysics, 21, 204, \dodoi{10.1088/1674-4527/21/8/204}

\bibitem[{Cameron(2011)}]{Cameron_2011}
Cameron, E. 2011, Publications of the Astronomical Society of Australia, 28, 128–139, \dodoi{10.1071/AS10046}

\bibitem[{{Cisternas} {et~al.}(2011){Cisternas}, {Jahnke}, {Inskip}, {Kartaltepe}, {Koekemoer}, {Lisker}, {Robaina}, {Scodeggio}, {Sheth}, {Trump}, {Andrae}, {Miyaji}, {Lusso}, {Brusa}, {Capak}, {Cappelluti}, {Civano}, {Ilbert}, {Impey}, {Leauthaud}, {Lilly}, {Salvato}, {Scoville}, \& {Taniguchi}}]{2011ApJ...726...57C}
{Cisternas}, M., {Jahnke}, K., {Inskip}, K.~J., {et~al.} 2011, \apj, 726, 57, \dodoi{10.1088/0004-637X/726/2/57}

\bibitem[{{Conselice}(2003)}]{2003ApJS..147....1C}
{Conselice}, C.~J. 2003, \apjs, 147, 1, \dodoi{10.1086/375001}

\bibitem[{{Conselice} {et~al.}(2000){Conselice}, {Bershady}, \& {Jangren}}]{2000ApJ...529..886C}
{Conselice}, C.~J., {Bershady}, M.~A., \& {Jangren}, A. 2000, \apj, 529, 886, \dodoi{10.1086/308300}

\bibitem[{{Darg} {et~al.}(2010){Darg}, {Kaviraj}, {Lintott}, {Schawinski}, {Sarzi}, {Bamford}, {Silk}, {Andreescu}, {Murray}, {Nichol}, {Raddick}, {Slosar}, {Szalay}, {Thomas}, \& {Vandenberg}}]{2010MNRAS.401.1552D}
{Darg}, D.~W., {Kaviraj}, S., {Lintott}, C.~J., {et~al.} 2010, \mnras, 401, 1552, \dodoi{10.1111/j.1365-2966.2009.15786.x}

\bibitem[{{Davis} {et~al.}(2022){Davis}, {Kaviraj}, {Hardcastle}, {Martin}, {Jackson}, {Kraljic}, {Malek}, {Peirani}, {Smith}, {Volonteri}, \& {Wang}}]{2022MNRAS.511.4109D}
{Davis}, F., {Kaviraj}, S., {Hardcastle}, M.~J., {et~al.} 2022, \mnras, 511, 4109, \dodoi{10.1093/mnras/stac068}

\bibitem[{{Dekany} {et~al.}(2020){Dekany}, {Smith}, {Riddle}, {Feeney}, {Porter}, {Hale}, {Zolkower}, {Belicki}, {Kaye}, {Henning}, {Walters}, {Cromer}, {Delacroix}, {Rodriguez}, {Reiley}, {Mao}, {Hover}, {Murphy}, {Burruss}, {Baker}, {Kowalski}, {Reif}, {Mueller}, {Bellm}, {Graham}, \& {Kulkarni}}]{2020PASP..132c8001D}
{Dekany}, R., {Smith}, R.~M., {Riddle}, R., {et~al.} 2020, \pasp, 132, 038001, \dodoi{10.1088/1538-3873/ab4ca2}

\bibitem[{{Dey} {et~al.}(2019){Dey}, {Schlegel}, {Lang}, {Blum}, {Burleigh}, {Fan}, {Findlay}, {Finkbeiner}, {Herrera}, {Juneau}, {Landriau}, {Levi}, {McGreer}, {Meisner}, {Myers}, {Moustakas}, {Nugent}, {Patej}, {Schlafly}, {Walker}, {Valdes}, {Weaver}, {Y{\`e}che}, {Zou}, {Zhou}, {Abareshi}, {Abbott}, {Abolfathi}, {Aguilera}, {Alam}, {Allen}, {Alvarez}, {Annis}, {Ansarinejad}, {Aubert}, {Beechert}, {Bell}, {BenZvi}, {Beutler}, {Bielby}, {Bolton}, {Brice{\~n}o}, {Buckley-Geer}, {Butler}, {Calamida}, {Carlberg}, {Carter}, {Casas}, {Castander}, {Choi}, {Comparat}, {Cukanovaite}, {Delubac}, {DeVries}, {Dey}, {Dhungana}, {Dickinson}, {Ding}, {Donaldson}, {Duan}, {Duckworth}, {Eftekharzadeh}, {Eisenstein}, {Etourneau}, {Fagrelius}, {Farihi}, {Fitzpatrick}, {Font-Ribera}, {Fulmer}, {G{\"a}nsicke}, {Gaztanaga}, {George}, {Gerdes}, {Gontcho}, {Gorgoni}, {Green}, {Guy}, {Harmer}, {Hernandez}, {Honscheid}, {Huang}, {James}, {Jannuzi}, {Jiang}, {Joyce}, {Karcher}, {Karkar}, {Kehoe}, {Kneib}, {Kueter-Young}, {Lan},
  {Lauer}, {Le Guillou}, {Le Van Suu}, {Lee}, {Lesser}, {Perreault Levasseur}, {Li}, {Mann}, {Marshall}, {Mart{\'\i}nez-V{\'a}zquez}, {Martini}, {du Mas des Bourboux}, {McManus}, {Meier}, {M{\'e}nard}, {Metcalfe}, {Mu{\~n}oz-Guti{\'e}rrez}, {Najita}, {Napier}, {Narayan}, {Newman}, {Nie}, {Nord}, {Norman}, {Olsen}, {Paat}, {Palanque-Delabrouille}, {Peng}, {Poppett}, {Poremba}, {Prakash}, {Rabinowitz}, {Raichoor}, {Rezaie}, {Robertson}, {Roe}, {Ross}, {Ross}, {Rudnick}, {Safonova}, {Saha}, {S{\'a}nchez}, {Savary}, {Schweiker}, {Scott}, {Seo}, {Shan}, {Silva}, {Slepian}, {Soto}, {Sprayberry}, {Staten}, {Stillman}, {Stupak}, {Summers}, {Sien Tie}, {Tirado}, {Vargas-Maga{\~n}a}, {Vivas}, {Wechsler}, {Williams}, {Yang}, {Yang}, {Yapici}, {Zaritsky}, {Zenteno}, {Zhang}, {Zhang}, {Zhou}, \& {Zhou}}]{2019AJ....157..168D}
{Dey}, A., {Schlegel}, D.~J., {Lang}, D., {et~al.} 2019, \aj, 157, 168, \dodoi{10.3847/1538-3881/ab089d}

\bibitem[{{Di Matteo} {et~al.}(2005){Di Matteo}, {Springel}, \& {Hernquist}}]{2005Natur.433..604D}
{Di Matteo}, T., {Springel}, V., \& {Hernquist}, L. 2005, \nat, 433, 604, \dodoi{10.1038/nature03335}

\bibitem[{{Draper} \& {Ballantyne}(2012)}]{2012ApJ...751...72D}
{Draper}, A.~R., \& {Ballantyne}, D.~R. 2012, \apj, 751, 72, \dodoi{10.1088/0004-637X/751/1/72}

\bibitem[{{Fan} {et~al.}(2014){Fan}, {Fang}, {Chen}, {Li}, {Lv}, {Knudsen}, \& {Kong}}]{2014ApJ...784L...9F}
{Fan}, L., {Fang}, G., {Chen}, Y., {et~al.} 2014, \apjl, 784, L9, \dodoi{10.1088/2041-8205/784/1/L9}

\bibitem[{{Freeman} {et~al.}(2013){Freeman}, {Izbicki}, {Lee}, {Newman}, {Conselice}, {Koekemoer}, {Lotz}, \& {Mozena}}]{2013MNRAS.434..282F}
{Freeman}, P.~E., {Izbicki}, R., {Lee}, A.~B., {et~al.} 2013, \mnras, 434, 282, \dodoi{10.1093/mnras/stt1016}

\bibitem[{{Frei} {et~al.}(1996){Frei}, {Guhathakurta}, {Gunn}, \& {Tyson}}]{1996AJ....111..174F}
{Frei}, Z., {Guhathakurta}, P., {Gunn}, J.~E., \& {Tyson}, J.~A. 1996, \aj, 111, 174, \dodoi{10.1086/117771}

\bibitem[{{Getachew-Woreta} {et~al.}(2022){Getachew-Woreta}, {Povi{\'c}}, {Masegosa}, {Perea}, {Beyoro-Amado}, \& {M{\'a}rquez}}]{2022MNRAS.514..607G}
{Getachew-Woreta}, T., {Povi{\'c}}, M., {Masegosa}, J., {et~al.} 2022, \mnras, 514, 607, \dodoi{10.1093/mnras/stac851}

\bibitem[{{Gonz{\'a}lez Delgado} {et~al.}(2015){Gonz{\'a}lez Delgado}, {Garc{\'\i}a-Benito}, {P{\'e}rez}, {Cid Fernandes}, {de Amorim}, {Cortijo-Ferrero}, {Lacerda}, {L{\'o}pez Fern{\'a}ndez}, {Vale-Asari}, {S{\'a}nchez}, {Moll{\'a}}, {Ruiz-Lara}, {S{\'a}nchez-Bl{\'a}zquez}, {Walcher}, {Alves}, {Aguerri}, {Bekerait{\'e}}, {Bland-Hawthorn}, {Galbany}, {Gallazzi}, {Husemann}, {Iglesias-P{\'a}ramo}, {Kalinova}, {L{\'o}pez-S{\'a}nchez}, {Marino}, {M{\'a}rquez}, {Masegosa}, {Mast}, {M{\'e}ndez-Abreu}, {Mendoza}, {del Olmo}, {P{\'e}rez}, {Quirrenbach}, \& {Zibetti}}]{2015A&A...581A.103G}
{Gonz{\'a}lez Delgado}, R.~M., {Garc{\'\i}a-Benito}, R., {P{\'e}rez}, E., {et~al.} 2015, \aap, 581, A103, \dodoi{10.1051/0004-6361/201525938}

\bibitem[{{Graham} {et~al.}(2019){Graham}, {Kulkarni}, {Bellm}, {Adams}, {Barbarino}, {Blagorodnova}, {Bodewits}, {Bolin}, {Brady}, {Cenko}, {Chang}, {Coughlin}, {De}, {Eadie}, {Farnham}, {Feindt}, {Franckowiak}, {Fremling}, {Gezari}, {Ghosh}, {Goldstein}, {Golkhou}, {Goobar}, {Ho}, {Huppenkothen}, {Ivezi{\'c}}, {Jones}, {Juric}, {Kaplan}, {Kasliwal}, {Kelley}, {Kupfer}, {Lee}, {Lin}, {Lunnan}, {Mahabal}, {Miller}, {Ngeow}, {Nugent}, {Ofek}, {Prince}, {Rauch}, {van Roestel}, {Schulze}, {Singer}, {Sollerman}, {Taddia}, {Yan}, {Ye}, {Yu}, {Barlow}, {Bauer}, {Beck}, {Belicki}, {Biswas}, {Brinnel}, {Brooke}, {Bue}, {Bulla}, {Burruss}, {Connolly}, {Cromer}, {Cunningham}, {Dekany}, {Delacroix}, {Desai}, {Duev}, {Feeney}, {Flynn}, {Frederick}, {Gal-Yam}, {Giomi}, {Groom}, {Hacopians}, {Hale}, {Helou}, {Henning}, {Hover}, {Hillenbrand}, {Howell}, {Hung}, {Imel}, {Ip}, {Jackson}, {Kaspi}, {Kaye}, {Kowalski}, {Kramer}, {Kuhn}, {Landry}, {Laher}, {Mao}, {Masci}, {Monkewitz}, {Murphy}, {Nordin}, {Patterson}, {Penprase},
  {Porter}, {Rebbapragada}, {Reiley}, {Riddle}, {Rigault}, {Rodriguez}, {Rusholme}, {van Santen}, {Shupe}, {Smith}, {Soumagnac}, {Stein}, {Surace}, {Szkody}, {Terek}, {Van Sistine}, {van Velzen}, {Vestrand}, {Walters}, {Ward}, {Zhang}, \& {Zolkower}}]{2019PASP..131g8001G}
{Graham}, M.~J., {Kulkarni}, S.~R., {Bellm}, E.~C., {et~al.} 2019, \pasp, 131, 078001, \dodoi{10.1088/1538-3873/ab006c}

\bibitem[{{Grogin} {et~al.}(2005){Grogin}, {Conselice}, {Chatzichristou}, {Alexander}, {Bauer}, {Hornschemeier}, {Jogee}, {Koekemoer}, {Laidler}, {Livio}, {Lucas}, {Paolillo}, {Ravindranath}, {Schreier}, {Simmons}, \& {Urry}}]{2005ApJ...627L..97G}
{Grogin}, N.~A., {Conselice}, C.~J., {Chatzichristou}, E., {et~al.} 2005, \apjl, 627, L97, \dodoi{10.1086/432256}

\bibitem[{{Hewlett} {et~al.}(2017){Hewlett}, {Villforth}, {Wild}, {Mendez-Abreu}, {Pawlik}, \& {Rowlands}}]{2017MNRAS.470..755H}
{Hewlett}, T., {Villforth}, C., {Wild}, V., {et~al.} 2017, \mnras, 470, 755, \dodoi{10.1093/mnras/stx997}

\bibitem[{{Ho}(2008)}]{2008ARA&A..46..475H}
{Ho}, L.~C. 2008, \araa, 46, 475, \dodoi{10.1146/annurev.astro.45.051806.110546}

\bibitem[{{Ho}(2009)}]{2009ApJ...699..638H}
---. 2009, \apj, 699, 638, \dodoi{10.1088/0004-637X/699/1/638}

\bibitem[{{Hopkins} \& {Hernquist}(2009)}]{2009ApJ...694..599H}
{Hopkins}, P.~F., \& {Hernquist}, L. 2009, \apj, 694, 599, \dodoi{10.1088/0004-637X/694/1/599}

\bibitem[{{Jackson} {et~al.}(2022){Jackson}, {Kaviraj}, {Martin}, {Devriendt}, {Noakes-Kettel}, {Silk}, {Ogle}, \& {Dubois}}]{2022MNRAS.511..607J}
{Jackson}, R.~A., {Kaviraj}, S., {Martin}, G., {et~al.} 2022, \mnras, 511, 607, \dodoi{10.1093/mnras/stac058}

\bibitem[{{Jarrett} {et~al.}(2011){Jarrett}, {Cohen}, {Masci}, {Wright}, {Stern}, {Benford}, {Blain}, {Carey}, {Cutri}, {Eisenhardt}, {Lonsdale}, {Mainzer}, {Marsh}, {Padgett}, {Petty}, {Ressler}, {Skrutskie}, {Stanford}, {Surace}, {Tsai}, {Wheelock}, \& {Yan}}]{2011ApJ...735..112J}
{Jarrett}, T.~H., {Cohen}, M., {Masci}, F., {et~al.} 2011, \apj, 735, 112, \dodoi{10.1088/0004-637X/735/2/112}

\bibitem[{{Kauffmann} {et~al.}(1993){Kauffmann}, {White}, \& {Guiderdoni}}]{1993MNRAS.264..201K}
{Kauffmann}, G., {White}, S.~D.~M., \& {Guiderdoni}, B. 1993, \mnras, 264, 201, \dodoi{10.1093/mnras/264.1.201}

\bibitem[{{Kauffmann} {et~al.}(2004){Kauffmann}, {White}, {Heckman}, {M{\'e}nard}, {Brinchmann}, {Charlot}, {Tremonti}, \& {Brinkmann}}]{2004MNRAS.353..713K}
{Kauffmann}, G., {White}, S. D.~M., {Heckman}, T.~M., {et~al.} 2004, \mnras, 353, 713, \dodoi{10.1111/j.1365-2966.2004.08117.x}

\bibitem[{{Kaviraj}(2014)}]{2014MNRAS.440.2944K}
{Kaviraj}, S. 2014, \mnras, 440, 2944, \dodoi{10.1093/mnras/stu338}

\bibitem[{{Kaviraj} {et~al.}(2025){Kaviraj}, {Lazar}, {Watkins}, {Laigle}, {Martin}, \& {Jackson}}]{2025MNRAS.538..153K}
{Kaviraj}, S., {Lazar}, I., {Watkins}, A.~E., {et~al.} 2025, \mnras, 538, 153, \dodoi{10.1093/mnras/staf233}

\bibitem[{{Kaviraj} {et~al.}(2019){Kaviraj}, {Martin}, \& {Silk}}]{2019MNRAS.489L..12K}
{Kaviraj}, S., {Martin}, G., \& {Silk}, J. 2019, \mnras, 489, L12, \dodoi{10.1093/mnrasl/slz102}

\bibitem[{{Kennicutt} {et~al.}(2008){Kennicutt}, {Lee}, {Funes}, {J.}, {Sakai}, \& {Akiyama}}]{2008ApJS..178..247K}
{Kennicutt}, Jr., R.~C., {Lee}, J.~C., {Funes}, J.~G., {et~al.} 2008, \apjs, 178, 247, \dodoi{10.1086/590058}

\bibitem[{{Kimbrell} {et~al.}(2021){Kimbrell}, {Reines}, {Schutte}, {Greene}, \& {Geha}}]{2021ApJ...911..134K}
{Kimbrell}, S.~J., {Reines}, A.~E., {Schutte}, Z., {Greene}, J.~E., \& {Geha}, M. 2021, \apj, 911, 134, \dodoi{10.3847/1538-4357/abec40}

\bibitem[{{Kocevski} {et~al.}(2023){Kocevski}, {Barro}, {McGrath}, {Finkelstein}, {Bagley}, {Ferguson}, {Jogee}, {Yang}, {Dickinson}, {Hathi}, {Backhaus}, {Bell}, {Bisigello}, {Buat}, {Burgarella}, {Casey}, {Cleri}, {Cooper}, {Costantin}, {Croton}, {Daddi}, {Fontana}, {Fujimoto}, {Gardner}, {Gawiser}, {Giavalisco}, {Grazian}, {Grogin}, {Guo}, {Arrabal Haro}, {Hirschmann}, {Holwerda}, {Huertas-Company}, {Hutchison}, {Iyer}, {Jones}, {Juneau}, {Kartaltepe}, {Kewley}, {Kirkpatrick}, {Koekemoer}, {Kurczynski}, {Le Bail}, {Long}, {Lotz}, {Lucas}, {Papovich}, {Pentericci}, {P{\'e}rez-Gonz{\'a}lez}, {Pirzkal}, {Rafelski}, {Ravindranath}, {Somerville}, {Straughn}, {Tacchella}, {Trump}, {Wilkins}, {Wuyts}, {Yung}, \& {Zavala}}]{2023ApJ...946L..14K}
{Kocevski}, D.~D., {Barro}, G., {McGrath}, E.~J., {et~al.} 2023, \apjl, 946, L14, \dodoi{10.3847/2041-8213/acad00}

\bibitem[{{Kormendy} \& {Ho}(2013)}]{2013ARA&A..51..511K}
{Kormendy}, J., \& {Ho}, L.~C. 2013, \araa, 51, 511, \dodoi{10.1146/annurev-astro-082708-101811}

\bibitem[{{Lazar} {et~al.}(2023){Lazar}, {Kaviraj}, {Martin}, {Laigle}, {Watkins}, \& {Jackson}}]{2023MNRAS.520.2109L}
{Lazar}, I., {Kaviraj}, S., {Martin}, G., {et~al.} 2023, \mnras, 520, 2109, \dodoi{10.1093/mnras/stad224}

\bibitem[{{Lazar} {et~al.}(2024){Lazar}, {Kaviraj}, {Watkins}, {Martin}, {Bichang'a}, \& {Jackson}}]{2024MNRAS.529..499L}
{Lazar}, I., {Kaviraj}, S., {Watkins}, A.~E., {et~al.} 2024, \mnras, 529, 499, \dodoi{10.1093/mnras/stae510}

\bibitem[{{Lee} {et~al.}(2009){Lee}, {Kennicutt}, {Funes}, {Sakai}, \& {Akiyama}}]{2009ApJ...692.1305L}
{Lee}, J.~C., {Kennicutt}, Jr., R.~C., {Funes}, S.~J. J.~G., {Sakai}, S., \& {Akiyama}, S. 2009, \apj, 692, 1305, \dodoi{10.1088/0004-637X/692/2/1305}

\bibitem[{{Lehmer} {et~al.}(2016){Lehmer}, {Basu-Zych}, {Mineo}, {Brandt}, {Eufrasio}, {Fragos}, {Hornschemeier}, {Luo}, {Xue}, {Bauer}, {Gilfanov}, {Ranalli}, {Schneider}, {Shemmer}, {Tozzi}, {Trump}, {Vignali}, {Wang}, {Yukita}, \& {Zezas}}]{2016ApJ...825....7L}
{Lehmer}, B.~D., {Basu-Zych}, A.~R., {Mineo}, S., {et~al.} 2016, \apj, 825, 7, \dodoi{10.3847/0004-637X/825/1/7}

\bibitem[{{Li} {et~al.}(2024){Li}, {Zhao}, \& {Bai}}]{2024RAA....24f5006L}
{Li}, X., {Zhao}, Y., \& {Bai}, J. 2024, Research in Astronomy and Astrophysics, 24, 065006, \dodoi{10.1088/1674-4527/ad3d13}

\bibitem[{{Lintott} {et~al.}(2011){Lintott}, {Schawinski}, {Bamford}, {Slosar}, {Land}, {Thomas}, {Edmondson}, {Masters}, {Nichol}, {Raddick}, {Szalay}, {Andreescu}, {Murray}, \& {Vandenberg}}]{2011MNRAS.410..166L}
{Lintott}, C., {Schawinski}, K., {Bamford}, S., {et~al.} 2011, \mnras, 410, 166, \dodoi{10.1111/j.1365-2966.2010.17432.x}

\bibitem[{{Lotz} {et~al.}(2004){Lotz}, {Primack}, \& {Madau}}]{2004AJ....128..163L}
{Lotz}, J.~M., {Primack}, J., \& {Madau}, P. 2004, \aj, 128, 163, \dodoi{10.1086/421849}

\bibitem[{{Lotz} {et~al.}(2008){Lotz}, {Davis}, {Faber}, {Guhathakurta}, {Gwyn}, {Huang}, {Koo}, {Le Floc'h}, {Lin}, {Newman}, {Noeske}, {Papovich}, {Willmer}, {Coil}, {Conselice}, {Cooper}, {Hopkins}, {Metevier}, {Primack}, {Rieke}, \& {Weiner}}]{2008ApJ...672..177L}
{Lotz}, J.~M., {Davis}, M., {Faber}, S.~M., {et~al.} 2008, \apj, 672, 177, \dodoi{10.1086/523659}

\bibitem[{{Malin}(1977)}]{1977AASPB..16...10M}
{Malin}, D.~F. 1977, AAS Photo Bulletin, 16, 10

\bibitem[{{Martin} {et~al.}(2018){Martin}, {Kaviraj}, {Devriendt}, {Dubois}, \& {Pichon}}]{2018MNRAS.480.2266M}
{Martin}, G., {Kaviraj}, S., {Devriendt}, J.~E.~G., {Dubois}, Y., \& {Pichon}, C. 2018, \mnras, 480, 2266, \dodoi{10.1093/mnras/sty1936}

\bibitem[{{Martin} {et~al.}(2020){Martin}, {Kaviraj}, {Hocking}, {Read}, \& {Geach}}]{2020MNRAS.491.1408M}
{Martin}, G., {Kaviraj}, S., {Hocking}, A., {Read}, S.~C., \& {Geach}, J.~E. 2020, \mnras, 491, 1408, \dodoi{10.1093/mnras/stz3006}

\bibitem[{{Mateo}(1998)}]{1998ARA&A..36..435M}
{Mateo}, M.~L. 1998, \araa, 36, 435, \dodoi{10.1146/annurev.astro.36.1.435}

\bibitem[{{Mezcua} \& {Dom{\'\i}nguez S{\'a}nchez}(2024)}]{2024MNRAS.528.5252M}
{Mezcua}, M., \& {Dom{\'\i}nguez S{\'a}nchez}, H. 2024, \mnras, 528, 5252, \dodoi{10.1093/mnras/stae292}

\bibitem[{{Mineo} {et~al.}(2012){Mineo}, {Gilfanov}, \& {Sunyaev}}]{2012MNRAS.426.1870M}
{Mineo}, S., {Gilfanov}, M., \& {Sunyaev}, R. 2012, \mnras, 426, 1870, \dodoi{10.1111/j.1365-2966.2012.21831.x}

\bibitem[{{Morelli} {et~al.}(2015){Morelli}, {Corsini}, {Pizzella}, {Dalla Bont{\`a}}, {Coccato}, \& {M{\'e}ndez-Abreu}}]{2015MNRAS.452.1128M}
{Morelli}, L., {Corsini}, E.~M., {Pizzella}, A., {et~al.} 2015, \mnras, 452, 1128, \dodoi{10.1093/mnras/stv1357}

\bibitem[{{Oh} {et~al.}(2011){Oh}, {Sarzi}, {Schawinski}, \& {Yi}}]{2011ApJS..195...13O}
{Oh}, K., {Sarzi}, M., {Schawinski}, K., \& {Yi}, S.~K. 2011, \apjs, 195, 13, \dodoi{10.1088/0067-0049/195/2/13}

\bibitem[{{Petrosian}(1976)}]{1976ApJ...209L...1P}
{Petrosian}, V. 1976, \apjl, 210, L53, \dodoi{10.1086/18230110.1086/182253}

\bibitem[{{Reines} {et~al.}(2020){Reines}, {Condon}, {Darling}, \& {Greene}}]{2020ApJ...888...36R}
{Reines}, A.~E., {Condon}, J.~J., {Darling}, J., \& {Greene}, J.~E. 2020, \apj, 888, 36, \dodoi{10.3847/1538-4357/ab4999}

\bibitem[{{Reines} {et~al.}(2013){Reines}, {Greene}, \& {Geha}}]{2013ApJ...775..116R}
{Reines}, A.~E., {Greene}, J.~E., \& {Geha}, M. 2013, \apj, 775, 116, \dodoi{10.1088/0004-637X/775/2/116}

\bibitem[{{Rodriguez-Gomez} {et~al.}(2019){Rodriguez-Gomez}, {Snyder}, {Lotz}, {Nelson}, {Pillepich}, {Springel}, {Genel}, {Weinberger}, {Tacchella}, {Pakmor}, {Torrey}, {Marinacci}, {Vogelsberger}, {Hernquist}, \& {Thilker}}]{2019MNRAS.483.4140R}
{Rodriguez-Gomez}, V., {Snyder}, G.~F., {Lotz}, J.~M., {et~al.} 2019, \mnras, 483, 4140, \dodoi{10.1093/mnras/sty3345}

\bibitem[{{Rosen} {et~al.}(2016){Rosen}, {Webb}, {Watson}, {Ballet}, {Barret}, {Braito}, {Carrera}, {Ceballos}, {Coriat}, {Della Ceca}, {Denkinson}, {Esquej}, {Farrell}, {Freyberg}, {Gris{\'e}}, {Guillout}, {Heil}, {Koliopanos}, {Law-Green}, {Lamer}, {Lin}, {Martino}, {Michel}, {Motch}, {Nebot Gomez-Moran}, {Page}, {Page}, {Page}, {Pakull}, {Pye}, {Read}, {Rodriguez}, {Sakano}, {Saxton}, {Schwope}, {Scott}, {Sturm}, {Traulsen}, {Yershov}, \& {Zolotukhin}}]{2016A&A...590A...1R}
{Rosen}, S.~R., {Webb}, N.~A., {Watson}, M.~G., {et~al.} 2016, \aap, 590, A1, \dodoi{10.1051/0004-6361/201526416}

\bibitem[{{Sartori} {et~al.}(2015){Sartori}, {Schawinski}, {Treister}, {Trakhtenbrot}, {Koss}, {Shirazi}, \& {Oh}}]{2015MNRAS.454.3722S}
{Sartori}, L.~F., {Schawinski}, K., {Treister}, E., {et~al.} 2015, \mnras, 454, 3722, \dodoi{10.1093/mnras/stv2238}

\bibitem[{{Scarlata} {et~al.}(2007){Scarlata}, {Carollo}, {Lilly}, {Sargent}, {Feldmann}, {Kampczyk}, {Porciani}, {Koekemoer}, {Scoville}, {Kneib}, {Leauthaud}, {Massey}, {Rhodes}, {Tasca}, {Capak}, {Maier}, {McCracken}, {Mobasher}, {Renzini}, {Taniguchi}, {Thompson}, {Sheth}, {Ajiki}, {Aussel}, {Murayama}, {Sanders}, {Sasaki}, {Shioya}, \& {Takahashi}}]{2007ApJS..172..406S}
{Scarlata}, C., {Carollo}, C.~M., {Lilly}, S., {et~al.} 2007, \apjs, 172, 406, \dodoi{10.1086/516582}

\bibitem[{{Schade} {et~al.}(1995){Schade}, {Lilly}, {Crampton}, {Hammer}, {Le Fevre}, \& {Tresse}}]{1995ApJ...451L...1S}
{Schade}, D., {Lilly}, S.~J., {Crampton}, D., {et~al.} 1995, \apjl, 451, L1, \dodoi{10.1086/309677}

\bibitem[{{Shirazi} \& {Brinchmann}(2012)}]{2012MNRAS.421.1043S}
{Shirazi}, M., \& {Brinchmann}, J. 2012, \mnras, 421, 1043, \dodoi{10.1111/j.1365-2966.2012.20439.x}

\bibitem[{{Smethurst} {et~al.}(2024){Smethurst}, {Beckmann}, {Simmons}, {Coil}, {Devriendt}, {Dubois}, {Garland}, {Lintott}, {Martin}, \& {Peirani}}]{2024MNRAS.52710855S}
{Smethurst}, R.~J., {Beckmann}, R.~S., {Simmons}, B.~D., {et~al.} 2024, \mnras, 527, 10855, \dodoi{10.1093/mnras/stad1794}

\bibitem[{{Snyder} {et~al.}(2015{\natexlab{a}}){Snyder}, {Lotz}, {Moody}, {Peth}, {Freeman}, {Ceverino}, {Primack}, \& {Dekel}}]{2015MNRAS.451.4290S}
{Snyder}, G.~F., {Lotz}, J., {Moody}, C., {et~al.} 2015{\natexlab{a}}, \mnras, 451, 4290, \dodoi{10.1093/mnras/stv1231}

\bibitem[{{Snyder} {et~al.}(2015{\natexlab{b}}){Snyder}, {Torrey}, {Lotz}, {Genel}, {McBride}, {Vogelsberger}, {Pillepich}, {Nelson}, {Sales}, {Sijacki}, {Hernquist}, \& {Springel}}]{2015MNRAS.454.1886S}
{Snyder}, G.~F., {Torrey}, P., {Lotz}, J.~M., {et~al.} 2015{\natexlab{b}}, \mnras, 454, 1886, \dodoi{10.1093/mnras/stv2078}

\bibitem[{{Springel} {et~al.}(2005){Springel}, {Di Matteo}, \& {Hernquist}}]{2005MNRAS.361..776S}
{Springel}, V., {Di Matteo}, T., \& {Hernquist}, L. 2005, \mnras, 361, 776, \dodoi{10.1111/j.1365-2966.2005.09238.x}

\bibitem[{{Stern} {et~al.}(2012){Stern}, {Assef}, {Benford}, {Blain}, {Cutri}, {Dey}, {Eisenhardt}, {Griffith}, {Jarrett}, {Lake}, {Masci}, {Petty}, {Stanford}, {Tsai}, {Wright}, {Yan}, {Harrison}, \& {Madsen}}]{2012ApJ...753...30S}
{Stern}, D., {Assef}, R.~J., {Benford}, D.~J., {et~al.} 2012, \apj, 753, 30, \dodoi{10.1088/0004-637X/753/1/30}

\bibitem[{{Treister} {et~al.}(2012){Treister}, {Schawinski}, {Urry}, \& {Simmons}}]{2012ApJ...758L..39T}
{Treister}, E., {Schawinski}, K., {Urry}, C.~M., \& {Simmons}, B.~D. 2012, \apjl, 758, L39, \dodoi{10.1088/2041-8205/758/2/L39}

\bibitem[{{Urrutia} {et~al.}(2008){Urrutia}, {Lacy}, \& {Becker}}]{2008ApJ...674...80U}
{Urrutia}, T., {Lacy}, M., \& {Becker}, R.~H. 2008, \apj, 674, 80, \dodoi{10.1086/523959}

\bibitem[{{van der Walt} {et~al.}(2014){van der Walt}, {Sch{\"o}nberger}, {Nunez-Iglesias}, {Boulogne}, {Warner}, {Yager}, {Gouillart}, {Yu}, \& {scikit-image Contributors}}]{2014PeerJ...2..453V}
{van der Walt}, S., {Sch{\"o}nberger}, J.~L., {Nunez-Iglesias}, J., {et~al.} 2014, PeerJ, 2, e453, \dodoi{10.7717/peerj.453}

\bibitem[{{van Zee}(2000)}]{2000AJ....119.2757V}
{van Zee}, L. 2000, \aj, 119, 2757, \dodoi{10.1086/301378}

\bibitem[{{Villforth} {et~al.}(2014){Villforth}, {Hamann}, {Rosario}, {Santini}, {McGrath}, {van der Wel}, {Chang}, {Guo}, {Dahlen}, {Bell}, {Conselice}, {Croton}, {Dekel}, {Faber}, {Grogin}, {Hamilton}, {Hopkins}, {Juneau}, {Kartaltepe}, {Kocevski}, {Koekemoer}, {Koo}, {Lotz}, {McIntosh}, {Mozena}, {Somerville}, \& {Wild}}]{2014MNRAS.439.3342V}
{Villforth}, C., {Hamann}, F., {Rosario}, D.~J., {et~al.} 2014, \mnras, 439, 3342, \dodoi{10.1093/mnras/stu173}

\bibitem[{{Villforth} {et~al.}(2017){Villforth}, {Hamilton}, {Pawlik}, {Hewlett}, {Rowlands}, {Herbst}, {Shankar}, {Fontana}, {Hamann}, {Koekemoer}, {Pforr}, {Trump}, \& {Wuyts}}]{2017MNRAS.466..812V}
{Villforth}, C., {Hamilton}, T., {Pawlik}, M.~M., {et~al.} 2017, \mnras, 466, 812, \dodoi{10.1093/mnras/stw3037}

\bibitem[{{Walmsley} {et~al.}(2023){Walmsley}, {G{\'e}ron}, {Kruk}, {Scaife}, {Lintott}, {Masters}, {Dawson}, {Dickinson}, {Fortson}, {Garland}, {Mantha}, {O'Ryan}, {Popp}, {Simmons}, {Baeten}, \& {Macmillan}}]{2023MNRAS.526.4768W}
{Walmsley}, M., {G{\'e}ron}, T., {Kruk}, S., {et~al.} 2023, \mnras, 526, 4768, \dodoi{10.1093/mnras/stad2919}

\bibitem[{{Ward} {et~al.}(2022){Ward}, {Gezari}, {Nugent}, {Bellm}, {Dekany}, {Drake}, {Duev}, {Graham}, {Kasliwal}, {Kool}, {Masci}, \& {Riddle}}]{2022ApJ...936..104W}
{Ward}, C., {Gezari}, S., {Nugent}, P., {et~al.} 2022, \apj, 936, 104, \dodoi{10.3847/1538-4357/ac8666}

\bibitem[{{Wright} {et~al.}(2010){Wright}, {Eisenhardt}, {Mainzer}, {Ressler}, {Cutri}, {Jarrett}, {Kirkpatrick}, {Padgett}, {McMillan}, {Skrutskie}, {Stanford}, {Cohen}, {Walker}, {Mather}, {Leisawitz}, {Gautier}, {McLean}, {Benford}, {Lonsdale}, {Blain}, {Mendez}, {Irace}, {Duval}, {Liu}, {Royer}, {Heinrichsen}, {Howard}, {Shannon}, {Kendall}, {Walsh}, {Larsen}, {Cardon}, {Schick}, {Schwalm}, {Abid}, {Fabinsky}, {Naes}, \& {Tsai}}]{2010AJ....140.1868W}
{Wright}, E.~L., {Eisenhardt}, P. R.~M., {Mainzer}, A.~K., {et~al.} 2010, \aj, 140, 1868, \dodoi{10.1088/0004-6256/140/6/1868}

\bibitem[{{Wuyts} {et~al.}(2011){Wuyts}, {F{\"o}rster Schreiber}, {van der Wel}, {Magnelli}, {Guo}, {Genzel}, {Lutz}, {Aussel}, {Barro}, {Berta}, {Cava}, {Graci{\'a}-Carpio}, {Hathi}, {Huang}, {Kocevski}, {Koekemoer}, {Lee}, {Le Floc'h}, {McGrath}, {Nordon}, {Popesso}, {Pozzi}, {Riguccini}, {Rodighiero}, {Saintonge}, \& {Tacconi}}]{2011ApJ...742...96W}
{Wuyts}, S., {F{\"o}rster Schreiber}, N.~M., {van der Wel}, A., {et~al.} 2011, \apj, 742, 96, \dodoi{10.1088/0004-637X/742/2/96}

\bibitem[{{Yao} {et~al.}(2023){Yao}, {Song}, {Kong}, {Fang}, {Zhang}, \& {Chen}}]{2023ApJ...954..113Y}
{Yao}, Y., {Song}, J., {Kong}, X., {et~al.} 2023, \apj, 954, 113, \dodoi{10.3847/1538-4357/ace7b5}

\bibitem[{{Zhang} {et~al.}(2012){Zhang}, {Hunter}, {Elmegreen}, {Gao}, \& {Schruba}}]{2012AJ....143...47Z}
{Zhang}, H.-X., {Hunter}, D.~A., {Elmegreen}, B.~G., {Gao}, Y., \& {Schruba}, A. 2012, \aj, 143, 47, \dodoi{10.1088/0004-6256/143/2/47}

\bibitem[{{Zhao} {et~al.}(2022){Zhao}, {Li}, {Shangguan}, {Zhuang}, \& {Ho}}]{2022ApJ...925...70Z}
{Zhao}, Y., {Li}, Y.~A., {Shangguan}, J., {Zhuang}, M.-Y., \& {Ho}, L.~C. 2022, \apj, 925, 70, \dodoi{10.3847/1538-4357/ac375b}

\bibitem[{{Zheng} {et~al.}(2017){Zheng}, {Wang}, {Ge}, {Mao}, {Li}, {Li}, {Mo}, {Goddard}, {Bundy}, {Li}, {Nair}, {Lin}, {Long}, {Riffel}, {Thomas}, {Masters}, {Bizyaev}, {Brownstein}, {Zhang}, {Law}, {Drory}, {Roman Lopes}, \& {Malanushenko}}]{2017MNRAS.465.4572Z}
{Zheng}, Z., {Wang}, H., {Ge}, J., {et~al.} 2017, \mnras, 465, 4572, \dodoi{10.1093/mnras/stw3030}

\bibitem[{{Zhong} {et~al.}(2022){Zhong}, {Inoue}, {Yamanaka}, \& {Yamada}}]{2022ApJ...925..157Z}
{Zhong}, Y., {Inoue}, A.~K., {Yamanaka}, S., \& {Yamada}, T. 2022, \apj, 925, 157, \dodoi{10.3847/1538-4357/ac3edb}

\end{thebibliography}
\bibliographystyle{aasjournal}
\end{document}